\newif\ifseparatesupp
\newif\ifseparatehead
\separatesuppfalse 
\separateheadtrue 

\documentclass[acmtog, authorversion, review=false, anonymous=false]{acmart}

\AtBeginDocument{%
	}

\usepackage[type={CC},modifier={by},version={4.0}]{doclicense}

\setcopyright{cc}
\setcctype{by}

\acmJournal{TOG}
\acmYear{} \acmVolume{1} \acmNumber{1} \acmArticle{}
\acmMonth{1} \acmPrice{} \acmDOI{10.1145/3796224}

\usepackage{siunitx}
\usepackage{bm}
\usepackage{xcolor, soul}
\usepackage{multirow, makecell}
\usepackage{pifont}%
\usepackage{tabularx}
\usepackage{microtype}
\usepackage{mdframed}

\usepackage[hang,flushmargin]{footmisc}
\usepackage{wasysym}

\newcolumntype{Y}{>{\centering\arraybackslash}X}
\newcommand{\cmark}{\ding{51}}%

\newcommand{\insertfig}{\includegraphics[width=\linewidth]{images/teaser_wide.pdf}}
\makeatletter
\apptocmd{\@maketitle}{\centering\insertfig}{}{}%
\makeatother



\renewcommand{\tableautorefname}{Table}

\definecolor{red}{rgb}{0.85,0,0.2}
\definecolor{green}{rgb}{0,0.65,0.2}
\definecolor{orange}{rgb}{0.96,0.43,0}

\definecolor{inputcol}{rgb}{0.88,0.52,0.3}
\definecolor{outputcol}{rgb}{0.85,0,0}
\definecolor{urlcol}{RGB}{163, 78, 45}

\newcommand{\sms}{\,}
\newcommand{\obj}[1]{\textsl{\textsf{#1}}}

\newcommand{\best}[1]{\textbf{#1}}
\newcommand{\secbest}[1]{\fontseries{sb}\selectfont{\textit{#1}}}

\newcommand{\mone}{$\text{C}_\text{g}$}
\newcommand{\mtwo}{$\text{C}_\text{s}$}
\newcommand{\mthree}{P}
\newcommand{\mfour}{$\text{P}_{\text{e}}$}

\renewcommand{\bf}[1]{\textbf{#1}}

\begin{document}
	\title{MAROON: A Dataset for the Joint Characterization of Near-Field High-Resolution Radio-Frequency and Optical Depth Imaging Techniques}
	
	\author{Vanessa Wirth}
	\email{vanessa.wirth@fau.de}
	\orcid{1234-5678-9012}
	\affiliation{%
		\institution{Friedrich-Alexander-Universit{\"a}t Erlangen-N{\"u}rnberg}
		\city{Erlangen}
		\country{Germany}
	}
	
	\author{Johanna Br{\"a}unig}
	\email{johanna.braeunig@fau.de}
	\affiliation{%
		\institution{Friedrich-Alexander-Universit{\"a}t Erlangen-N{\"u}rnberg}
		\city{Erlangen}
		\country{Germany}}
	
	\author{Nikolai Hofmann}
	\email{nikolai.hofmann@fau.de}
	\affiliation{%
		\institution{Friedrich-Alexander-Universit{\"a}t Erlangen-N{\"u}rnberg}
		\city{Erlangen}
		\country{Germany}}
	
	\author{Martin Vossiek}
	\email{martin.vossiek@fau.de}
	\affiliation{%
		\institution{Friedrich-Alexander-Universit{\"a}t Erlangen-N{\"u}rnberg}
		\city{Erlangen}
		\country{Germany}}
	
	\author{Tim Weyrich}
	\email{tim.weyrich@fau.de}
	\authornotemark[1]
	\affiliation{%
		\institution{Friedrich-Alexander-Universit{\"a}t Erlangen-N{\"u}rnberg}
		\city{Erlangen}
		\country{Germany}}
	\affiliation{%
		\institution{University College London}
		\city{London}
		\country{UK}}
	
	\author{Marc Stamminger}
	\email{marc.stamminger@fau.de}
	\authornotemark[1]
	\affiliation{%
		\institution{Friedrich-Alexander-Universit{\"a}t Erlangen-N{\"u}rnberg}
		\city{Erlangen}
		\country{Germany}
	}

	\renewcommand{\shortauthors}{Vanessa Wirth, Johanna Bräunig, Martin Vossiek, Tim Weyrich, and Marc Stamminger}
	
	\authorsaddresses{}

	\begin{abstract}
		Utilizing the complementary strengths of wavelength-specific range or depth sensors is crucial for robust computer-assisted tasks such as autonomous driving.
		Despite this, there is still little research done at the intersection of optical depth sensors and radars operating close range, where the target is decimeters away from the sensors.
		Together with a growing interest in high-resolution imaging radars operating in the near field, the question arises how these sensors behave in comparison to their traditional optical counterparts.
		In this work, we take on the unique challenge of jointly characterizing depth imagers from both, the optical and radio-frequency domain using a multimodal spatial calibration.
		We collect data from four depth imagers, with three optical sensors of varying operation principle and an imaging radar.
		We provide a comprehensive evaluation of their depth measurements with respect to distinct object materials, geometries, and object-to-sensor distances.
		Specifically, we reveal scattering effects of partially transmissive materials and investigate the response of radio-frequency signals.
		All object measurements are made public in form of a multimodal dataset, called MAROON, which can be accessed at: {\color{urlcol}\url{https://vwirth.github.io/maroon}}.
	\end{abstract}


	
	\begin{teaserfigure}
		\centering
		\setlength{\unitlength}{1cm}
		\begin{picture}(20, 5)(0,0)
		\includegraphics[width=\linewidth]{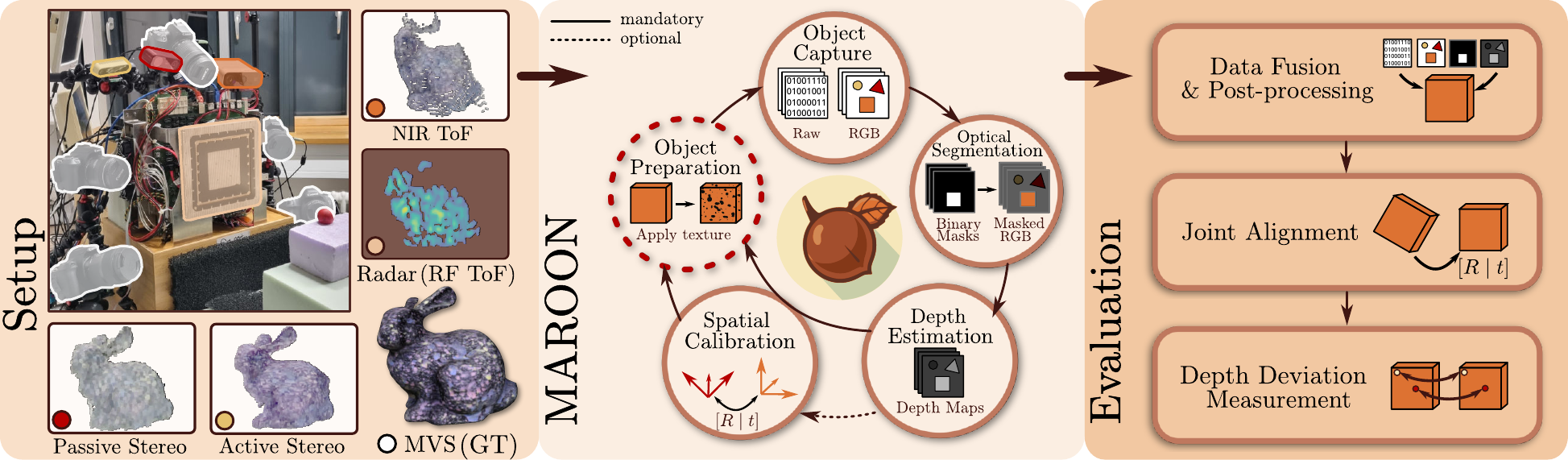}
		\put(-17.75,2.9){\fontsize{9}{9}\selectfont\rotatebox{90}{(Sec.~\ref{sec:setup})}}
		\put(-11.60,2.9){\fontsize{9}{9}\selectfont\rotatebox{90}{(Sec.~\ref{sec:dataset})}}
		\put(-5.4,2.9){\fontsize{9}{9}\selectfont\rotatebox{90}{(Sec.~\ref{sec:evaluation})}}
		\end{picture}
		\caption{\label{fig:setup}%
			Recent developments for near-field imaging radars enabled the acquisition of high-resolution depth images, and the sensors are now increasingly gaining attention as complementary modalities to optical depth sensing. Direct comparisons from our MAROON dataset, however, highlight significant differences between radar and optical reconstructions. This work employs the collected multimodal data of four depth imagers, depicted on the \textit{left}, to systematically characterize these fundamental differences together with  sensor-specific findings in a joint evaluation framework.%
		}
		\Description[Three separate panels depict the sensor setup (left), a sequence chart for data acquisition (center), and a sequence chart for data evaluation (right)]{The left panel depicts a photo of the sensor setup, with corresponding sensor reconstructions arranged around this photo; they show passive stereo, active stereo, near-infrared Time of Flight and radio-frequency Time of Flight reconstructions. 
			The center panel illustrates a sequence chart that shows the steps necessary for data acquisition and post-processing -- all of which are further described in Section 5.
			The right panel illustrates a sequence of three steps for data evaluation: data fusion and post-processing, joint alignment of the data, and the measurement of depth deviation from the jointly aligned data.}
	\end{teaserfigure}

	\maketitle

	{\let\thefootnote\relax 
		\footnotetext{* Both authors contributed equally to this research.}
	} %

	\section{Introduction}
	
	Real-world computer-assisted tasks, for instance in robotics and tracking applications, frequently require the immediate assessment of spatial information to accurately reason about the environment at a specific point in time, which has led to the development of several single-view range and depth sensors.
	For autonomous driving, it has been shown that utilizing multimodal depth sensing techniques from both the optical (lidar) and radio-frequency (radar) domain can lead to superior performance and robustness in computer-assisted tasks~\cite{velasco_2020}.
	Due to its environment, the autonomous driving industry has traditionally concentrated on far-field range sensing, with an unambiguous range of several meters and beyond.
	As recent high-resolution radio-frequency technologies utilize the concept of \textit{radar imaging} to produce 3D information in form of a depth map\,---\,similar to optical depth or RGB-D cameras\,---\,they also become more popular in close range, where the target of interest is up to a few decimeters away from the sensor;
	however, a comprehensive and detailed characterization of these radar imaging technologies, which frequently operate in the radar's near field, is yet to be realized.

	As part of this work, we devised a dataset \bf{MAROON} (\bf{M}ultimodal \bf{A}ligned \bf{R}adio and \bf{O}ptical frequency \bf{O}bject Reconstructions in the \bf{N}ear Field) (cf.\ \autoref{sec:maroon}) that enables studying of different sensor modalities in direct comparison. As is immediately visible in \autoref{fig:setup} (\textit{left}), the reconstructions of near-field imaging radars appear fundamentally different in comparison to their well-researched counterparts in the optical domain.

	A key advantage of radar is that it is insensitive to environmental light and can penetrate, for instance, fabric and dust.
	Following the success of Google's project Soli~\cite{lien_2016} for gesture sensing, radars were utilized in close range for the detection of vital signs~\cite{vilesov_2022}, activity recognition~\cite{braeunig_2023_2}, people tracking~\cite{fusion_zewge_2019} and human body reconstruction~\cite{fusion_chen_2022, fusion_chen_2023}.
	With the growing trend towards larger antenna apertures to achieve high-resolution imaging~\cite{automotive_schwarz_2022, fusion_chen_2023}, radars will more frequently operate in the near field, as determined by the Fraunhofer boundary condition~\cite{selvan_2017}. 
	At the same time, characteristics of near-field radar are generally under-researched.
	\begin{figure}[t]
		\centering
		\includegraphics[width=0.95\linewidth]{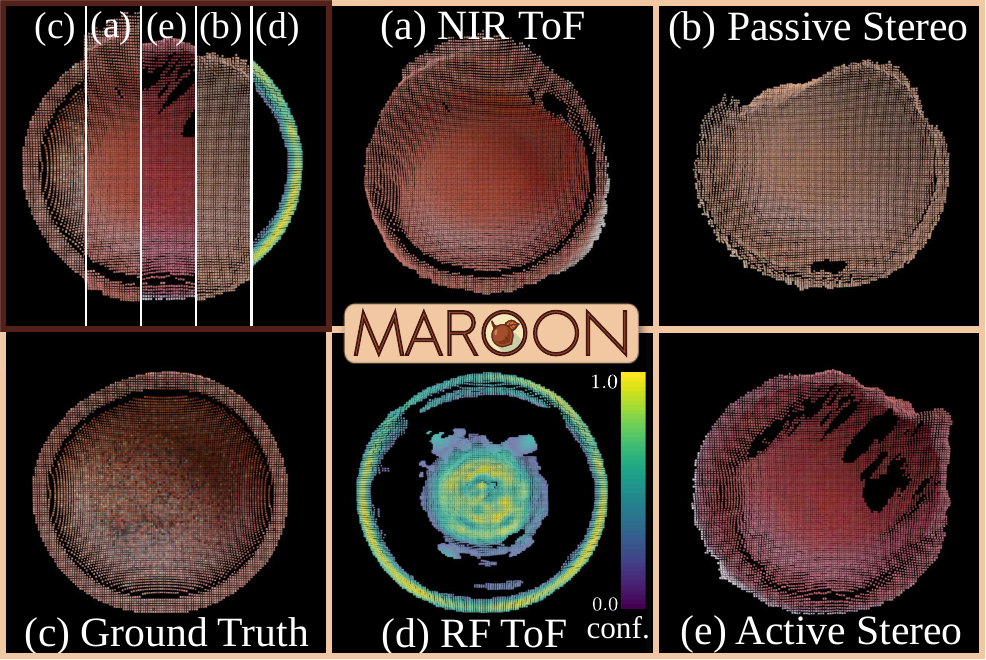}
		\caption{\label{fig:maroon}%
			Example data of the \obj{Plunger} object from the MAROON dataset. In the \textit{upper left}, all reconstructions are spatially aligned with respect to the RF ToF coordinate system. The RF ToF colorscale encodes the normalized reconstruction confidence (cf. \autoref{sec:mimo_fscw}).\vspace{-6ex}}
		\Description[Six subfigures display point clouds of the sensor reconstructions, the ground truth, and the fused sensor and ground-truth data in a joint coordinate system]{Six subfigures of point clouds are displayed, four of which refer to the sensor reconstructions, where the dense optical point clouds are colored using the corresponding RGB data. The radio-frequency time-of-flight point cloud only displays a fraction of the ground-truth surface and is color-coded by reconstruction confidence.}
	\end{figure}
	
	Drawing on prior research about wavelength-specific strengths and weaknesses, this paper addresses the unique challenge of characterizing various optical depth-imaging techniques alongside a high-resolution multiple-input multiple-output (MIMO) imaging radar in the near field.
	The latter is interchangeably referred to as a radio-frequency (RF) Time-of-Flight (ToF) sensor.
	To this end, we mutually calibrated sensors of four different depth sensing technologies, that is active and passive stereo, near-infrared (NIR) amplitude-modulated continuous wave (AMCW) ToF, and RF frequency-stepped continuous wave (FSCW) ToF in the millimeter-wave range.

	There is a notable lack of multimodal datasets suitable for close-range applications, and, to our knowledge, this work is the first to incorporate imaging radars in this research area. With this in mind, we captured the MAROON dataset of various household objects and construction materials, of which example data is shown in \autoref{fig:maroon}.
	Utilizing a high-resolution MIMO imaging radar, with a spatial resolution currently far beyond prevalent RF imaging sensors, we captured this dataset with a multitude of key objectives:
	\begin{enumerate}
		\item To evaluate sensor-specific reconstructions, considering various object materials, geometries, and distances to the sensors.
		\item To establish a public data base for multimodal reconstruction research in close-range applications, bridging the radio-frequency and optical domains.
		\item To characterize the under-researched effects of millimeter waves in near-field imaging radars, e.g. object materials in the radio-frequency domain, akin to studies on bi-directional reflectance distribution functions (BRDFs) in optics. 
		\item To improve radio-frequency signal simulations by supplying raw data for modeling high- and lower-resolution radar architectures and comparing synthetic signals with real measurements and a ground truth.
	\end{enumerate}		
	
	Together with the dataset, we developed a joint sensor evaluation framework that measures reconstruction differences between sensors and a ground truth using different metrics, providing supplementary visualizations tailored to identify sensor-specific trends across multiple objects.
	By analyzing these trends, we identified ToF scattering effects in partially transmissive materials and examined RF ToF reconstructions, which are typically less complete than those from optical sensors.
	
	Moreover, we utilize the multimodal data of MAROON in two example applications:
	first, we show that the dataset serves as foundation for addressing inverse rendering problems in the radio-frequency domain.
	Taking up on concurrent work~\cite{hofmann_2025}, we determine object-specific material properties, which are crucial in high-fidelity radar simulation. 	
	Second, the variety of challenging objects in our dataset serves as a benchmark for developing novel multimodal reconstruction algorithms, as demonstrated in~\cite{wirth_2025}.
	We extend this benchmark by additional experiments, varying the sensor configuration.

	\noindent To summarize, our contributions include:
	\begin{itemize}
		\item A novel multimodal dataset, MAROON, comprising common objects in the near field, captured using a jointly calibrated setup of three optical depth sensors, a high-resolution imaging radar, and high-quality multi-view reconstruction for ground-truth geometry. We release this dataset alongside the raw radar measurements to facilitate exploration of various signal reconstruction and filtering techniques: {\color{urlcol}\url{https://vwirth.github.io/maroon}}.
		
		\item A detailed analysis of trends and sensor-specific effects emerging from that dataset. This includes aspects of different object materials, geometries, and distances to the sensors, signal response and reconstruction quality of imaging radars, as well as ToF scattering effects of partially transmissive materials.
		\item  Two applications of the dataset: inverse rendering for material characterization and high-speed multimodal reconstruction.

	\end{itemize}

	\section{Related Work}
	While a considerable amount of literature exists on optical and RF depth sensors in isolation, no directly related work on the joint characterization of these two domains has been identified. 
	Instead, the first two sections comprise an overview of existing research about sensor characteristics self-contained within a single frequency domain.
	We further address the sensor fusion of optical and RF sensors, since in that research direction the complementary strengths of the sensors are utilized as well.

	\subsection{Optical Depth Sensing}
	Depth cameras %
	have been characterized with respect to a number of different aspects, and related work can be broadly classified into three categories: the sensor technologies, the capture environments, and the methods of comparison used to evaluate their performance.
	
	Considering the sensor technologies, metrological research has been conducted in terms of optical ToF~\cite{optics_xiong_2017, zanuttigh_2016} and active stereo~\cite{optics_giancola_2018, optics_wang_2021}.
	Furthermore, working principles of passive stereo sensors have been widely addressed in computer vision algorithms~\cite{szeliski_2022}.
	Similar to our work, Chiu et al.~\shortcite{optics_chiu_2019} and Halmetschlager-Funek et al.~\shortcite{optics_halmetschlager_2019} jointly characterize ToF and active stereo.
	
	With respect to the capture environments, related work examined the effects of object material~\cite{optics_xiong_2017, optics_halmetschlager_2019, optics_giancola_2018, optics_hansard_2012}, color~\cite{optics_xiong_2017, optics_giancola_2018, optics_hansard_2012}, texture~\cite{optics_xiong_2017, optics_hansard_2012} and distance to objects~\cite{optics_halmetschlager_2019}.
	Furthermore, environmental lighting conditions~\cite{optics_halmetschlager_2019} and multi-path effects~\cite{optics_giancola_2018} were investigated.
	Specifically for ToF sensors, Wu et al.~\shortcite{wu_2012} analyze multi-path effects originating from subsurface scattering and interreflections.
	
	Moreover, we discuss related frameworks for jointly characterizing sensors. 
	Halmetschlager-Funek et al.~\shortcite{optics_halmetschlager_2019} compare individually estimated depth values against manual measurements.
	Chiu et al.~\shortcite{optics_chiu_2019} and Giancola et al.~\shortcite{optics_giancola_2018} align the 3D data captured from sensors with real or synthetic ground-truth data, respectively. 
	Most similar to ours, Hansard et al.~\shortcite{optics_hansard_2012} analyze ToF and structured light sensors using a spatial calibration and investigated object material, color, geometry, and texture using ground-truth data obtained from a structured light scanner.

	\subsection{Radio Wave Propagation and Range Sensing}
	So far, radio-frequency depth sensors (radar) were characterized in isolation. A more fundamental research direction examines the propagation of electromagnetic waves, which is the basis for RF ToF sensors.
	The general RF propagation behavior under different materials and geometries is measured by a parameter known as the radar cross section (RCS)~\cite{radar_knott_2004}.
	The RCS approximates the returned ratio of a transmitted radio signal and was measured in relation to a variety of materials~\cite{radar_knott_2004, radar_semkin_2020}, as well as in the context of humans~\cite{radar_yoshana_2020, radar_marchetti_2018}.
	Orthogonal research of Zhadobov et al.~\shortcite{zhadobov_2011} investigates the interaction of radio waves and human skin with respect to electromagnetic, thermal and biological aspects.
	
	Moreover, studies of individual radar technologies have been conducted.
	\v{C}opi\v{c} Pucihar et al.~\shortcite{radar_pucihar_2022} evaluate the recognition of hand gestures using millimeter-wave radars in the presence of various materials.
	Wei et al.~\shortcite{radar_wei_2021} characterize imaging radars with respect to the geometry of metal objects in the context of security scanning.
	Furthermore, Bhutani et al.~\shortcite{radar_bhutani_2022} examine millimeter-wave radars at different frequencies, whereas Jha et al.~\shortcite{radar_jha_2018} analyze differences in their radiation between the near field and the far field.
	Sun et al.~\shortcite{radar_sun_2022} provide an overview of MIMO radars for autonomous driving, together with the characterization of their wave forms.
	Lastly, Ahmed~\shortcite{sherif_2021} presents millimeter-wave MIMO radar imaging systems in the context of security screening.
	To the best of our knowledge, no comprehensive characterization in conjunction with optical technologies has been done so far. 
	Additionally, the existing efforts have been limited in scope with regard to RF depth sensing in the near field.

	\subsection{Fusion of RF and Optical Sensors in Close Range}
	Knowledge about complementary strengths is important for both, sensor characterization and sensor fusion. While significant research efforts have been devoted to the field of autonomous driving\,---\,where radar sensors primarily operate in their respective far field\,---\,research on multimodal sensor fusion in close range is very limited and mostly focused on capturing humans.
	
	Zewge et al.~\shortcite{fusion_zewge_2019} perform people tracking with a $4 \times 3$ MIMO radar and an active stereo camera.
	Similarly, Lee et al.~\shortcite{lee_2023} propose a method for human pose estimation, which utilizes the data acquired from two $4 \times 3$ MIMO radars synchronized with a monocular RGB camera.
	Both works do not utilize radar imaging methods due to the limited resolution.
	More similar to ours,
	Chen et al.~\shortcite{fusion_chen_2023} use a high resolution $48 \times 48$ MIMO radar and an RGB camera for human body reconstruction.
	
	Furthermore, we address related datasets.
	Lim et al.~\shortcite{fusion_teck-yian_2021} introduce RaDICaL, an indoor and outdoor dataset of multiple people and objects, captured with a $4 \times 3$ MIMO imaging radar and an active stereo camera.
	In the context of human body reconstruction, Chen et al.~\shortcite{fusion_chen_2022}  propose the mmBody benchmark that was captured with a $48 \times 48$ MIMO radar and an RGB camera.

	\section{Preliminaries}
	
	As different research communities partially differ in their terminology, this paper pursues a unified terminology, summarized in the table below and used in the remainder of this paper.
	
\begin{figure}[htbp]
		\centering
		\includegraphics[width=\linewidth]{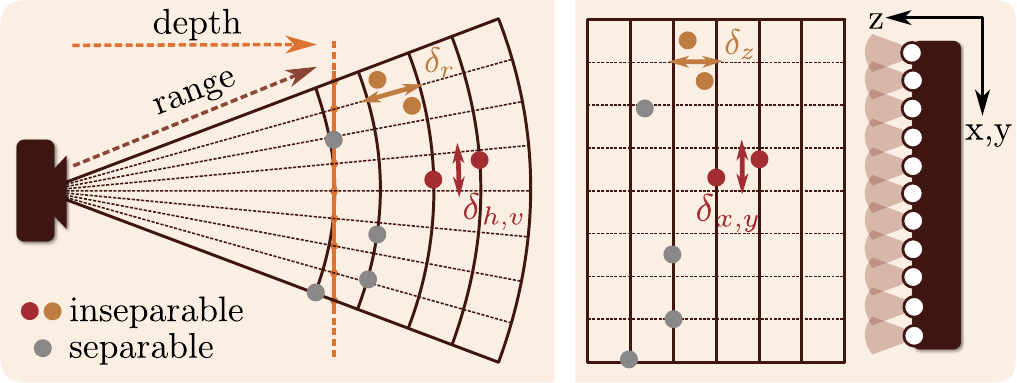}
		\caption{\label{fig:range_depth}%
			Visualization of the effects caused by limited spatial resolution for multiple point targets. Optical sensors (\textit{left}) %
			divide spatial resolution into range resolution, $\delta_r$, and perspective pixel resolution, $\delta_{h,v}$. Contrary to that, near-field imaging radars (\textit{right}) refer to range $\delta_z$ and cross-range $\delta_{x,y}$ resolution. We assume $\delta_z \approx \delta_r$ and $\delta_{x,y} \approx \delta_{h,v}$ for the sensor center, yet emphasize the conceptual difference between range and depth.
		}
		\Description[Two side-by-side panels, where spatial resolution variables are displayed as the extents of 2D lines, which are aligned with the cone-shaped perspective camera frustum (left) and the grid-shaped orthogonal frustum of an imaging radar (right)]{Two side-by-side panels display separate and inseparate point targets within the  cone-shaped perspective view frustum of a camera (left) and the grid-shaped orthogonal view frustum of an imaging radar (right). The limited spatial resolution, which classifies some of the visualized point targets as inseparable, is visualized by line extents that lie within the 2D line segments of the respective sensor frustums. While range is illustrated by an arrow that follows perspective in the optical camera frustum, the depth arrow aligns with the orthogonal axis to the line of sight.} 
	\end{figure}

	\mdfdefinestyle{mystyle}{leftmargin=0.0cm,rightmargin=0.0cm,
		innerleftmargin=0.1cm,
		innerrightmargin=0.1cm,
		innertopmargin=0.1cm,
		innerbottommargin=0.1cm}
	\noindent\begin{mdframed}[style=mystyle]%
		\setlength{\parindent}{0pt}%
		\bf{Depth Imager.} A sensor that, directly or indirectly, captures a depth image  $\boldsymbol{D}$ of resolution $W\times H$, where each pixel $(u,v)$ contains a depth value $d$ measured along the axis perpendicular to the image plane. 
		The depth may be indirectly measured from range and pixel position.
		We show the difference between range and depth in \autoref{fig:range_depth}.
		
		\bf{Transmitter and Receiver.} Optical receivers are small cells of image sensors, with a direct mapping to pixels. 
		Transmitters are commonly LEDs or projectors.
		RF sensors have transmitting (TX) and receiving (RX) antennas.
	\end{mdframed}
	\noindent\begin{mdframed}[style=mystyle]%
		\setlength{\parindent}{0pt}%
		\bf{Sensor.} Describes all physical parts required for depth sensing and their spatial arrangement.
		Optical sensors typically consist of one or two cameras. 
		Active sensors contain an additional illumination unit.
		RF sensors usually have one or more antenna arrays in different arrangements.
		
		\bf{Depth Image Resolution $\mathbf{W \times H}$.} The number of depth samples computed from the incoming signal.
		In cameras, the depth samples are directly computed for each pixel, i.e., each receiver.	
		MIMO imaging radars apply signal post-processing to compute depth from the signal diversity at different receiver positions.

		Hence, the image resolution is not directly affected by the number of receivers but by the signal processing parameters such as the voxel density (cf. \autoref{sec:mimo_fscw}).
		While, in theory, the depth image resolution can be indefinitely high, in practice it is limited by the spatial resolution.
		
		\bf{Spatial Resolution $\boldsymbol{\delta}$.} 
		We refer to spatial resolution as the minimum distance between two points in space that can be resolved from the received signal.
		Lower spatial resolution means higher minimum resolvable distance, so more incorrect measurements are made when points become inseparable, as seen in \autoref{fig:range_depth}.
		Spatial resolution is a theoretical measure, and external factors such as the sensor design can affect its \textit{effective} resolution.
		The resolution of near-field MIMO imaging radars is defined with respect to the three orthogonal axes $x,y,z$, where $z$ refers to the depth and $x, y$ are parallel to the antenna aperture.
		Sharing the same terminology as far-field RF ToF sensors that measure range, resolution is divided into range $\delta_z$ and cross-range~\cite{sherif_2014} $\delta_{x,y}$ %
		resolution, respectively.
		Optical sensors define either depth or range resolution, where the latter is denoted as $\delta_r$.
		Due to perspective, depth differs from range; within this work, we consider resolution at the sensor center, where $\delta_z \approx \delta_r$ (cf. \autoref{fig:range_depth}).
		Furthermore, optical sensors refer to pixel resolution along the perspective horizontal ($\delta_{h}$) and vertical ($\delta_v$) axes, respectively, of which we assume $\delta_{h,v}\approx \delta_{x,y}$ holds in the center.	
		
	\end{mdframed}

	\begin{figure*}[htbp]
		\centering
		\setlength{\unitlength}{1cm}
		\begin{picture}(14.03, 4)(0,0)
		\centering
		\includegraphics[width=0.8\linewidth]{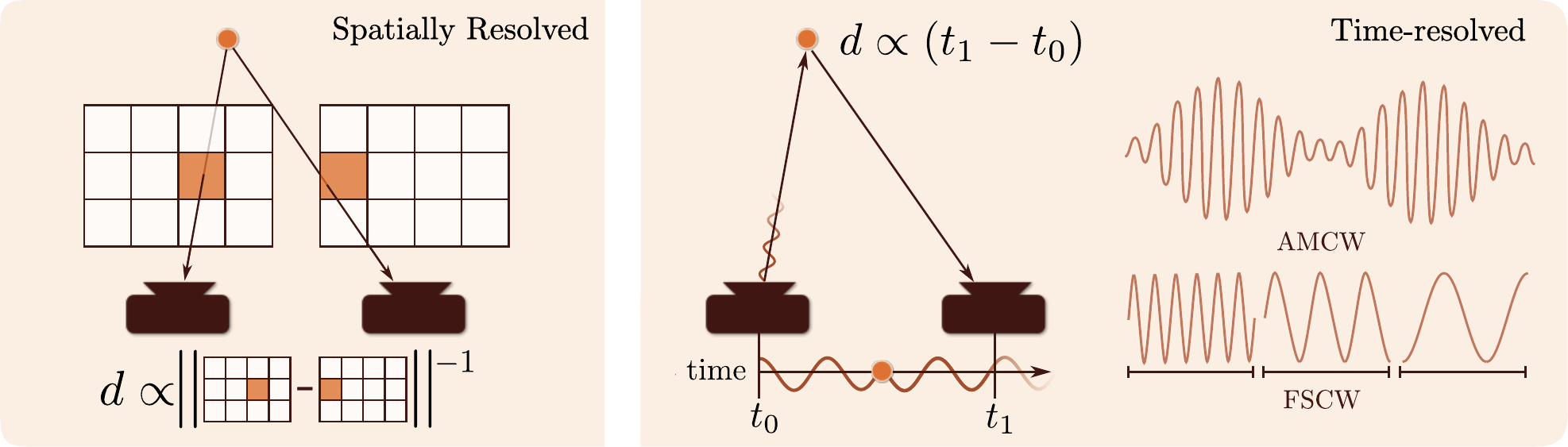}
		\put(-10.55,3.4){\fontsize{8}{8}\selectfont(\autoref{sec:spatially_resolved})}
		\put(-1.95,3.4){\fontsize{8}{8}\selectfont(\autoref{sec:time_resolved})}
		\end{picture}
		\caption{\label{fig:depth_sensing}%
			Overview of the two depth sensing categories considered in this work. Spatially resolved methods compute the depth from disparity in the pixel positions. Time-resolved methods measure the depth through the round-trip propagation time of the received continuous wave (CW) signal. The types of wave forms utilized in our experiments are amplitude-modulated continuous wave (AMCW) and frequency-stepped continuous wave (FSCW).}
		\Description{The left panel for spatially resolved methods shows how the position of a 3D object shifts across pixel grids when seen from two different sensor viewpoints, with arrows linking the object to its projected pixel coordinates. The difference between these pixel coordinates describes pixel disparity, which is inversely proportional to depth, as further illustrated. For time-resolved methods on the right, a signal traveling to a 3D object and back is shown along a time axis with two timesteps for the time of emission and reception, respectively. The depth is proportional to the difference between these timesteps. Furthermore, the two signal modulations are illustrated by wavy sinusodal signal shapes. For AMCW, these shapes smoothly vary in their height, i.e., their amplitude. For FSCW, three separate signal shapes from left to right are visualized, where the sinusoidal pattern of each repeats at a constant amplitude and frequency. Between these shapes from left to right, the frequency becomes lower, i.e., the sinusoidal pattern is scaled to contain less repetitions over the same period of time.}
	\end{figure*}

	\section{Working Principles of Depth Imagers}
	In order to gain insight into the fundamental differences between optical and RF sensors, the first section characterizes wavelength-specific signal propagation. 
	This is followed by an outline of the hardware design choices that are made for optical and RF depth imagers.
	After this, the working principles of the sensor technologies that are used in our experiments are discussed.

	\subsection{Characterization of Wavelength} 
	\label{sec:wavelength_characterization}
	Depth imagers are susceptible to the received, and optionally transmitted, signal wavelength. 
	The wavelength affects both, the interaction of the signal with matter and the design of the sensor hardware and depth sensing algorithms.
	In this section we elaborate on both aspects, with a particular focus on the near-infrared light spectrum and the millimeter-wave (mmWave) radio-frequency spectrum.
	
	\paragraph{Signal Interaction} NIR signals have a wavelength in the nanometer range. 
	Given their high energy and strong interaction with matter, signal reflection or absorption is common, with scattering and non-diffractive phenomena dominating across most materials. Indirect effects on interactions with matter, therefore, often play a subordinate role, such that short propagation paths can be expected.
	Moreover, NIR light is pervasive in the environment, rendering optical technologies susceptible to external interference.
	
	As suggested by their name, the wavelength of mmWave signals is longer by comparison. 
	The low energy and reduced interaction with matter result in lower absorption and reflection, while there is a higher chance of a signal being transmitted through material. 
	Specifically, the penetration depth of millimeter waves through matter is dependent on material parameters, such as, the resistivity and permittivity. 
	For instance, the signals of security scanners can penetrate fabric but are primarily reflected on contact with metal objects~\cite{sherif_2021}.
	Furthermore, diffraction is more common with millimeter waves. 
	This allows waves to bend around objects. 
	Due to the aforementioned phenomena, the propagation paths of signals from active RF sensors are typically longer than of signals from optical sensors.
	Lastly, mmWave depth imagers operate with reduced external interference, as there are few natural microwave sources in the environment.
	
	\paragraph{Wavelength-specific Hardware} Due to the wavelength, the sensor design of RF sensors is inherently different from that of optical sensors.
	As stated by the general formulation of the Rayleigh criterion, the focus capacity and, hence, angular resolution $\omega$ of a sensor is limited by the signal wavelength $\lambda$ and the size of the sensor's aperture $L$~\cite{hasch_2012}:
	\begin{equation}
	\omega_{x,y} = 1.22 \frac{\lambda}{L_{x,y}}\;.
	\label{eq:rayleigh}
	\end{equation}
	Optical sensors utilize camera lenses to refract the received signal, which enable a precise focus onto nanometer-sized pixels and, at the same time, exhibit a high angular resolution.
	In the context of the mmWave domain, a camera analogue can be conceptualized as a single-input multiple-output (SIMO) radar, i.e., a sensor comprising a single transmitter and multiple receivers.
	As indicated by \autoref{eq:rayleigh}, mmWave sensors have a considerably lower angular resolution than optical sensors.
	Thus, high-resolution SIMO radars require comparably large antenna arrays with large lenses, which has proven to be impractical.
	Instead, high-resolution RF imaging sensors often are  synthetic aperture radars (SAR), which use digital beamforming to focus.
	They utilize the angle diversity from distinct transmitter and receiver positions, which form a virtual aperture of size $L$, to increase the angular resolution~\cite{bliss_2003} and require fewer antennas compared to SIMO systems.
	The majority of near-field SAR radars is implemented with MIMO arrays, that is, with multiple transmitters and receivers.

	\subsection{Depth Sensing Methods} 
	\label{sec:depth_sensing_categorization}
	In this section, we address the working principle of both optical and RF-based depth sensing methods used in our experiments.
	The content is organized in two categories: spatially resolved and time-resolved depth sensing, which are both depicted in \autoref{fig:depth_sensing}.
	
	\subsubsection{Spatially Resolved Depth Sensing}
	\label{sec:spatially_resolved}
	Spatially resolved depth imagers compute the point-wise depth from the respective pixel position in the image.
	In the following, we particularly address passive or active stereo(scopy) sensors.
	
	Passive stereo sensors commonly utilize two cameras with a known relative spatial position to identify surface points in their respective images, a process known as \textit{correspondence} or \textit{stereo matching}~\cite{szeliski_2022}.
	Given a correspondence pair of two pixels, the respective depth of this surface point is computed from their disparity.
	The quality of the correspondence matches affects the depth and accuracy of the results.
	Ambiguities in correspondence can arise due to textureless regions, poor lighting, motion or lens blur.
	Similarly, stereo matching can fail in terms of view-dependent effects or partial surface occlusions from one of the two receivers.

	Active stereo sensors assist correspondence finding with an illumination unit that projects a pattern onto the target, usually in the NIR range, captured by the two cameras. The signal-multiplexed~\cite{zanuttigh_2016} pattern 
	supports epipolar correspondence matching in addition to shading and texture cues, improving depth quality in textureless regions and low light. However, challenges include pattern distortions and signal oversaturation at the NIR receiver in bright conditions. Further details on spatially resolved depth sensors are provided in \ifseparatesupp Section~S11 \else \autoref{sec:spatially_resolved_extended}\fi~of the supplementary material.

	\subsubsection{Time-Resolved Sensors (Time-of-Flight)}
	\label{sec:time_resolved}
	Time-of-Flight is an active depth sensing method, in which depth is derived from the round-trip propagation time that it takes for a signal to be transmitted and received.
	The majority of ToF sensors utilized in the near field employ continuous wave (CW) signal modulations, which measure time based on the relative phase shift $\Delta \varphi$ between the transmitted and received signal.
	The depth is derived from the range $r$, which is measured by~\cite{zanuttigh_2016}:
	\begin{equation}
	\label{eq:tof}
	r = \text{c} \frac{\Delta \varphi}{4 \pi f}\;.
	\end{equation}
	The signal frequency is denoted as $f$, and $\text{c}$ is the speed of light in vacuum, which closely matches that of light in air.
	For further details about the operating principle, we refer to the supplementary \ifseparatesupp Section~S12\else \autoref{sec:time_resolved_extended}\fi.
	ToF technologies employ a simplified model for range sensing, which assumes that targets are weak scatterers~\cite{sherif_2014}, with each signal reflecting directly from the first target. As a result, these technologies are sensitive to multi-path interference. In \autoref{sec:discussion_scattering}, we identify partially transmissive materials as a major cause of such interference. 
	Furthermore, the unambiguous range, in which $\Delta \varphi$ can be correctly resolved, is limited to the periodicity of the sinusoidal CW signal.
	To extend this range, the carrier signal can be modulated over time. 
	Noting that various modulation schemes exist, e.g., frequency-modulated continuous wave (FMCW) modulation, we use ToF sensors with AMCW and FSCW signal modulations, which are illustrated in \autoref{fig:depth_sensing}. 
	Up next, we will discuss the operating principles of these depth sensing methods.
	
	\paragraph{NIR AMCW Time-of-Flight}
	\label{sec:nir_amcw}
	AMCW ToF algorithms modulate the amplitude $A$ of a carrier signal over time $t$ using a repetitive modulation signal $s_{\text{m}}$ such that the transmitted signal $s_{\text{t}}$ can be described as:
	\begin{equation}
	s_{\text{t}}(t) = \underbrace{s_{\text{m}}(t)}_{\mbox{\small\raisebox{0.5ex}{\smash{$A$}}}} \cdot \cos(2\pi tf + \phi_{\text{c}})\;.
	\end{equation}
	A constant phase offset is described by $\phi_{\text{c}}$.
	To extract the phase shift from the received signal, it is demodulated at receiving time to yield $m_{\text{r}}$. It is then cross-correlated with a so-called signal hypothesis $s_{\text{h}}$, which is commonly chosen as the currently transmitted signal $s_{\text{t}}$, such that the output of this correlation, $c_{\text{r}}$, describes the signal similarity from which the relative phase shift $x$ %
	is inferred~\cite{horaud_2016}:
	\begin{equation}
	c_{\text{r}}(x) = \int_{-T_{\text{m}}/2}^{+T_{\text{m}}/2} m_{\text{r}}(t) s_{\text{h}}(t - x) dt\;,
	\end{equation}
	where $T_{\text{m}}$ is the period of the modulation signal.
	Commonly, $s_{\text{h}}$ is chosen as the currently transmitted signal $s_{\text{t}}$ such that $c_{\text{r}}$ describes the signal similarity from which the relative phase shift to $m_{\text{r}}$ is inferred.
	Extracting this shift requires solving a multivariable equation system with parameters such as the received amplitude and external illumination.
	To achieve this, $c_{\text{r}}$ and, consequently, $m_{\text{r}}$ are commonly sampled at four points within $T_{\text{m}}$ (four-bucket-method)~\cite{optics_giancola_2018}.
	During the acquisition of those samples, AMCW ToF is affected by environmental changes, such as varying external NIR illumination and motion.
	Moreover, oversaturation of the NIR receiver may cause invalid signal responses.

	\paragraph{MIMO FSCW Time-of-Flight}
	\label{sec:mimo_fscw}
	FSCW ToF sensors model the frequency of the transmitting signal as a function of time.
	Given the frequency band $\text{b} = f_{\text{max}} - f_{\text{min}}$, they iteratively send $\text{N}_f$ signals of one frequency in steps of $\Delta f = \text{b} / (\text{N}_f-1)$~\cite{braeunig_2023_fsk}. 
	More specifically, the transmitted signal $s_{\text{t}}$ of one capture can be described as:
	\begin{align}
	s_{\text{t}}(t) &= A \cdot \cos ( 2\pi t f_{\text{m}} (n  )  + \phi_{\textit{c}} )\text{ with } n =  \left\lfloor t / \Delta t  \right\rfloor
	\label{eq:fscw_wave}
	\\
	f_{\text{m}}(n) &= f_{\text{min}} + \left( n \bmod{\text{N}_f} \right) \Delta f\;.
	\end{align}
	We denote the time window of one frequency step as $\Delta t$.
	Time-division multiplexing (TDM) avoids signal interference and facilitates the separation of the received signal into its originating transmitter and frequency components.
	SAR signal processing computes the depth $d$ and the pixel position $(u,v)$ from the phase shift and the angular diversity originating from multiple transmitting and receiving positions.
	For MIMO imaging radars, the state-of-the-art algorithm of \textit{backprojection}~(BP)~\cite{wolf_1969, sherif_2021} computes confidence values about a target's presence in 3D space.
	This is achieved on the basis of local feature distributions within a volume based on the integrated signal of each RX-TX antenna pair.
	Similar to the four-bucket-method for AMCW ToF, the BP algorithm computes a correlation between the received demodulated signal $m_{\text{r}}$ and a signal hypothesis $s_{\text{h}}$:
	\begin{equation}
	c_{\text{BP}}(\underbrace{x,y,z}_{\boldsymbol{p}}) = \sum_{n=1}^{\text{N}_f} \sum_{i=1}^{\text{N}_{\text{RX}}} \sum_{j=1}^{\text{N}_{\text{TX}}} m_{\text{r}}(f_{n}, \boldsymbol{r}_i, \boldsymbol{t}_j) s_{\text{h}}(f_{n}, \boldsymbol{r}_i, \boldsymbol{t}_j, \boldsymbol{p})\;.
	\label{eq:backproj}
	\end{equation}
	
	A hypothesis is made on the basis of the transmitted signal, which is assumed to reflect at a point $\boldsymbol{p}\in \mathbb{R}^3$ in the sensor coordinate system, commonly sampled from a voxel grid of size $N_v = N_x \times N_y \times N_z$.
	The demodulated received signal, $m_{\text{r}}$, varies in transmit frequency $f_n = f_{\text{m}}(n)$, transmitter position $\boldsymbol{t}_j  \in \mathbb{R}^3$, and receiver position $\boldsymbol{r}_i \in \mathbb{R}^3$.
	The numbers of transmitters and receivers are denoted as $\text{N}_{\text{TX}}$, and $\text{N}_{\text{RX}}$, respectively.
	Generally, hypotheses are made by assuming the signal propagation is following the Born approximation~\cite{sherif_2014}.
	The result of the above equation is a complex phasor $c_{\text{BP}}$, calculated from $m_{\text{r}}$ and $s_{\text{h}}$, which are analytic signals in complex notation. 
	To compute a 2D depth map from the 3D voxel grid, an orthogonal \textit{maximum (intensity) projection}~\cite{braeunig_2023_fsk} is performed for each pixel $(u,v) = (x,y)$ along the cross-depth axes of the voxel grid:
	\begin{equation}
	d(u,v) = \underset{z}{\text{argmax}} \lVert  c_{\text{BP}}(x,y,z) \rVert_2 = \underset{z}{\text{argmax}}\ \kappa(x,y,z)\;.
	\label{eq:maxproj}
	\end{equation}
	The letter $\kappa$ denotes the so-called \textit{confidence} of a target's presence, as visualized in \autoref{fig:maroon}. 
	Besides projection, the confidence values are used as thresholds for depth filtering, that is, to distinguish target depth from sidelobes and background noise.
	As $\kappa$ directly relates to $c_{\text{BP}}$, it depends on both, the received phase and amplitude.
	Besides the depth, reasons for varying amplitude and phase over different object materials and geometries are manifold, and further insights will be given in \autoref{sec:discussion_radar_signal}.
	As a result, it is challenging to generalize the depth filtering process for unknown objects.
	
	Similar to NIR AMCW ToF, a MIMO FSCW ToF sensor is sensitive to environmental changes while capturing multiple signal samples.
	
	\newcommand\undertheset[2]{\ensuremath{\mathop{\kern\z\mbox{#2}}\limits_{\mbox{\scriptsize #1}}}}
	\newcommand{\textunderset}[2]{\begin{tabular}[t]{@{}c@{}}#2\\[-1em]\small#1\end{tabular}}

	\begin{table*}[]
		\centering
		\caption{\label{table:sensors}%
			Overview of the sensors and their parameters used in our experiments. Rows with \textbf{*} indicate derived information not directly given by the manufacturer. Depth processing times were computed on a system with an NVIDIA GeForce RTX 3080 graphics card (10GB VRAM) and an Intel Xeon W-1390P (3.50 GHz) processor. Note that due to its fundamentally different operating principle, modeling the field of view of the QAR50 similar to a camera is a very simplified approximation, and we refer to the supplementary \ifseparatesupp Section~S13.1 \else \autoref{sec:radar_field_of_view} \fi~for further details. $\delta_x$ and $\delta_y$ of camera-based systems is approximately determined from the per-pixel field of view. Spatial resolution formulae are provided in \ifseparatesupp Section~S11 and Section~S12 \else \autoref{sec:spatially_resolved_extended} and \autoref{sec:time_resolved_extended} \fi~of the supplementary material. Due to missing data for the Azure Kinect, $\delta_z$ is assumed to be theoretically higher than the depth accuracy given in~\cite{kinect_bamji_2018}.}
		\begin{tabular}{@{}llcccc@{}} 
			\toprule
			\multicolumn{2}{l}{} & \multirowcell{2}{\textbf{ZED X Mini (2.2~mm)}  \\ \cite{stereolabs_zed} }&  \multirowcell{2}{\textbf{Realsense D435i} \\\cite{intel_rs}} & \multirowcell{2}{\textbf{Azure Kinect} \\\cite{microsoft_kinect}} & \multirowcell{2}{\textbf{QAR50 (Submodule)}\\\cite{rs_qar} } \\
			\multicolumn{2}{l}{} & & & &  \\
			\midrule
			\multicolumn{2}{l}{\textbf{Manufacturer}} & Stereolabs & Intel & Microsoft & Rohde \& Schwarz \\
			\midrule
			\multicolumn{2}{l}{\textbf{Depth Sensing Technology}} & Passive Stereoscopy &  Active Stereoscopy & Time-of-Flight (NIR) & Time-of-Flight (RF) \\
			\midrule
			\multicolumn{2}{l}{\textbf{Arrangement}} & 
			\begin{minipage}{0.1\textwidth}
				\vspace*{1.1mm}
				\centering\includegraphics[width=\linewidth]{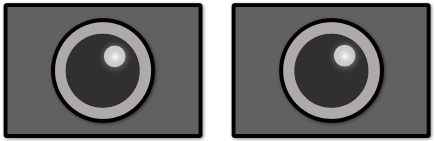}\\
				\vspace*{0.5mm}
				\centering \textunderset{\text{5~cm}}{$\xleftarrow[\quad]{}\!\xrightarrow[\quad]{}$}\\
				
			\end{minipage}  
			&  \begin{minipage}{0.155\textwidth}
				\vspace*{1.1mm}
				\centering\includegraphics[width=\linewidth]{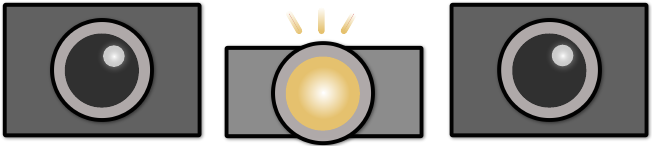}\\
				\vspace*{0.5mm}  
				\centering \textunderset{\text{5~cm}}{$\xleftarrow[\quad\quad \ ]{}\!\xrightarrow[\quad\quad\ ]{}$}\\
				
			\end{minipage}  
			& \begin{minipage}{0.1\textwidth}
				\centering\includegraphics[width=\linewidth]{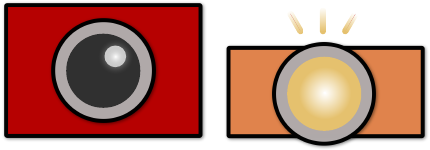}\\
				\vspace*{1.7mm} 
				\centering \color{inputcol}SI\color{outputcol}MO
			\end{minipage} & \begin{minipage}{0.1\textwidth}
				\vspace*{0.1mm}
				\centering
				\includegraphics[width=0.51\linewidth]{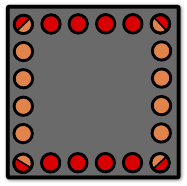}\\ 
				\centering \color{inputcol}MI\color{outputcol}MO~\color{black}(square)\\
				\vspace*{0.4mm}  
			\end{minipage}   \\
			
			\midrule
			\multicolumn{2}{l}{\textbf{Capture Frame Rate}} & 30~fps & 30~fps & 30~fps & $\approx$ 70~fps* \\
			\midrule
			\multicolumn{2}{l}{\textbf{Depth Processing Time*}} & < 33~ms & < 33~ms & < 33~ms & $\approx$ 78~s \\
			\midrule
			\multirow{5}*{\textbf{Transmitters}} & Type &  $-$ & Laser Projector & LED Array & TDM Antenna Array \\
			& Array Size &  $-$ &  $-$ & $-$ & 2$\times$47 \color{inputcol}TX \color{black} $\updownarrow$ \\
			& Wavelength&  $-$ & 840--860~nm & 860~nm & 3.6-4.2~mm \\
			& Frequency&  $-$ & $\approx$ 353~THz & $\approx$ 353~THz & 72-82~GHz \\
			& Signal Modulation &  $-$ &  Spatial Multiplexing & AMCW & FSCW \scriptsize($\text{N}_f=128$) \\
			\midrule
			\multirow{3}*{\textbf{Receivers}} & Type & Image sensor & Image sensor & Image Sensor & Antenna Array \\
			& Array Size &  2$\times$1928$\times$1208~px & 2$\times$1280$\times$800~px & 1024$\times$1024~px & 2$\times$47 \color{outputcol}RX \color{black} $\leftrightarrow$  \\
			& Spatial Size\textbf{*} & 2$\times$5.8$\times$3.6~mm & 2$\times$3.8$\times$2.4~mm & 3.6$\times$3.6~mm & 2$\times$13.8~cm \\
			& Field of view & $110^\circ \times 80^\circ$& $87^\circ \times 58^\circ$ & $75^\circ \times 65^\circ$& $\approx 53^\circ \times 53^\circ$\,*\\
			\midrule
			\multicolumn{2}{l}{\textbf{Depth Image Resolution}} & 1920$\times$1080~px & 1280$\times$720~px & 640$\times$576~px & 301$\times$301~px \\
			\midrule
			\multirowcell{3}{\textbf{Spatial} \\ \textbf{Resolution*} \\ $\boldsymbol{\delta_x \times \delta_y \times \delta_z}$} & 30~cm & 0.30$\times$0.39$\times$1.34~mm & 0.36$\times$0.42$\times$0.21~mm & 0.61$\times$0.59$\times \leq$ 2.0~mm & 4.08$\times$4.08$\times$11.08~mm \\
			& 40~cm & 0.40$\times$0.52$\times$2.38~mm & 0.47$\times$0.56$\times$0.38~mm & 0.82$\times$0.79$\times \leq$ 2.0~mm & 5.38$\times$5.38$\times$12.44~mm \\
			& 50~cm & 0.50$\times$0.65$\times$3.72~mm & 0.59$\times$0.70$\times$0.59~mm & 1.02$\times$0.98$\times \leq$ 2.0~mm & 6.69$\times$6.69$\times$13.23~mm\\
			\bottomrule
		\end{tabular}
	\end{table*}

	\section{The MAROON Dataset}
	\label{sec:maroon}
	The capture of the MAROON dataset allows for a comprehensive analysis with respect to the characteristics of the four previously described depth sensing techniques.
	
	To accomplish this, we collected a diverse set of common household and construction objects. We ensured having a broad variety of materials and geometries, with varying complexity, which we selected based on prior knowledge of the sensor operating principles (see~\autoref{sec:depth_sensing_categorization}). The selection aimed to identify challenging objects for reconstruction, highlighting the limitations of current depth imagers and providing a valuable data resource for improving these, e.g. through integration of multimodal sensor data.
	
	In the course of this section, we outline the capture setup and data acquisition pipeline, depicted in \autoref{fig:setup}, to aid future research on the publicly accessible data.
	Four single-view depth sensors are used in our experiments: Stereolabs ZED X~\cite{stereolabs_zed} (Passive Stereo), Intel Realsense D435i~\cite{intel_rs} (Active Stereo), Microsoft Azure Kinect~\cite{microsoft_kinect} (NIR ToF), and a submodule of Rohde \& Schwarz's QAR50~\cite{rs_qar} (RF ToF).

	\subsection{Sensor Setup}
	\label{sec:setup}
	Our sensor setup consists of four mounted single-view depth sensors and a ground-truth (GT) optical multi-view stereo (MVS) system comprising five calibrated DSLR cameras, which are depicted on the \textit{left} in \autoref{fig:setup}.
	While all single-view sensors are designed to achieve an optimal balance between depth quality and acquisition time, the MVS system employs an offline reconstruction process that is specifically optimized for depth quality.
	In summary, eight cameras are mounted on tripods and arranged around the MIMO imaging radar on a desk, thereby maximizing the area of intersection of each sensor's field of view, to ensure similar object visibility.
	All sensors and the GT system are time-synchronized, either through hardware or software, to capture the object at the same moment.
	
	Prior to capture, the object is positioned in the center of the squared radar aperture, and approximately at the center of the joint field-of-view intersection, propped up with boards crafted from styrofoam\,---\,a material that is considered to be nearly fully penetrated by the RF signal\,---\,to prevent external interference of radio waves from other sources in the vicinity, apart from the object of interest.
	For similar reasons, absorbers are placed behind the object of interest.
	Similarly, for optical sensors a loose black cloth, which is penetrated by RF signals, is suspended in front of the absorbers to visually occlude the room's background.
	The sensor settings are chosen with respect to a trade-off between fair sensor comparability and practical applicability (see supp. mat.). 
	An overview of the chosen settings and relevant sensor parameters, is given in \autoref{table:sensors}.

	\subsection{Data Acquisition Pipeline}
	\label{sec:dataset}
	The MAROON dataset comprises static and \textit{quasi-static}, i.e., with slow, minimal motion as in case for human hands, targets of differing materials and geometries, captured at multiple distances from all sensors simultaneously.
	With respect to the order of steps described in \autoref{fig:setup} (\textit{middle}), we will now continue to elaborate on the details of the acquisition pipeline.

	\paragraph{Spatial Calibration}
	We spatially aligned the coordinate systems of each depth imager using the calibration method in~\cite{wirth_calibration}.
	In this method, four respective spherical objects of styrofoam and metal, tailored to the visibility of optical and RF sensors, are captured.
	In the sensor-specific reconstructions, these spheres are automatically located and jointly aligned using spatial registration.
	This approach enables a direct comparison of the object reconstructions in a metrical space.
	Calibration errors are expected to be in 1--2~mm range with respect to the Chamfer distance, in analogy with the evaluation scheme used by
	Wirth et al.~\shortcite{wirth_calibration}.
	
	The five DSLR cameras of the MVS system are treated as a unified sensor with a common coordinate system, which is spatially calibrated with that of the depth imagers.
	The camera extrinsic and intrinsic parameters of the MVS system are determined from images capturing a conventional optical calibration target with a checkerboard pattern.
	For this, we use the commercial software provided by Agisoft Metashape.
	Remaining calibration errors exhibit a root mean square reprojection error of 0.38~px, averaged over all camera calibrations performed during the dataset capture.

	\paragraph{Object Preparation and Capture}
	The reconstruction method of MVS is similar to passive stereo imagers.
	Hence, inaccurate reconstructions can be the result when dealing with textureless and view-dependent object materials.
	To circumvent this limitation, we implement a distinct capture process for a subset of particularly challenging objects to generate more reliable GT reconstructions.
	After the object has been captured once by all sensors (including MVS), a thin multicolored speckle pattern is applied using water colors that assists the correspondence finding of the subsequent, additional MVS-only capture. In order to ensure exact alignment between that GT reconstruction and other imaging modalities, the speckle is applied in situ without moving the object.
	
	In total, each object is recorded at three different distances to the MIMO imaging radar of 30~cm, 40~cm, and 50~cm, respectively.
	The remaining depth imagers are situated behind the radar. 
	Their corresponding object-to-sensor distance is determined from the distance to the radar and from the relative position between each optical sensor and the imaging radar, which is given by the calibration parameters.
	Based on the Euclidean norm of the mean translation across all calibrations conducted, we report an additional object-to-sensor distance of +8~cm (Azure Kinect), +6~cm (Realsense D435i), and +5~cm (ZED X Mini), respectively.
	We record 20 frames for each optical sensor and 10 radar frames.
	In total, we capture 45 objects and list further statistics about the dataset in \autoref{table:statistics}.

	\paragraph{Optical Segmentation}
	To perform an accurate object-centric sensor evaluation, it is essential to isolate the estimated object depth from the background.
	For optical systems, we acquire segmentation masks by performing a semi-automatic foreground-background segmentation.
	Given that all depth imagers capture RGB images\,---\,either for depth estimation or via a separate calibrated camera\,---\,we first segment the RGB images using manually defined object labels in conjunction with Grounded-SAM~\cite{ren_2023}. This generates a binary segmentation mask of the object, $\boldsymbol{M}$, where all valid pixels $(u,v)$ are included in $\boldsymbol{M}^{+}(u,v) = \{ \boldsymbol{M}(u,v) > 0 \}$.
	We then manually correct failure cases in the resulting segmentation masks.
	The same procedure is employed to MVS images to produce masked GT reconstructions.
	For the imaging radar, the voxel volume of the BP algorithm (\autoref{eq:backproj}) is constrained to enclose only the object of interest. In this way, segmentation masks are automatically determined from the valid pixels remaining after depth estimation.

	\paragraph{Reconstruction and Depth Estimation}
	
	MAROON offers raw sensor data, along with intermediate and final reconstruction output, stored in various data representations depending on each depth imager.
	
	For optical depth sensors, we store RGB images, auxiliary data such as infrared measurements, and depth maps, which are obtained using the corresponding signal processing algorithms provided by the manufacturer.
	
	The imaging radar captures raw measurements in form of a tensor of $N_{\text{RX}} \times N_{\text{TX}} \times N_{f}$ complex numbers, where $N_{\text{RX}} = N_{\text{TX}} = 94$ and $N_{f} = 128$.
	They are stored alongside the volumetric output produced after backprojection, as well as post-processed 2D depth and confidence maps.
	Using the raw tensor, we perform the BP algorithm on a $301\times 301\times 201$ voxel grid, with voxel centers uniformly sampled within a $30\times 30\times 20$~$\text{cm}^3$ volume around the object center, yielding volumetric data that is stored as intermediate output.
	Subsequently, we apply maximum projection (\autoref{eq:maxproj}) to acquire a 2D projection of the depth as well as a 2D confidence map.
	Using the latter, we filter out depth values according to a threshold of $-14$~dB relative to the maximum value. As mentioned in~\autoref{sec:mimo_fscw}, such thresholding is challenging for unknown objects. We chose this threshold empirically over all objects in the dataset, aiming at a good balance between pruning of noise and retention of object details, and provide an ablation study with different thresholds in \ifseparatesupp Section~S14.1 \else \autoref{sec:radar_depth_filtering} \fi~of the supplementary material. We encourage interested readers to experiment with different thresholds, using the raw radar data available in our dataset.
	After thresholding, the filtered result is stored as an orthographic depth map.

	The ground-truth MVS setup captures five RGB images, which are stored alongside post-processed depth images and a mesh representation of the object, after performing reconstruction using Agisoft Metashape. Metashape (formerly Photoscan) commonly has a reconstruction accuracy in sub-millimeter range for similar capture en\-vi\-ron\-ments \cite{mousavi_2018, remondino_2014}.
	After reconstruction, we finally apply Laplacian smoothing.
	\begin{table}[!h]
		\centering
		\caption{\label{table:statistics}%
			Statistics of the MAROON dataset. Assuming that all captured objects are static, the number of total frames include duplicate captures, possibly varying in random depth noise, while the unique frames only contain one capture per object of each sensor.}
		\begin{tabular}{@{}lc@{}} 
			\toprule
			Statistics & MAROON \\
			\midrule
			\# objects & 45 \\
			\# static objects & 41 \\
			\# quasi-static objects & 4 \\
			\# prepared speckled objects & 14 \\
			\# captures (\# objects $\times$ 3 distances) & 135 \\
			\# total / unique optical depth frames & 8100 / 405 \\
			\# total / unique RF depth frames & 1350 / 135 \\	
			\bottomrule
		\end{tabular}
	\end{table}
    \vspace{-2.5ex} %
	\section{Evaluation}
	\label{sec:evaluation}
	In this section, we compare the reconstructions produced by the four presented depth imagers with a ground-truth reconstruction in a common metric space and describe the metrics used in this process. 
	Subsequently, the results of these methods are presented.
	
	We note that in this section the results are objectively presented, reserving further interpretations for  \autoref{sec:discussion}, where they will be discussed with specific attention to partially transmissive media (\autoref{sec:discussion_scattering}) and focusing on the RF signal response (\autoref{sec:discussion_radar_signal}).

	\subsection{Metrics}
	\label{sec:methods}
	
	First, we average valid depth values of each sensor across 10 frames for static objects to incorporate temporal characteristics and reduce random noise.
	We do not average quasi-static objects, of which their reconstruction did not require the application of speckles, and instead take the first frame, as it is closest to the point in time where the GT captures without the speckle pattern have been taken.
	
	Using the extrinsic calibration parameters (see supp. mat.), we subsequently transform the masked GT reconstruction, $\boldsymbol{R}_g$, into each sensor space $s$, yielding $\boldsymbol{R}_g^s$.
	We use the notation $\boldsymbol{R}^*$ to indicate a transformation to sensor space $*$.

	Next, for each object, we compute the point-wise deviation between a sensor and the transformed GT reconstruction with respect to two metrics: one-sided Chamfer distance and one-sided projective error.
	The one-sided Chamfer distance, $\text{C}$, is computed per point $\boldsymbol{p} \in \mathbb{R}^3$ in the source point cloud $\boldsymbol{P} \in \mathbb{R}^{N \times 3}$ with respect to the distance to the nearest point $\boldsymbol{q} \in \mathbb{R}^3$ in the destination point cloud $\boldsymbol{Q} \in \mathbb{R}^{M \times 3}$:
	
	\begin{equation}
	\text{C}_{\boldsymbol{p}}(\boldsymbol{Q}) =  \underset{\boldsymbol{q} \in \boldsymbol{Q}}{\text{min}} \lVert \boldsymbol{p} - \boldsymbol{q} \lVert_2\;.
	\end{equation}
	The one-sided projective depth error $\text{P}$ is computed per pixel $(u,v)$ of two depth maps $\boldsymbol{D} \in \mathbb{R}^{W\times H}$ and $\boldsymbol{F} \in \mathbb{R}^{W\times H}$ of a common image plane:
	\begin{equation}
	\text{P}_{u, v}(\boldsymbol{D}, \boldsymbol{F}) = \lvert \boldsymbol{D}(u,v) - \boldsymbol{F}(u,v) \rvert\;.
	\end{equation} 
	For both, $\text{C}$ and $\text{P}$, the respective subscripts $\boldsymbol{p}$ and $u,v$ are used as placeholders to denote the points and pixels used in the metric computation.
	
	Since the one-sided Chamfer distances are sensitive to the point cloud density, we uniformly sample the points from the sensor and the GT with respect to a common image pixel grid.
	We achieve this by, first, computing a simulated depth map $\widehat{\boldsymbol{D}}_{g}^s$ in the image space of the sensor $s$ and, second, reconstructing $\widehat{\boldsymbol{R}}^s_g$ from this depth map, using the inverse camera parameters to project it into 3D (see supp. mat.).
	The simulated depth map is computed by rasterizing a triangulated representation of $\boldsymbol{R}^s_g$ with respect to $\boldsymbol{T}_s$.
	In this way, we also discard points in optical sensors $\boldsymbol{R}^s_g$ that are not visible in the view of $s$.
	The resulting depth map $\widehat{\boldsymbol{D}}_{g}^s$ is additionally used to measure the projective error.
	To summarize, we compute:
	\framebox{	
		\begin{minipage}[t]{0.453\textwidth}
			\vspace{0.1pt}
			\begin{enumerate}
				\itemindent=-7pt
				\item[\textbf{\mone{}}] \textbf{Chamfer distance ground truth.} $\forall \boldsymbol{g} \in \widehat{\boldsymbol{R}}^s_g:\text{C}_{\boldsymbol{g}}(\boldsymbol{R}_s)$
				\item[\textbf{\mtwo{}}] \textbf{Chamfer distance sensor.} $ \forall \boldsymbol{s} \in \boldsymbol{R}_s: \text{C}_{\boldsymbol{s}}(\widehat{\boldsymbol{R}}^s_g)$ 
				\item[\textbf{\mthree{}} ] \textbf{Projective error.} $\forall (u,v) \in \boldsymbol{M}^{+}(u,v): \text{P}_{u,v}(\boldsymbol{D}_s,  \widehat{\boldsymbol{D}}_g^s)$ for ${\boldsymbol{M} = \boldsymbol{M}_s \cap \widehat{\boldsymbol{M}}_g^s}$, where $\boldsymbol{M}_s = \{ \boldsymbol{D}_s > 0 \}$ and $\widehat{\boldsymbol{M}}_g^s = \{ \widehat{\boldsymbol{D}}_g^s > 0 \}$ describe the intersection masks of valid pixels from the sensor and the projected GT, respectively. 
				\item[\textbf{\mfour{}}] \textbf{Projective error with erosion.} $\forall (u,v) \in \boldsymbol{M}_e^+(u,v): \text{P}_{u,v}(\boldsymbol{D}_s, \widehat{\boldsymbol{D}}_g^s)$ for ${\boldsymbol{M}_e = \boldsymbol{M}_s \cap f(\widehat{\boldsymbol{M}}_g^s)}$, where $f(\widehat{\boldsymbol{M}}_g^s)$ is a function performing mask erosion using a kernel size of $K \times K$ pixels. The size $K \in [0,20]$ is semi-manually selected for each object and sensor, and included together with other evaluation metadata in the release of our dataset.
			\end{enumerate}
			
		\end{minipage}
	}

	\begin{table}[h!]
		\caption{\label{table:metric_meaning}%
			Categorization of the presented metrics with respect to their sensitivities. In addition to depth, a 3D error evaluates errors along the cross-depth axes.}
		\begin{tabular}{@{}ccccc@{}} 
			\toprule
			\multirowcell{2}{Metric \\ Type} & \multirowcell{2}{Silhouette \\ Noise} & \multirowcell{2}{Missing \\ Surfaces} & \multirowcell{2}{3D Error} & \multirowcell{2}{Depth Error} \\
			&&&& \\
			\midrule
			\textbf{\mone{}} &\,---\,& \cmark & \cmark &\,---\,\\
			\textbf{\mtwo{}} & \cmark &\,---\,& \cmark &\,---\,\\
			\textbf{\mthree{}} & \cmark &\,---\,&\,---\,& \cmark\\
			\textbf{\mfour{}} &\,---\,&\,---\,&\,---\,& \cmark\\
			\bottomrule
		\end{tabular}
	\end{table}

	\subsection{Results}
	\label{sec:results}
	In this section, we first present the evaluation results, quantified using four complementary metrics: \mone{}, \mtwo{}, \mthree{}, and \mfour{}.
	Each metric is sensitive to different aspects, as detailed in \autoref{table:metric_meaning}.
	We begin by presenting the depth deviation in relation to various objects and different object-to-sensor distances. Given that near-field imaging radars are less explored compared to optical depth sensors, we dedicate the latter part of this section to the radio-frequency signal response.

	\subsubsection{Depth Deviation}	
	\label{sec:obj_dist_spec_dev}
	
	In \autoref{table:metrics}, we list the mean $\mu$ and standard deviation $\sigma$ of each metric type with respect to 12 selected objects from the MAROON dataset. 
	These objects were positioned at an object-to-sensor distance of 30~cm.
	We provide object images and a comprehensive evaluation of all 45 objects in \ifseparatesupp Figure~S20 \else \autoref{fig:object_images} \fi~of the supplementary material.
	For completeness, we give a brief overview of the overall statistics by investigating the number of best and worst results across all objects for $\mu$, respectively:
	\begin{itemize}
		\item RF ToF: performs worst in \mone{}
		\item NIR ToF: performs worst in \mtwo{}, \mthree{}, \mfour{}
		\item Active Stereo: performs best in \mone{}, \mtwo{}, \mthree{}, \mfour{}
		\item Passive Stereo: performs neither best or worst
	\end{itemize}
	To give an intuition on relative depth deviations between sensors, we present the median, mean, and standard deviation, denoted as $\widetilde{\mu}/\mu~(\pm \sigma)$, calculated across the differences in metric values for all pairwise sensor combinations.
	The results for each metric type are:
	\begin{itemize}
		\item \mone{}:~$0.23~\text{cm}/0.48~\text{cm}~(\pm 0.61~\text{cm})$
		\item \mtwo{}:~$0.19~\text{cm}/1.06~\text{cm}~(\pm 3.53~\text{cm})$
		\item \mthree{}:~$0.34~\text{cm}/1.58~\text{cm}~(\pm 4.36~\text{cm})$
		\item \mfour{}:~$0.31~\text{cm}/1.78~\text{cm}~(\pm 5.14~\text{cm})$
	\end{itemize}
	Additionally, we illustrate the distribution of the depth deviation across all objects for varying placement distances of 30~cm, 40~cm, and 50~cm in \ifseparatesupp Figure~S14 \else \autoref{fig:metrics_distance} \fi~of the supplementary material.
	
	\begin{table*}[!htbp]
		\caption{\label{table:metrics}%
			We measure the depth deviation with respect to \protect{}\mone{}\protect{}, \protect{}\mtwo{}\protect{}, \protect{}\mthree{}\protect{}, and \protect{}\mfour{}\protect{}, which we list in the form $(\mu \pm \sigma)$, consisting of the mean $\mu$ and standard deviation $\sigma$ in centimeters, computed over the entire metric domain, respectively. The best results among all sensors of one metric type are highlighted in \bf{bold} and the worst results are \underline{underlined}. The results are discussed in \autoref{sec:discussion_objects_specific}.}
		\begin{tabularx}{\textwidth}{@{}lcYYYYYY@{}} 
			\toprule 
			& \multirowcell{2}{Metric \\ Type} & \multirowcell{2}{ Cardboard }  & \multirowcell{2}{ Sponge }  & \multirowcell{2}{ Scrubber }  & \multirowcell{2}{ Plushie }  & \multirowcell{2}{ Tape Dispenser }  & \multirowcell{2}{ Statue }  \\ 
			&&&&&&& \\ 
			\hline
			RF ToF & \multirowcell{4}{\mone{}} 
			&  0.13 ($\pm$  0.06) &  \underline{1.52} ($\pm$  \underline{0.97}) &  0.58 ($\pm$  0.29) &  \underline{0.81} ($\pm$  \underline{0.47}) &  0.31 ($\pm$  0.23) &  0.27 ($\pm$  0.25)   \\ 
			NIR ToF & &  0.10 ($\pm$  0.06) &  0.79 ($\pm$  0.45) &  \underline{0.64} ($\pm$  \underline{0.34}) &  0.49 ($\pm$  0.21) &  \underline{0.84} ($\pm$  \underline{0.30}) &  \underline{0.32} ($\pm$  \underline{0.28})   \\ 
			Active Stereo & &  \textbf{0.08} ($\pm$  \textbf{0.05}) &  \textbf{0.18} ($\pm$  0.17) &  0.20 ($\pm$  0.16) &  \textbf{0.13} ($\pm$  \textbf{0.14}) &  0.16 ($\pm$  0.14) &  0.16 ($\pm$  \textbf{0.13})   \\ 
			Passive Stereo & &  \underline{0.24} ($\pm$  \underline{0.11}) &  0.26 ($\pm$  \textbf{0.15}) &  \textbf{0.14} ($\pm$  \textbf{0.11}) &  0.19 ($\pm$  0.21) &  \textbf{0.15} ($\pm$  \textbf{0.11}) &  \textbf{0.13} ($\pm$  \textbf{0.13})   \\ 
			\hline
			RF ToF & \multirowcell{4}{\mtwo{}} 
			&  0.15 ($\pm$  0.08) &  0.54 ($\pm$  0.40) &  \underline{0.97} ($\pm$  \underline{0.61}) &  \underline{2.21} ($\pm$  \underline{2.05}) &  0.55 ($\pm$  \underline{0.77}) &  \textbf{0.17} ($\pm$  \textbf{0.11})   \\ 
			NIR ToF & &  0.16 ($\pm$  \underline{0.18}) &  \underline{0.79} ($\pm$  \underline{0.42}) &  0.59 ($\pm$  0.30) &  0.53 ($\pm$  0.28) &  \underline{1.16} ($\pm$  0.60) &  \underline{0.43} ($\pm$  0.42)   \\ 
			Active Stereo & &  \textbf{0.08} ($\pm$  \textbf{0.05}) &  \textbf{0.17} ($\pm$  \textbf{0.15}) &  \textbf{0.17} ($\pm$  \textbf{0.11}) &  \textbf{0.12} ($\pm$  0.38) &  \textbf{0.17} ($\pm$  \textbf{0.17}) &  0.19 ($\pm$  \underline{0.76})   \\ 
			Passive Stereo & &  \underline{0.30} ($\pm$  0.13) &  0.33 ($\pm$  0.18) &  0.19 ($\pm$  0.13) &  0.23 ($\pm$  \textbf{0.27}) &  0.22 ($\pm$  0.20) &  0.18 ($\pm$  0.69)   \\ 
			\hline 
			RF ToF & \multirowcell{4}{\mthree{}} 
			&  0.13 ($\pm$  0.14) &  \underline{2.73} ($\pm$  \underline{2.47}) &  \underline{1.28} ($\pm$  \underline{0.76}) &  \underline{3.64} ($\pm$  \underline{3.23}) &  0.67 ($\pm$  \underline{1.08}) &  \textbf{0.20} ($\pm$  \textbf{0.26})   \\ 
			NIR ToF & &  0.19 ($\pm$  \underline{0.25}) &  1.32 ($\pm$  0.58) &  0.91 ($\pm$  0.47) &  0.85 ($\pm$  \textbf{0.52}) &  \underline{1.57} ($\pm$  0.70) &  0.77 ($\pm$  3.13)   \\ 
			Active Stereo & &  \textbf{0.10} ($\pm$  \textbf{0.12}) &  \textbf{0.29} ($\pm$  \textbf{0.42}) &  \textbf{0.27} ($\pm$  \textbf{0.34}) &  \textbf{0.24} ($\pm$  1.25) &  \textbf{0.24} ($\pm$  \textbf{0.35}) &  0.90 ($\pm$  5.17)   \\ 
			Passive Stereo & &  \underline{0.36} ($\pm$  0.17) &  0.47 ($\pm$  0.51) &  0.29 ($\pm$  0.38) &  0.35 ($\pm$  0.54) &  0.29 ($\pm$  \textbf{0.35}) &  \underline{1.43} ($\pm$  \underline{7.09})   \\ 
			\hline
			RF ToF & \multirowcell{4}{\mfour{}} 
			&  0.12 ($\pm$  0.13) &  \underline{2.93} ($\pm$  \underline{2.60}) &  \underline{1.35} ($\pm$  \underline{0.66}) &  \underline{3.61} ($\pm$  \underline{3.23}) &  0.66 ($\pm$  \underline{1.07}) &  0.20 ($\pm$  \underline{0.27})   \\ 
			NIR ToF & &  0.08 ($\pm$  0.10) &  1.69 ($\pm$  \textbf{0.23}) &  1.16 ($\pm$  \textbf{0.27}) &  0.82 ($\pm$  0.38) &  \underline{1.70} ($\pm$  0.89) &  \underline{0.25} ($\pm$  \textbf{0.13})   \\ 
			Active Stereo & &  \textbf{0.06} ($\pm$  \textbf{0.08}) &  \textbf{0.42} ($\pm$  0.50) &  \textbf{0.22} ($\pm$  \textbf{0.27}) &  \textbf{0.16} ($\pm$  0.30) &  \textbf{0.15} ($\pm$  \textbf{0.21}) &  0.16 ($\pm$  0.21)   \\ 
			Passive Stereo & &  \underline{0.38} ($\pm$  \underline{0.14}) &  0.55 ($\pm$  0.56) &  0.24 ($\pm$  0.29) &  0.18 ($\pm$  \textbf{0.26}) &  0.20 ($\pm$  0.22) &  \textbf{0.10} ($\pm$  \textbf{0.13})   \\ 
			\bottomrule
		\end{tabularx} 
		\newline 
		\vspace*{0.15 cm} 
		\newline 
		\begin{tabularx}{\textwidth}{@{}lcYYYYYY@{}} 
			\toprule
			& \multirowcell{2}{Metric \\ Type} & \multirowcell{2}{ S1 Hand Open }  & \multirowcell{2}{ Hand Printed \\ Flat }  & \multirowcell{2}{ Mirror }  & \multirowcell{2}{ Candle }  & \multirowcell{2}{ Flowerpot \\ (Transparent) }  & \multirowcell{2}{ V1 Metal Plate }  \\ 
			&&&&&&& \\ 
			\hline
			RF ToF & \multirowcell{4}{\mone{}} 
			&  \underline{0.36} ($\pm$  \underline{0.38}) &  \underline{0.71} ($\pm$  \underline{0.78}) &  \textbf{0.87} ($\pm$  \textbf{0.26}) &  1.50 ($\pm$  \underline{1.12}) &  1.31 ($\pm$  \underline{1.21}) &  0.12 ($\pm$  \textbf{0.05})   \\ 
			NIR ToF & &  0.31 ($\pm$  0.14) &  0.25 ($\pm$  0.12) &  \underline{3.77} ($\pm$  \underline{1.97}) &  \underline{2.04} ($\pm$  0.40) &  \underline{2.73} ($\pm$  1.03) &  \underline{0.77} ($\pm$  \underline{0.42})   \\ 
			Active Stereo & &  \textbf{0.12} ($\pm$  \textbf{0.09}) &  \textbf{0.09} ($\pm$  \textbf{0.07}) &  2.13 ($\pm$  1.52) &  \textbf{0.26} ($\pm$  \textbf{0.29}) &  \textbf{0.74} ($\pm$  \textbf{0.53}) &  \textbf{0.08} ($\pm$  0.06)   \\ 
			Passive Stereo & &  0.20 ($\pm$  0.16) &  0.21 ($\pm$  0.37) &  2.31 ($\pm$  1.61) &  1.64 ($\pm$  0.78) &  2.01 ($\pm$  0.83) &  0.13 ($\pm$  0.07)   \\ 
			\hline 
			RF ToF & \multirowcell{4}{\mtwo{}} 
			&  0.22 ($\pm$  0.15) &  0.17 ($\pm$  0.13) &  \textbf{0.91} ($\pm$  \textbf{0.14}) &  \underline{5.57} ($\pm$  \underline{2.78}) &  1.86 ($\pm$  \underline{2.41}) &  0.13 ($\pm$  \textbf{0.06})   \\ 
			NIR ToF & &  \underline{0.38} ($\pm$  \underline{0.26}) &  \underline{0.29} ($\pm$  0.20) &  \underline{33.31} ($\pm$  9.07) &  1.71 ($\pm$  0.49) &  \underline{3.10} ($\pm$  1.22) &  \underline{0.81} ($\pm$  \underline{0.43})   \\ 
			Active Stereo & &  \textbf{0.13} ($\pm$  \textbf{0.10}) &  \textbf{0.09} ($\pm$  \textbf{0.06}) &  30.21 ($\pm$  \underline{14.59}) &  \textbf{0.25} ($\pm$  \textbf{0.26}) &  \textbf{1.27} ($\pm$  1.78) &  \textbf{0.09} ($\pm$  0.07)   \\ 
			Passive Stereo & &  0.26 ($\pm$  0.22) &  0.18 ($\pm$  \underline{0.34}) &  27.02 ($\pm$  11.33) &  1.28 ($\pm$  0.65) &  1.86 ($\pm$  \textbf{0.93}) &  0.16 ($\pm$  0.11)   \\ 
			\hline 
			RF ToF & \multirowcell{4}{\mthree{}} 
			&  \textbf{0.22} ($\pm$  \textbf{0.25}) &  \textbf{0.16} ($\pm$  \textbf{0.20}) &  \textbf{0.93} ($\pm$  \textbf{0.12}) &  \underline{7.41} ($\pm$  \underline{3.79}) &  2.74 ($\pm$  \underline{3.66}) &  0.11 ($\pm$  \textbf{0.12})   \\ 
			NIR ToF & &  \underline{0.52} ($\pm$  0.43) &  0.33 ($\pm$  0.29) &  37.84 ($\pm$  14.84) &  2.78 ($\pm$  \textbf{0.35}) &  \underline{5.24} ($\pm$  2.04) &  \underline{0.95} ($\pm$  \underline{0.48})   \\ 
			Active Stereo & &  \textbf{0.22} ($\pm$  \underline{1.25}) &  \textbf{0.16} ($\pm$  1.30) &  \underline{39.66} ($\pm$  \underline{24.75}) &  \textbf{0.42} ($\pm$  0.49) &  \textbf{2.08} ($\pm$  2.30) &  \textbf{0.10} ($\pm$  0.13)   \\ 
			Passive Stereo & &  0.35 ($\pm$  0.41) &  \underline{1.73} ($\pm$  \underline{8.95}) &  30.82 ($\pm$  14.01) &  2.10 ($\pm$  0.98) &  3.50 ($\pm$  \textbf{1.37}) &  0.19 ($\pm$  0.15)   \\ 
			\hline 
			RF ToF & \multirowcell{4}{\mfour{}} 
			&  0.22 ($\pm$  0.25) &  0.16 ($\pm$  \underline{0.20}) &  \textbf{0.93} ($\pm$  \textbf{0.12}) &  \underline{7.37} ($\pm$  \underline{3.85}) &  2.76 ($\pm$  \underline{3.66}) &  0.10 ($\pm$  0.11)   \\ 
			NIR ToF & &  \underline{0.51} ($\pm$  0.27) &  \underline{0.30} ($\pm$  \textbf{0.09}) &  39.68 ($\pm$  6.57) &  2.75 ($\pm$  \textbf{0.15}) &  \underline{6.18} ($\pm$  1.79) &  \underline{0.79} ($\pm$  \underline{0.39})   \\ 
			Active Stereo & &  \textbf{0.16} ($\pm$  \textbf{0.24}) &  \textbf{0.08} ($\pm$  0.10) &  \underline{43.84} ($\pm$  \underline{20.28}) &  \textbf{0.31} ($\pm$  0.44) &  \textbf{2.52} ($\pm$  1.77) &  \textbf{0.07} ($\pm$  \textbf{0.09})   \\ 
			Passive Stereo & &  0.25 ($\pm$  \underline{0.34}) &  0.17 ($\pm$  0.15) &  35.96 ($\pm$  7.83) &  2.15 ($\pm$  0.65) &  4.36 ($\pm$  \textbf{0.71}) &  0.15 ($\pm$  \textbf{0.09})   \\ 
			\bottomrule
		\end{tabularx} 
	\end{table*}
	
	\begin{figure*}[!htbp]
		\centering
		\includegraphics[width=\linewidth]{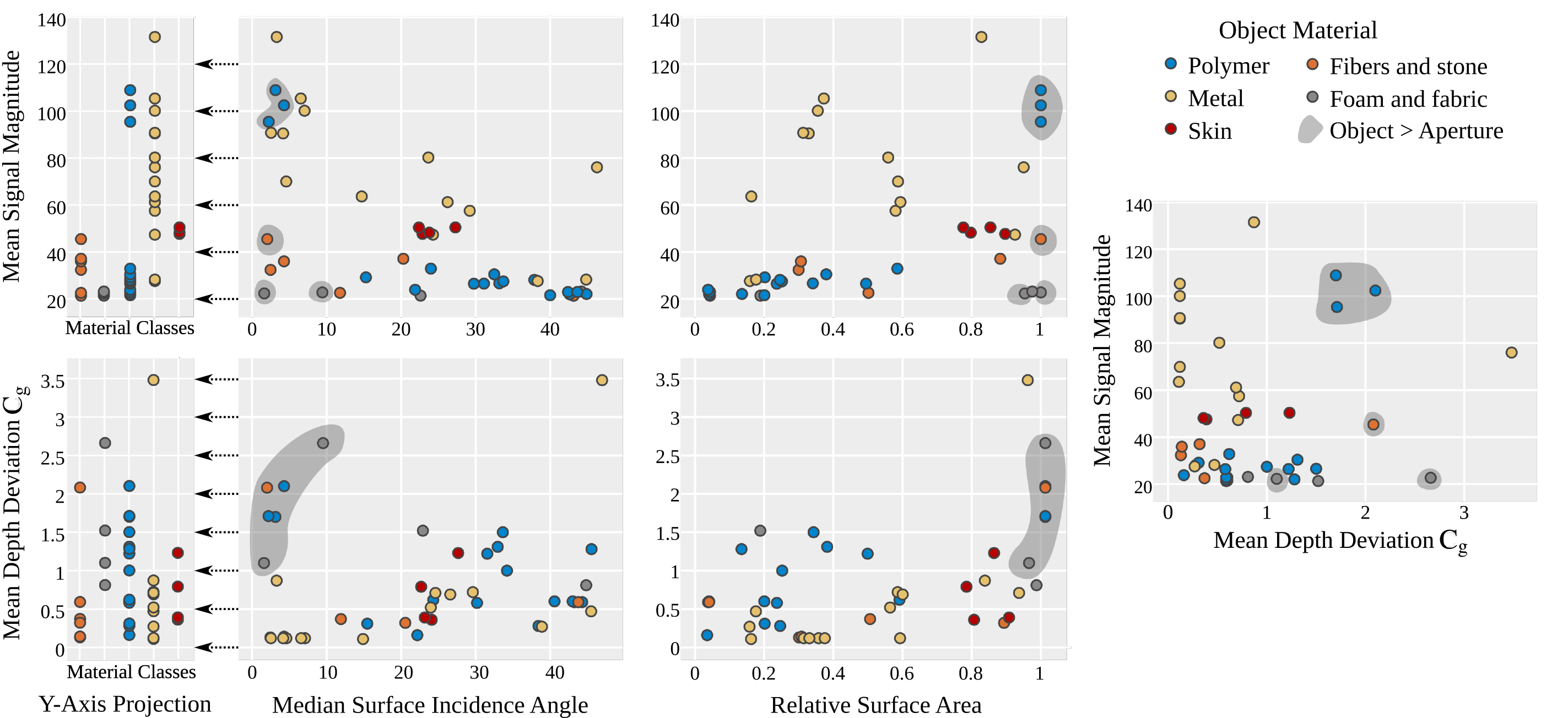}
		\caption{\label{fig:radar_intensity}%
			On the \textit{left}, object material, geometry (median surface incidence angle), and size (relative surface area) are put in relation to received signal response (mean signal magnitude, \textit{top row}) and mean depth deviation (\textit{bottom row}). On the \textit{right}, both quantities are directly compared to each other. Measurements, where large objects appear outside the radar's antenna aperture, are highlighted in gray regions, as they exhibit higher depth deviations compared to the ground-truth reconstructions, which may extend beyond this aperture; this is attributed to the comparably small field of view and the surface reflection characteristics with respect to radio waves (see supp. mat.).
			The results are discussed in \autoref{sec:discussion_radar_signal}.}
		\Description{The image contains three groups of scatter plots, where two of these are depicted in separate rows on the left side.
			All plots show colored dots that correspond to individual object measurements, where each color represents a different object material such as metal, polymer, fibers and stone, foam and fabric, and skin. 
			Each row displays three plots, respectively, which share a common y-axis for mean signal magnitude (top) and mean depth deviation with respect to $C_g$ (bottom), while varying the x-axis with respect to material (grouped by material class on the x-axis), geometry, and size, as described in the caption. On the right side, one single scatter plot is visualized. Gray shaded clusters highlight groups of points that behave differently from the rest, indicating cases where the size of the objects exceed the extents of the antenna aperture.}
	\end{figure*}
	
	\subsubsection{RF Signal Response}
	\label{sec:signal_resp}
	The point-wise confidence of the backprojection algorithm for RF ToF is considerably affected by the signal amplitude;
	however, disentangling amplitude from phase in the presence of signal interference imposes the same challenges that arise from recovering the phase shift itself.
	To avoid inducing additional bias through the assumptions made in signal processing, we investigate the unprocessed signal response of the MIMO imaging radar across multiple objects.
	For each object at 30~cm object-to-sensor distance, we compute the mean absolute value out of all complex phasors received from the raw signal, averaged over 10 frames.
	We refer to this quantity as \textit{signal (phasor) magnitude}, emphasizing the difference from signal amplitude.
	
	Differentiating between signal response and reconstruction quality, we examine their relationships with respect to object material, geometry, and size. 
	Our findings are shown on the \textit{left} of \autoref{fig:radar_intensity}, where we visualize these relationships for signal magnitude (\textit{upper row}) and mean depth deviation (\textit{bottom row}) in isolation.
	On the \textit{right}, we further investigate correlations between signal magnitude and mean depth deviation.

	Object materials are categorized into six classes, with detailed information available in \ifseparatesupp Table~S9 \else \autoref{table:material_classes} \fi~of the supplementary material.
	The goal is to highlight material differences on a coarse level, noting the large object variety that still persists within one material class. 
	
	The object geometry is quantified by the median angle in degrees between the point-wise surface normals of the GT reconstruction and the depth direction (along the $z$-axis) of the imaging radar.
	As we positioned the objects to align their primary orientation with the viewing direction of the planar square-shaped antenna aperture\,---\,which particularly becomes important for flat objects\,---\,the median angle mainly reflects geometric complexity, with objects having a higher surface incidence angle showing larger portions oriented away.
	An extension of \autoref{fig:radar_intensity} (\textit{left}) is available in \ifseparatesupp Section~S16 \else \autoref{sec:discussion_geometry_extended} \fi~of the supplementary material, visualizing the correlation of the median angle as well as per-angle measurements with respect to the depth deviation of all four sensors.
	
	Object size is determined by the relative surface area compared to the radar antenna aperture.
	It is computed from the fraction of the object's 2D axis-aligned bounding box $A$ (in the $x$- and $y$-axis) inside of the 2D axis-aligned bounding box $B$ of the antenna array by using the formula: $(A \cap B) / B$.

	\begin{figure*}[htbp]
		\centering
		\includegraphics[width=\linewidth]{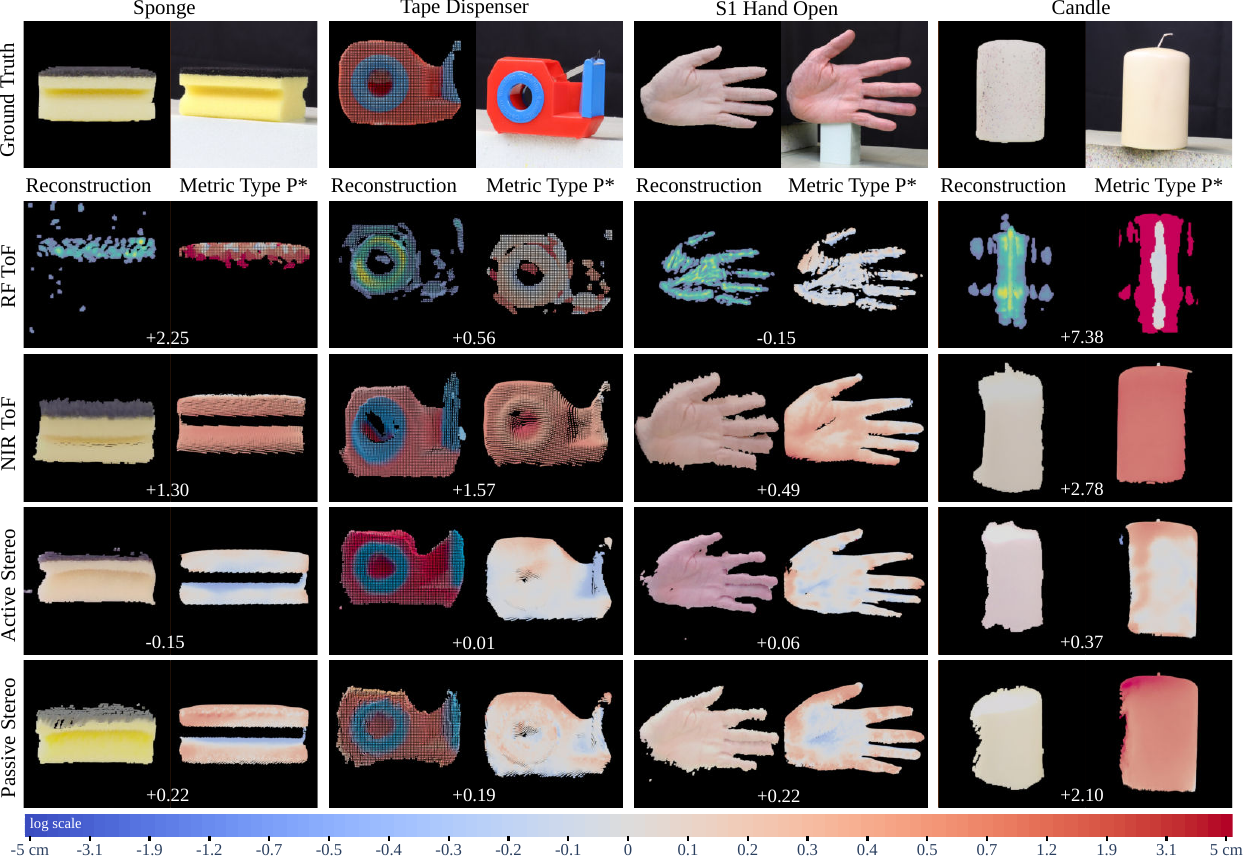}
		\caption{\label{fig:scattering}%
			For selected objects, we show the reconstructed point clouds (\textit{left}) next to their deviation from to the MVS reconstruction (\textit{right}). The signed depth deviation \mthree{}* is given for each pixel $(u,v)$ in centimeters. All measurements in the domain $\boldsymbol{M}^{+}(u,v)$ are projected onto the GT reconstruction and mapped to color using a combination of a symmetrical logarithmic scale\ and linear mapping between $[-0.5,0.5]$ centimeters. The mean deviation of \mthree{}* is quantified in centimeters below each sensor measurement.}
		\Description{The image is arranged as a grid comparing several objects, including the Sponge, Tape Dispenser, S1 Hand Open, and the Candle. The top row shows photographs of each object as a visual reference. Below, from top to bottom, each row displays the RGB-colored 3D point cloud reconstructions of RF ToF, NIR ToF, Active Stereo and Passive Stereo, respectively. Within the same row and next to each reconstruction is a color-mapped error point cloud, highlighting where the reconstructed shape is closer to or farther from the ground-truth shape. Colors range from blue to red, indicating negative to positive depth deviation, with numeric mean values shown beneath each reconstruction. Overall, the figure visually contrasts how accurately different sensors capture object shape, surface detail, and depth. The individual observations for each depth imager are discussed in Section 7.2.}
        \vspace{2ex} %
	\end{figure*}
	
	\section{Discussion}
	\label{sec:discussion}
	
	In this section, we provide a general discussion of the previously reported results related to depth deviation, followed by two focused discussions of time-of-flight sensor effects that offer complementary perspectives on these results. Regarding the latter, we investigate effects of partially transmissive media and explore RF ToF as a particularly under-explored sensor technology, focusing on the received signal response in relation to depth deviation.
	
	We note that a comprehensive sensor characterization, highlighting common trends across all 45 objects, is provided in \ifseparatesupp Section~S15.1 \else \autoref{sec:discussion_general_trends} \fi~of the supplementary material, where we discuss the interpretation of metrics, the depth deviation over varying distances, as well as relative depth deviations between sensors.
	
	\subsection{Discussion of Object-specific Depth Deviation}
	\label{sec:discussion_objects_specific}
	
	The following section will analyze the objects in \autoref{table:metrics} in regard to their relative depth deviations over one or multiple sensors.

	\paragraph{Radio-Frequency Time-of-Flight} For RF ToF, we find that the least deviation relative to the mean of all metrics occurs with planar object geometries (\obj{V1\sms Metal\sms Plate}, \obj{Cardboard}), followed by more complex shapes (\obj{Statue}, \obj{S1\sms Hand\sms Open}, \obj{Hand\sms Printed\sms Flat}). 
	For a deeper discussion, see \autoref{sec:discussion_radar_signal}.
	
	Objects made of foam (\obj{Sponge}), thin plastic (\obj{Scrubber}, \obj{Flowerpot}), fabric (\obj{Plushie}), and paraffin wax (\obj{Candle}), exhibit the highest depth deviations due to a large fraction of the transmitted RF signal not being immediately reflected. In the case of \obj{Mirror}, RF penetrates the first (glass) surface and images the silver coating behind, leading to an offset in the depth reconstruction.
	
	\paragraph{Near-infrared Time-of-Flight} For similar reasons, NIR ToF shows large depth deviations for visually transparent objects like \obj{Flowerpot}, \obj{Candle}, \obj{Sponge}, and \obj{Tape\sms Dispenser}. Both RF ToF and NIR ToF are susceptible to multi-path effects; however, our experiments suggest these effects do not occur for the same objects. Further examination of wavelength-specific multi-path effects, with a particular focus on partially transmissive materials, will be discussed in \autoref{sec:discussion_scattering}.
	
	Additional sources of high depth deviation for NIR ToF include thin structures (\obj{Scrubber}), which reduce the sensor's effective spatial resolution. Highly reflective objects (\obj{Metal\sms Plate}) may cause sensor oversaturation, while perfectly specular materials (\obj{Mirror}) yield depth values from the first weak scatterer after perfect reflection.

	\paragraph{Active and Passive Stereo Sensors} For the active stereo sensor, we observe higher depth deviations for textureless and partially transmissive materials (\obj{Sponge}, \obj{Candle}). Similar to NIR ToF, the uniqueness of the active NIR light pattern can be compromised by multi-path effects. The passive stereo camera is particularly sensitive to textureless objects (\obj{Sponge}, \obj{Candle}).

	\subsection{Discussion of Time-of-Flight Sensors: Partially Transmissive Media}
	\label{sec:discussion_scattering}
	
	As previously discussed in \autoref{sec:time_resolved}, both NIR and RF ToF sensors assume direct reflection and thus are susceptible to internal reflections, such that multi-path effects within the scene may lead to missing or incorrect reconstructions.
	In this analysis, following the nomenclature by Nayar et al.~\shortcite{nayar_2006}, we classify radiance transport that involves a single signal bounce between sender and receiver as \emph{direct} (as, within the sensor’s spatial resolution, it interacts with the scene at one surface point only), and all other types of transport as \emph{global} (involving multiple scattering or diffraction events within and between objects).
	Due to their significant difference in wavelength relative to scene features, global radiance transport takes very different forms for each modality. In the case of NIR, representative forms of global transport include inter-reflections, half-transparent surfaces, and subsurface scattering within the object material. Global transport at radio frequencies, on the other hand, is dominated by diffraction and reflections that reshape and redirect the wave front as it interacts with multiple scene elements, and by multiple superimposed responses akin partial transmittance at different depths.
	
	In the remainder, we will now study the four selected objects in \autoref{fig:scattering}.
	In addition to \autoref{table:metrics}, this figure visualizes depth deviations using a signed version \mthree{}* of metric \mthree{}, color-encoded on a symmetrical logarithmic scale (\texttt{SymLogNorm}\footnote{\url{https://matplotlib.org/3.8.4/api/_as_gen/matplotlib.colors.SymLogNorm.html}}), with a linear mapping between $[-0.5,0.5]$ centimeters. The supplementary \ifseparatesupp Section~S14.3\else \autoref{sec:additional_results}\fi~includes signed versions of \mthree{} and \mfour{} for all MAROON objects.
	
	\paragraph{Near-infrared Time-of-Flight} In the NIR domain, the most prominent effect of global transport occurs for objects with strong internal scattering. Here, the ToF reconstructions exhibit systematic depth deviations of \mthree{}*, generally biased toward larger distances than the ground truth. This is consistent with the light traveling an additional distance due to scattering within the object before being remitted again, so that the observed propagation time of the actively transmitted signal is consistently longer than for a direct (local) reflection at the object surface. Examples in \autoref{fig:scattering} for internal scattering include subsurface scattering (\obj{S1\sms Hand\sms Open}, \obj{Sponge}, \obj{Candle}) and inter-reﬂections within hollow objects (\obj{Tape\sms Dispenser}).
	Extended path length due to subsurface scattering is an established effect, systematically measured by Lukinsone et al.~\shortcite{lukinsone_2020}. For human skin (e.g. \obj{S1\sms Hand\sms Open}), and for points of incidence and exitance one millimeter apart, Lukinsone et al.\ observe effective sub-surface path lengths of up to $26\pm3$~mm at 800 nm wavelength, which\,---\,in the context of a ToF sensor\,---\,would result in a systematic depth deviation of half that path length ($\approx +13$~mm). At the same time, however, for human skin a significant portion of the total remitted light stays very close to the point of incidence~\cite{jensen_2001}, suggesting that the bulk of the received signal experiences even smaller path length extensions, lending plausibility to our measured systematic depth deviation of $+4.9$~mm for \obj{S1\sms Hand\sms Open} to be due to subsurface scattering.
	
	\paragraph{Radio-Frequency Time-of-Flight} For RF ToF, only one object (\obj{Candle}) showed a systematic path length extension, suggesting that optical subsurface scattering cannot fully model RF interactions.
	In contrast to the NIR ToF measurements, the depth deviation for the \obj{Candle} object is non-uniform, with higher values near the edges due to variations in surface position and orientation that affect radiance transport. 
	
	Where the \obj{Candle} surface faces the antenna array, the received signal is dominated by direct reflections; where direct reflections reflect away from the array (nearer to the candle's silhouettes), mostly global transport is observed. In accordance with the results by \'Alvarez L\'opez et al.~\shortcite{alvarez_2018} the depth reconstruction in the parts with little direct reflection appear more distant than ground truth, which the authors attribute to the high relative permittivity $\varepsilon_r \approx 2.6$ of paraffin wax that extends the inferred path length under the assumption of speed of light in vacuum.

	\hfill \newline
	In summary, objects composed of partially transmissive media primarily yield\ systematic bias in ToF reconstructions, with estimated depths biased toward larger values than the ground truth. Nevertheless, the factors causing these distortions vary between optical and RF modalities.

	\subsection{Discussion of MIMO Radar: Signal Response and Depth Deviation}
	\label{sec:discussion_radar_signal}
	The\ results in \autoref{table:metrics} suggest that RF ToF reconstructions are generally less complete than those of optical sensors, as further illustrated in the \textit{second row} of \autoref{fig:scattering}, where depth deviations are lowest when surface orientations align with the antenna aperture.
	Initially, this seems to contradict the expectation that larger antenna apertures should capture more surface compared to cameras, given the variety of positions and viewing angles from the individual RX-TX antenna pairs; this advantage, however, seems to be mitigated by the fact that most object surface reflections appear to be specular~\cite{lu_2013}.
	This means that reflections at surfaces oblique to the aperture are only received by a small fraction of antennas, thereby weakly contributing to the overall signal response, potentially at the same level as noise.
	
	We discuss further sources of incomplete RF ToF reconstructions in the next sections, where we first
	analyze the raw signal response\,---\,without inducing additional bias from the reconstruction algorithms\,---\,and subsequently relate it to the quality of the measured depth after reconstruction.

	\subsubsection{Radio-Frequency Signal Response}
	In \autoref{fig:radar_intensity} (\textit{top left}), we presented the received signal magnitude across objects with varying material, geometry and size. Following the scatter plot order from left to right, we will now discuss common trends, noting that it remains challenging to disentangle the presented quantities, as the large variability across objects prevents us from isolating one quantity while keeping the others constant.

	\paragraph{Influence of Material} Metal and metallic-coated objects generally show higher signal magnitudes, which is consistent with previous studies~\cite{sherif_2021, sherif_2014}.
	With a considerably lower spread, large magnitudes are also observed for captures of human skin, which is highly reflective due to its rich water content~\cite{sherif_2014}.
	Object materials made of polymers, fibers and stone, or foam generally respond with much smaller signal magnitude.
	
	\hfill \newline
	Regarding \textit{object geometry} and \textit{size}, no significant global trends are observed; however, consistent patterns emerge within subsets of the same material, particularly in the metal and polymer classes, which have the highest number of samples. We will discuss these patterns within the next paragraphs.

	\paragraph{Influence of Geometry} 
	For objects of complex geometry, with a median surface incidence angle greater than $10^\circ$, large portions of their surface area face away from the antenna array, resulting in decreased signal responses compared to planar objects aligned closely with the antenna aperture (<$10^\circ$).
	The reflection direction of the transmitted signal depends on the surface normal's orientation. As the angle between this normal and the depth axis increases, the solid angle of the object relative to the planar square-shaped antenna aperture (cf. \autoref{table:sensors}) decreases.
	In other words, a decreasing area around the hemisphere of outgoing reflection directions is aligned with the approximate, $53^\circ$ field of view of the RX antennas, resulting in reduced signal energy reception, and thus radar cross-section~\cite{radar_knott_2004}.
	Superficially this resembles the well-known cosine law in radiometry, but the exact quantitative relationship depends on the object's location relative to the individual RX and TX antennas and is further modulated by the non-trivial radiation and signal lobes of the antennas.
	
	\paragraph{Influence of Size} Aside from object geometry, the received signal magnitude also appears to increase with object size for non-metal materials.
	Within these material classes, the highest signal magnitude is achieved for objects close to or even larger than the antenna aperture.
	As the latter also typically exhibits a low median surface incidence angle, it remains questionable, whether this observation can be attributed to object size or object geometry. To address this, we additionally visualize the relation between the two quantities in \ifseparatesupp Section~S17 \else \autoref{sec:discussion_signal_response_extended} \fi~of the supplementary material.
	Assuming that signal magnitude is proportional to the received energy, our findings correspond to the fact that, the reflected signal energy received at the RX antennas directly depends on the surface area of the irradiated object, in case the energy density is constant.

	\subsubsection{Radio-Frequency Depth Deviation}
	Following the previous section, we now summarize the results of \autoref{fig:radar_intensity} (\textit{bottom left}), where we relate the depth deviation to varying object material, size, and geometry in their respective scatter plots.

	\paragraph{Influence of Material} Similar to our findings for signal response, metal objects generally exhibit the lowest depth deviation with a relatively small spread compared to other material classes, indicating that object material influences reconstruction quality. 
	
	\paragraph{Influence of Size and Geometry} While we find no direct relationship of object size to depth deviation, the most notable trend is seen with varying object geometry, where the deviation increases alongside the median surface incidence angle across all material classes. 
	The backprojection algorithm assumes that similar energy amounts are received at a point across the majority of RX antennas. 
	Received energy diminishes for surfaces oriented away from the antenna geometry, leading to variations based on antenna positioning. This, in turn, can lead to reduced confidence in the measurements, causing valid data to be filtered out along with noise.
	
	The MIMO radar we use has a large aperture and a high number of RX-TX antenna pairs, suggesting that stronger orientation-dependent effects may be observed with typical lower-resolution devices. To explore this further, we simulated various down-sampled antenna architectures and present their respective depth deviation in comparison to the fully occupied antenna array in \ifseparatesupp Section~S17.1 \else \autoref{sec:ablation_antenna} \fi~of the supplementary material.
	Lastly, we note that in the results, instances of object self-shadowing, that would contribute to increased depth deviation, are infrequent. While the response of backprojection to partial occlusions is an important consideration, the overall impact on the results is minimized given the characteristics of the primarily smooth objects in our dataset.	\subsubsection{Relation between RF Signal Response and Depth Deviation}
	\autoref{fig:radar_intensity} visualizes potential correlations between the RF signal response and depth deviation in the \textit{right} scatter plot. 
	Focusing on the most prevalent material groups of polymer and metal objects, we generally find no direct relationship between signal magnitude and depth deviation.
	Polymer and metal objects have a large spread in the $x$- and $y$-axis, respectively, while the opposite axis has comparably low variation.
	The received signal magnitude may have a more significant influence on the reconstruction quality in less constrained scenarios, involving multiple objects and signal sources, where, for example, depth filtering becomes increasingly relevant. In our experiments, within the RF near field, we suggest that reconstruction quality is more closely related to the distribution of received signals across antennas, influenced primarily by the local antenna layout and characteristics.
	
	To conclude, the analysis of MIMO imaging radar reconstruction quality is not as straightforward as that of optical sensors.
	We find no direct relationship between the RF signal response and the corresponding depth deviation after reconstruction and filtering on a global scale. 
	Consistent patterns within object subsets suggest that the reconstruction quality of RF ToF sensors is primarily influenced by object geometry, while the impact of object material should not be overlooked, as it is a crucial factor for depth filtering.
	
	To disentangle material and geometry-dependent effects at surface level, we suggest that material characterization is the first essential step in addressing this challenge; we will pick up on this topic in the subsequent application section.
    
	\section{Applications of MAROON}
	In this section, we highlight two applications utilizing the data from our proposed dataset.
	First, we briefly summarize the insights revealed from previous experiments:
	\begin{itemize}
		\item There is no single sensor modality that would consistently outperform the others. Each sensor has unique strengths and weaknesses related to the object’s material, geometry, and distance from the sensor.
		\item NIR ToF displays systematic depth distortions due to effects of global radiance transport within partially transmissive media, whereas RF ToF reconstructions were mostly unaffected.
		\item RF ToF partially exhibits missing reconstructions compared to its optical counterpart, primarily due to the object geometry contributing to the reconstruction sparseness.
	\end{itemize}
	Given these insights, we anticipate that multimodal depth sensing amplifies the sensors' complementary strengths, hence providing notable benefits for \textit{close-range} applications\,---\,similar to the multi-sensor design employed in \textit{far-range} sensing for self-driving cars. 
	
	We provide two examples, where we build upon prior work and demonstrate the substantial role of the high-quality sensor co-localizations from MAROON for their successful implementation:
	first, we show how the dataset is utilized for achieving realistic RF simulations by adapting and extending the concurrent research of Hofmann et al.~\shortcite{hofmann_2025}. Leveraging a pre-release of our dataset, Hofmann et al. propose a differentiable ray tracing pipeline to determine the material parameters of our captured objects. 
	
	Second, we present extended experiments for a recently proposed multimodal high-speed radar reconstruction method, MM-2FSK~\cite{wirth_2025}, which utilizes only two frequencies as opposed to our employed 128 frequency-stepped backprojection. 
	We note that we also examine the depth deviation in less computationally intensive versions of backprojection, by varying the frequency configuration, as detailed in the ablation study included in \ifseparatesupp Section~S17.2 \else \autoref{sec:ablation_frequencies} \fi~of the supplementary material.

	\subsection{Material Characterization with Differentiable Ray Tracing}
	
	Accurate material modeling is vital not only for isolating the effects of mmWave signal interaction but also for high-fidelity radar simulation.
	Highly reflective materials, such as metals, produce vastly different radar returns compared to largely diffuse scatterers, such as objects made of wood or rubber.
	Therefore, Hofmann et al.~\shortcite{hofmann_2025} propose a data-driven approach to determine reflective properties under mmWave radiation, such as permittivity and permeability.
	
	They initialize their differentiable optimization pipeline by simulating radar returns with randomized material properties and compare the result to the real RF ToF sensor data from MAROON, while utilizing the MVS data as ground-truth geometry.
	Analogous to neural networks, where parameters are iteratively optimized using gradient descent, they continuously update the material properties of the object until the difference between the simulated and measured radar returns is minimal.
	Notably, the loss is computed on the raw phasor data, instead of reconstructed images, which increased both robustness and fidelity of the optimization due to bypassing artifacts introduced by the reconstruction algorithm.
	To facilitate a close match between the simulated and real phasor data, the radar gain and a small registration offset of $\frac{\lambda}{2}$ along each principal axis, where $\lambda$ is the longest wavelength emitted in a FSCW sequence, is optimized alongside the material properties~\cite{hofmann_2025}.
	This registration offset requires the error in the MVS data to be smaller than one wavelength to avoid getting stuck in local minima due to ambiguities from recurring wave patterns.
	Fortunately, we can safely assume this to be the case in MAROON with the calibration error ranging from 1--2~mm, as discussed in \autoref{sec:dataset}, which is half of the mean wavelength of $\approx$~4~mm in the worst case.
	\let\OldTablename\tablename       %
	\renewcommand{\tablename}{Fig.}   %
	\renewcommand{\tableautorefname}{Figure} 
	\setcounter{table}{6}
	\setcounter{figure}{7}
	\begin{table}[t]
		\centering
		\setlength{\tabcolsep}{1pt} %
		\renewcommand{\arraystretch}{0.5} %
		\begin{tabularx}{\linewidth}{cY|YY|YY|YY}
			& \small{MVS Data} & \small{Result 100\%} & \small{Target 100\%} & \small{Result 50\%} & \small{Target 50\%} & \small{Result 25\%} & \small{Target 25\%} \\
			
			\rotatebox[origin=l]{90}{\small{\hspace{2pt}\obj{Cardb.}}} &
			\includegraphics[width=\linewidth]{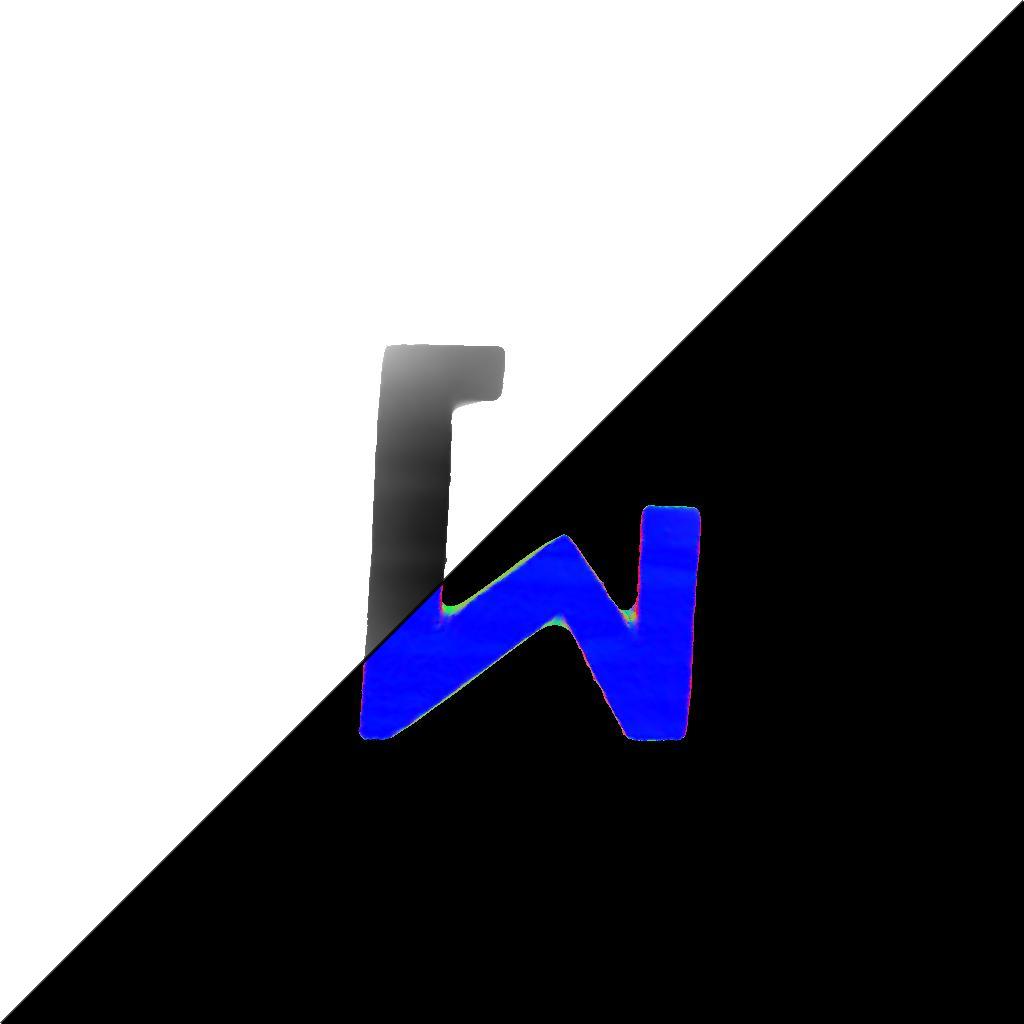} &
			\includegraphics[width=\linewidth]{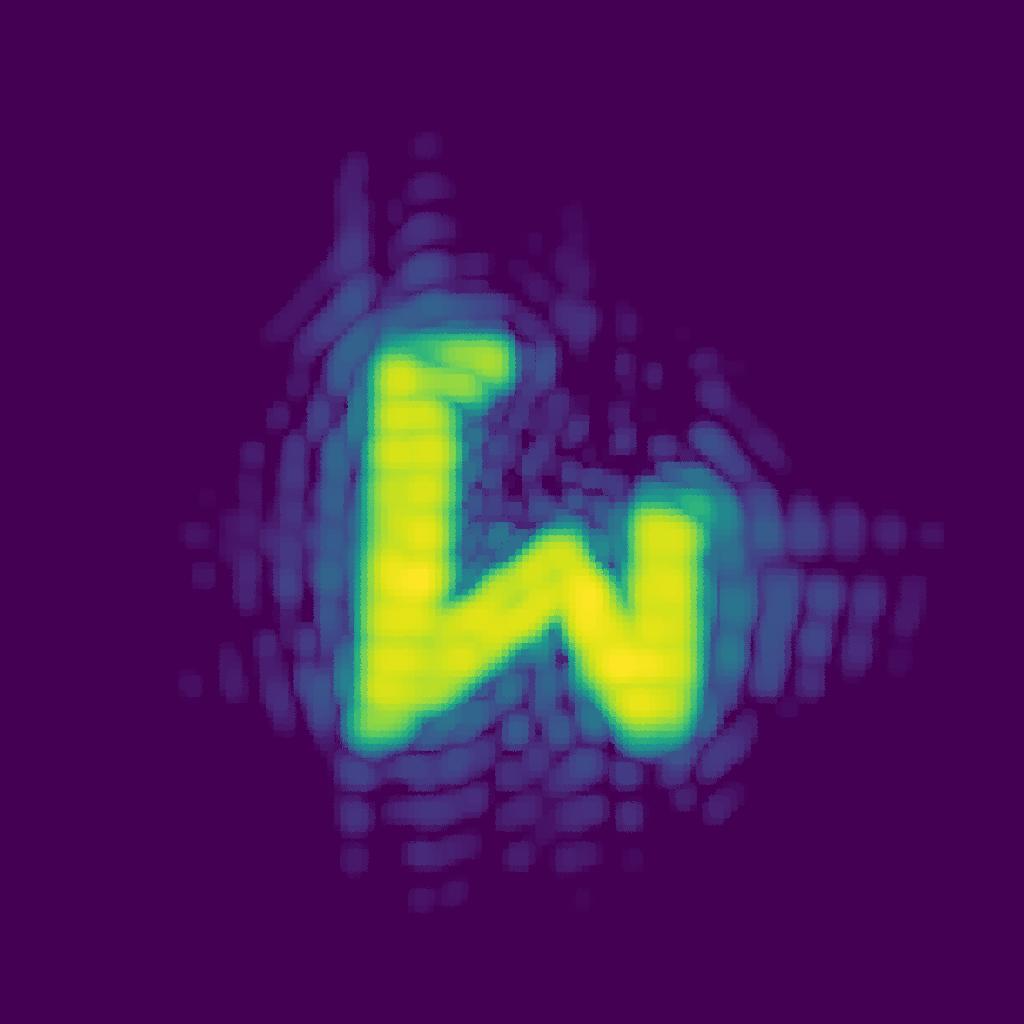} &
			\includegraphics[width=\linewidth]{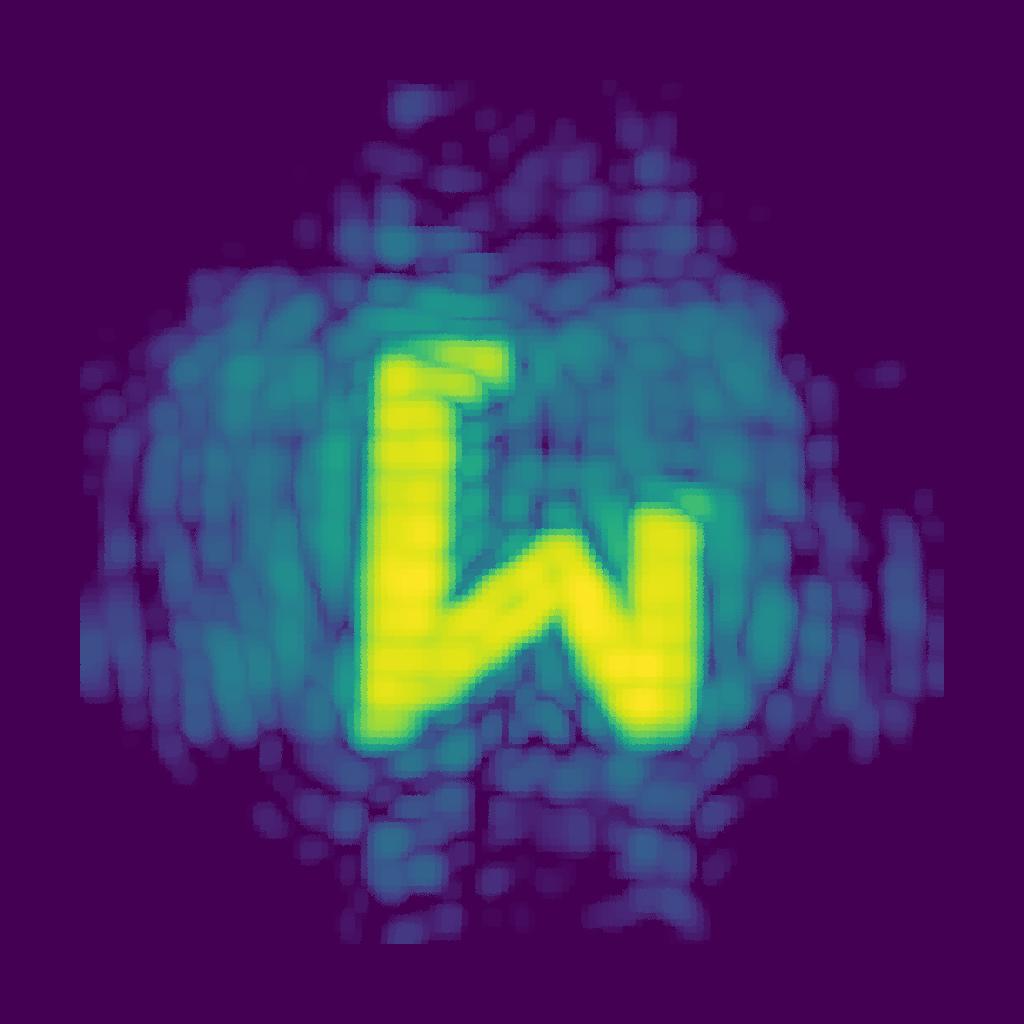} &
			\includegraphics[width=\linewidth]{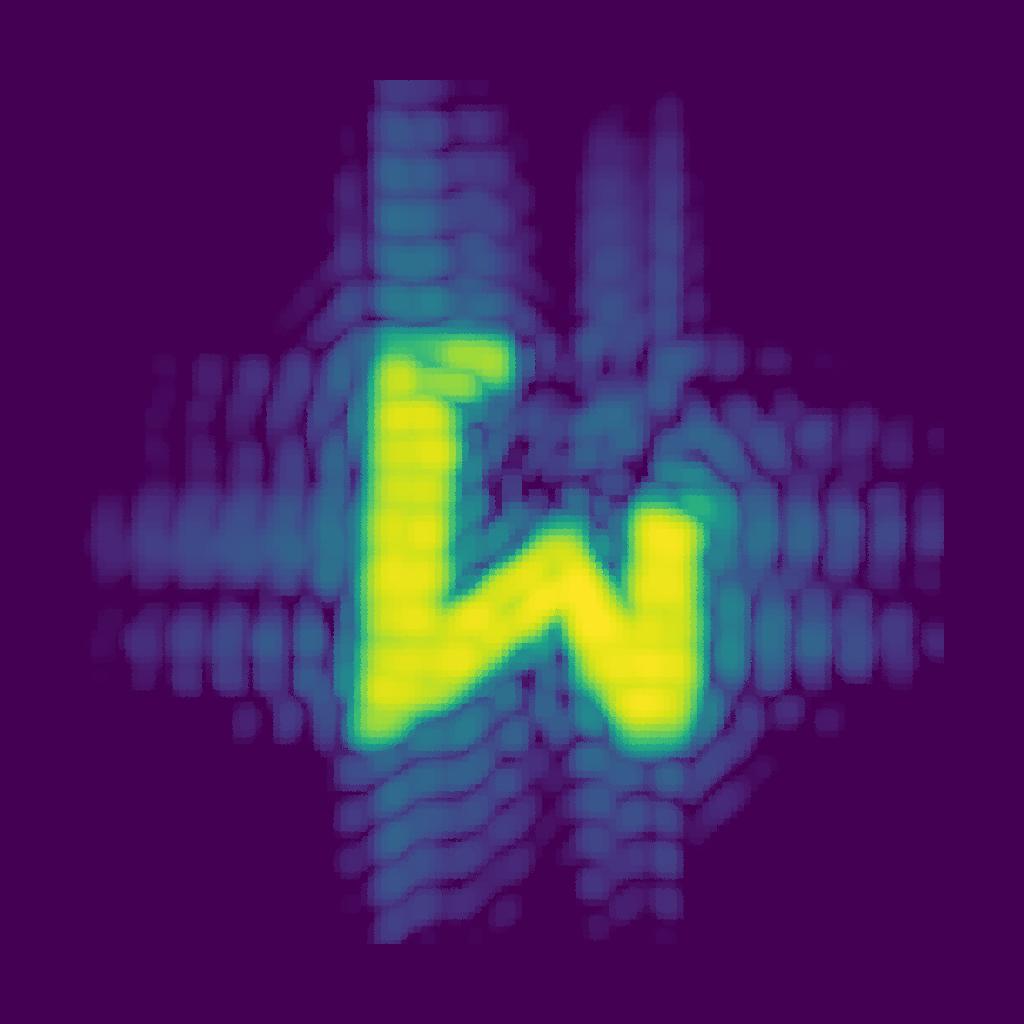} &
			\includegraphics[width=\linewidth]{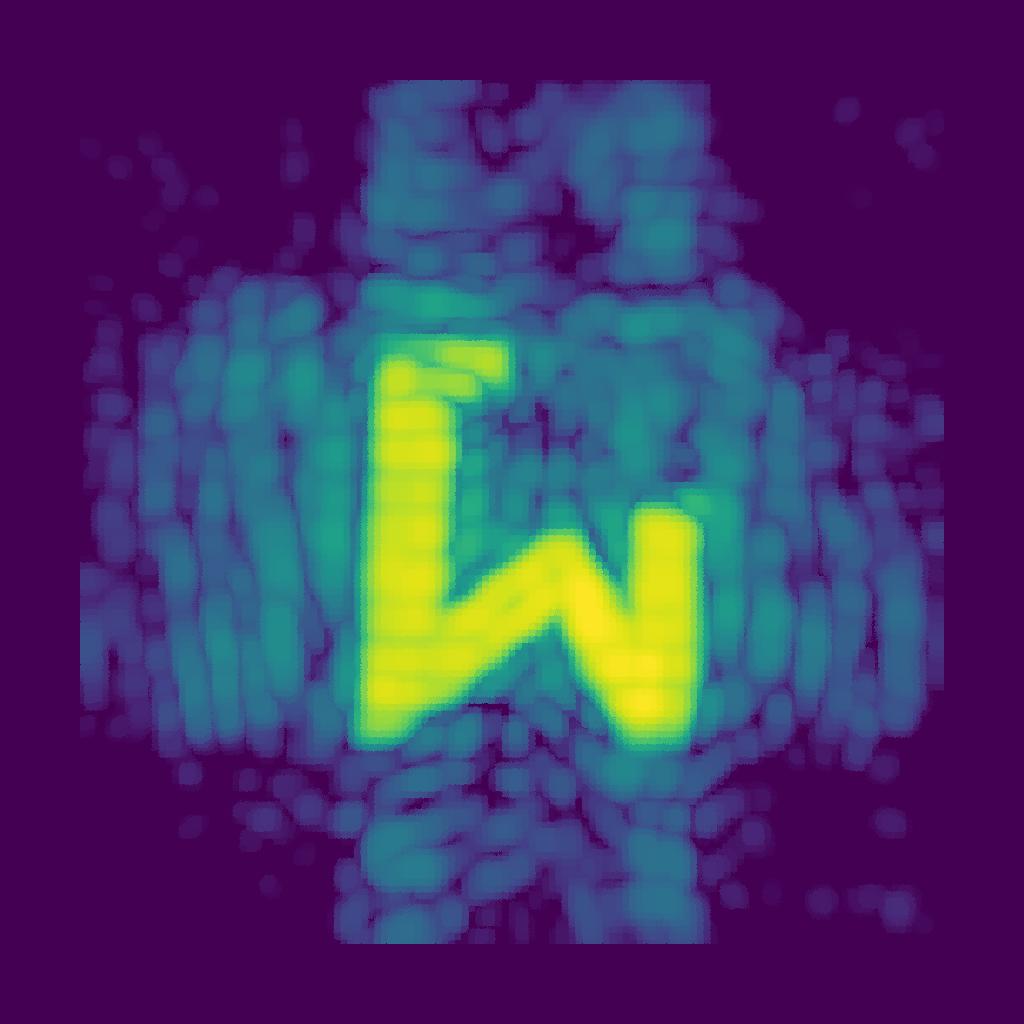} &
			\includegraphics[width=\linewidth]{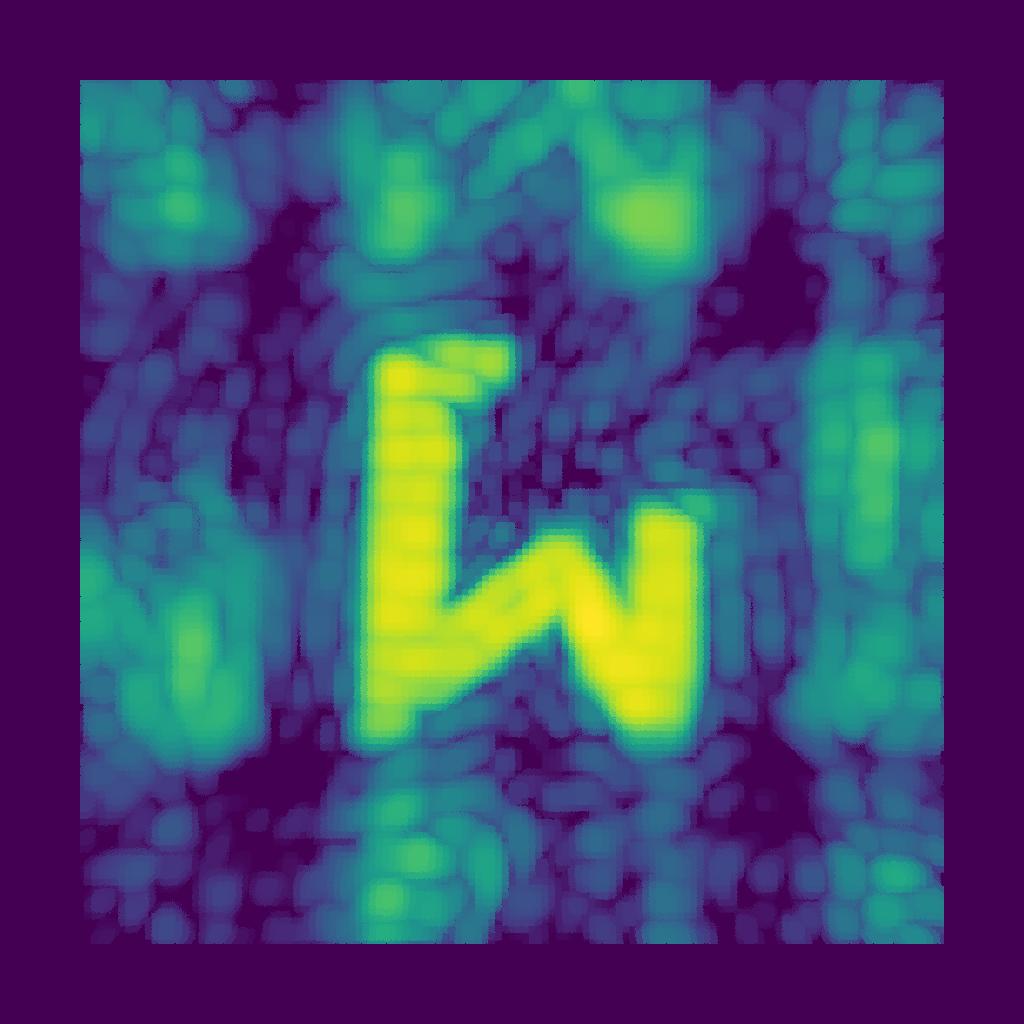} &
			\includegraphics[width=\linewidth]{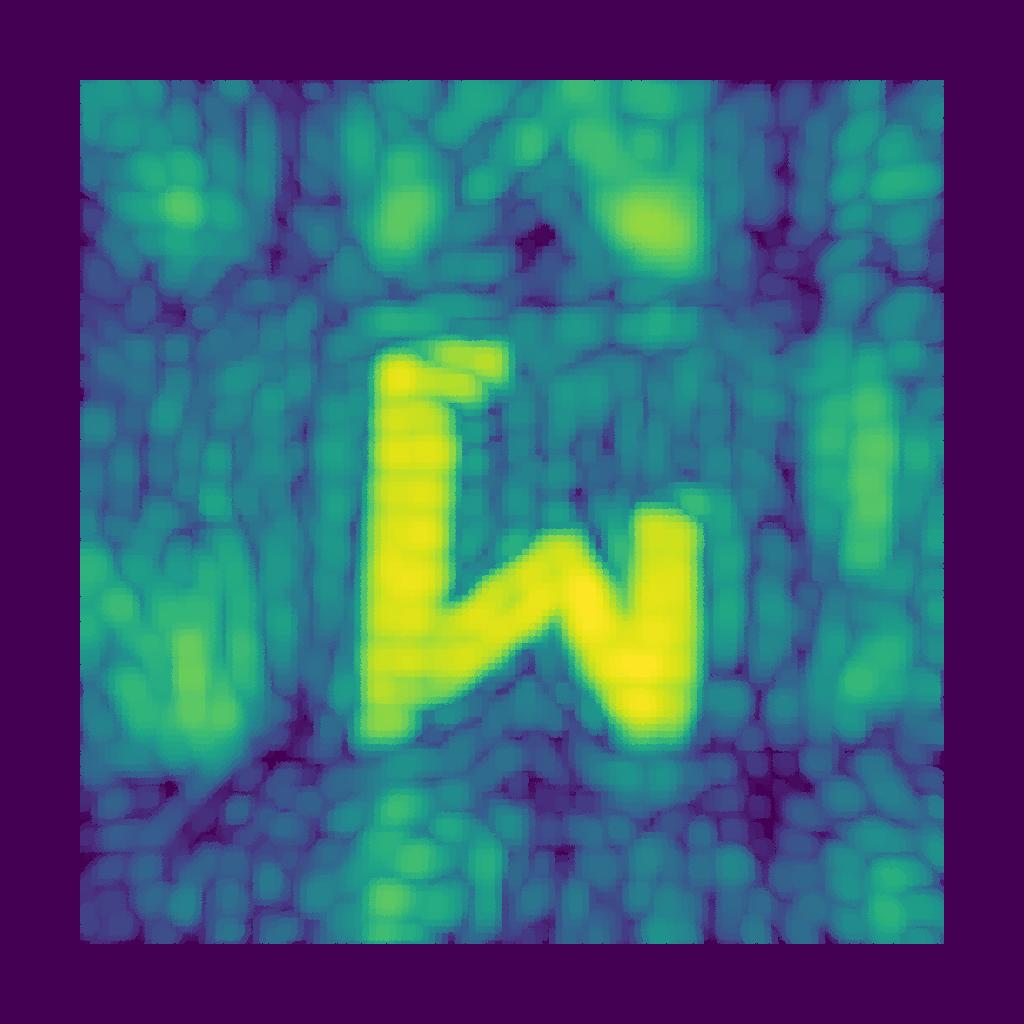} \\
			
			\rotatebox[origin=l]{90}{\small{\hspace{0pt}\obj{Plunger}}} &
			\includegraphics[width=\linewidth]{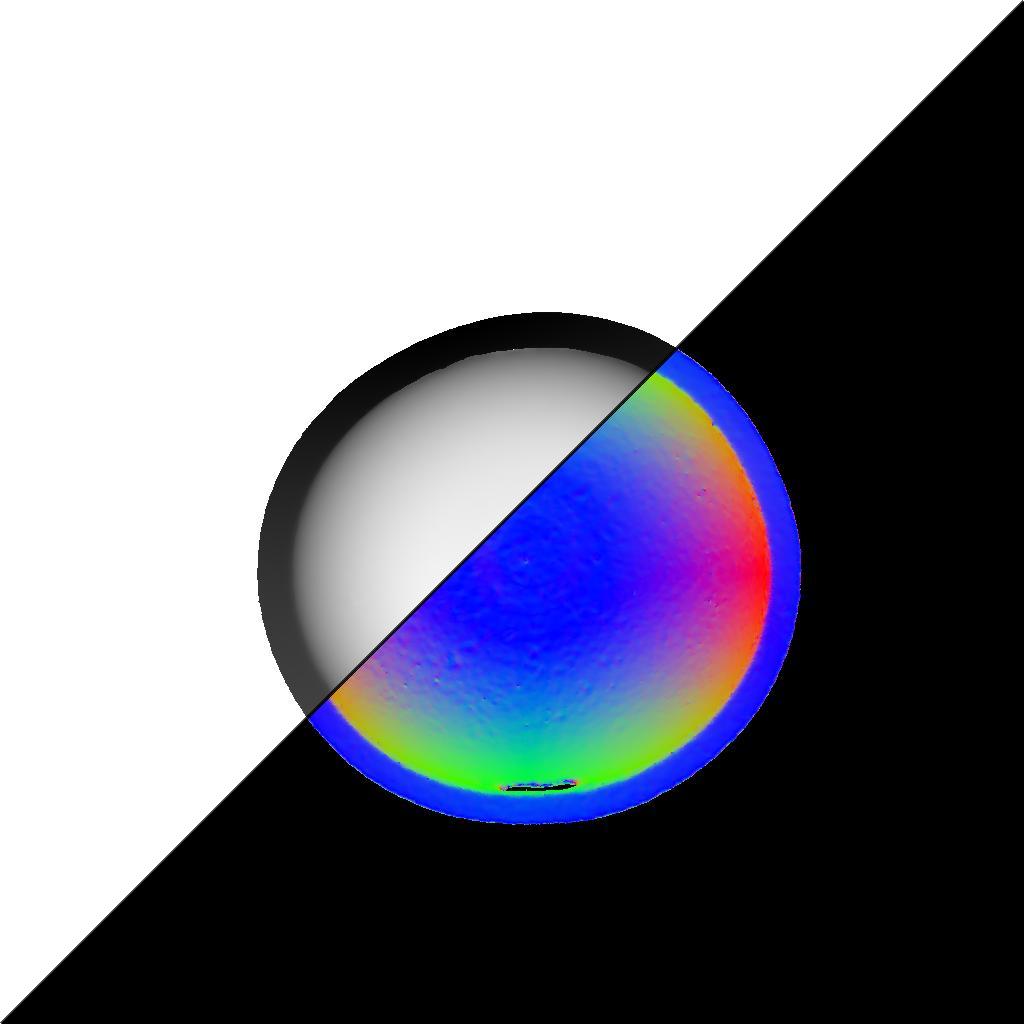} &
			\includegraphics[width=\linewidth]{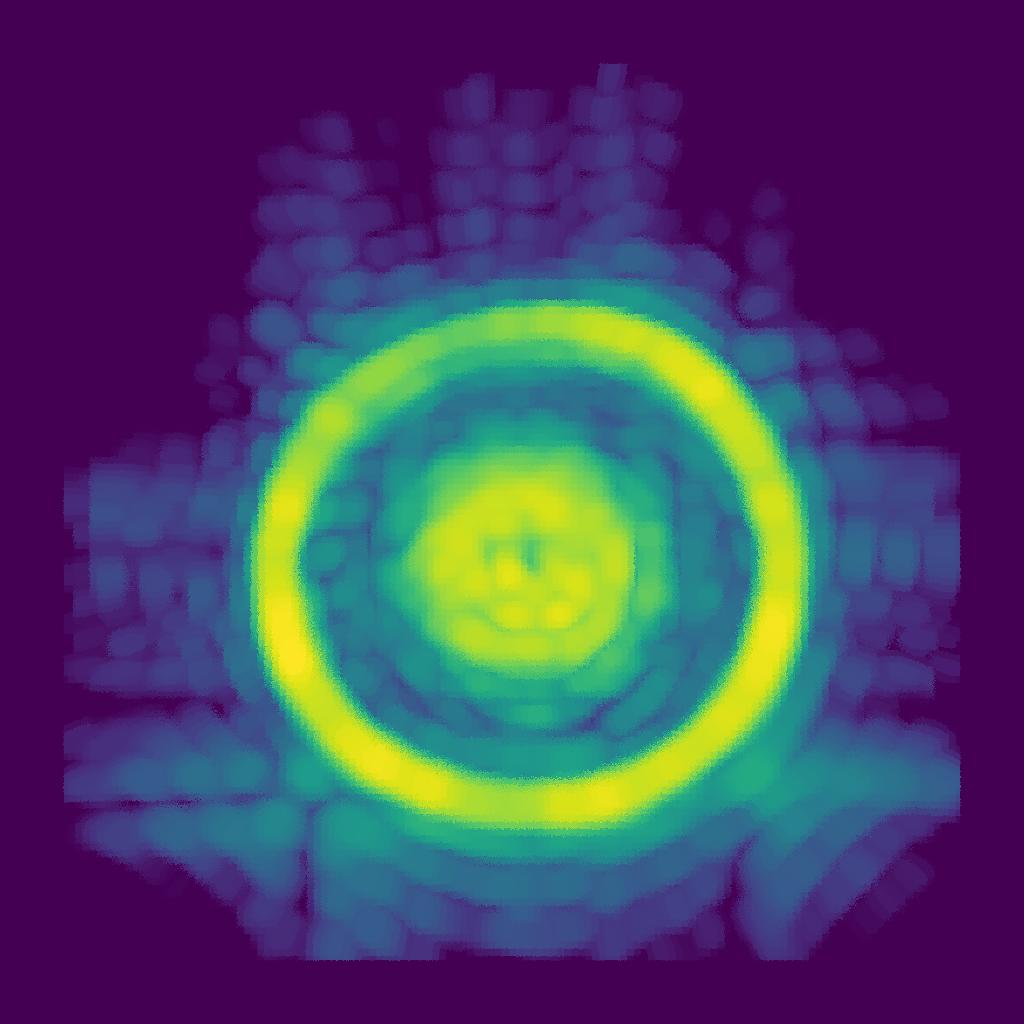} &
			\includegraphics[width=\linewidth]{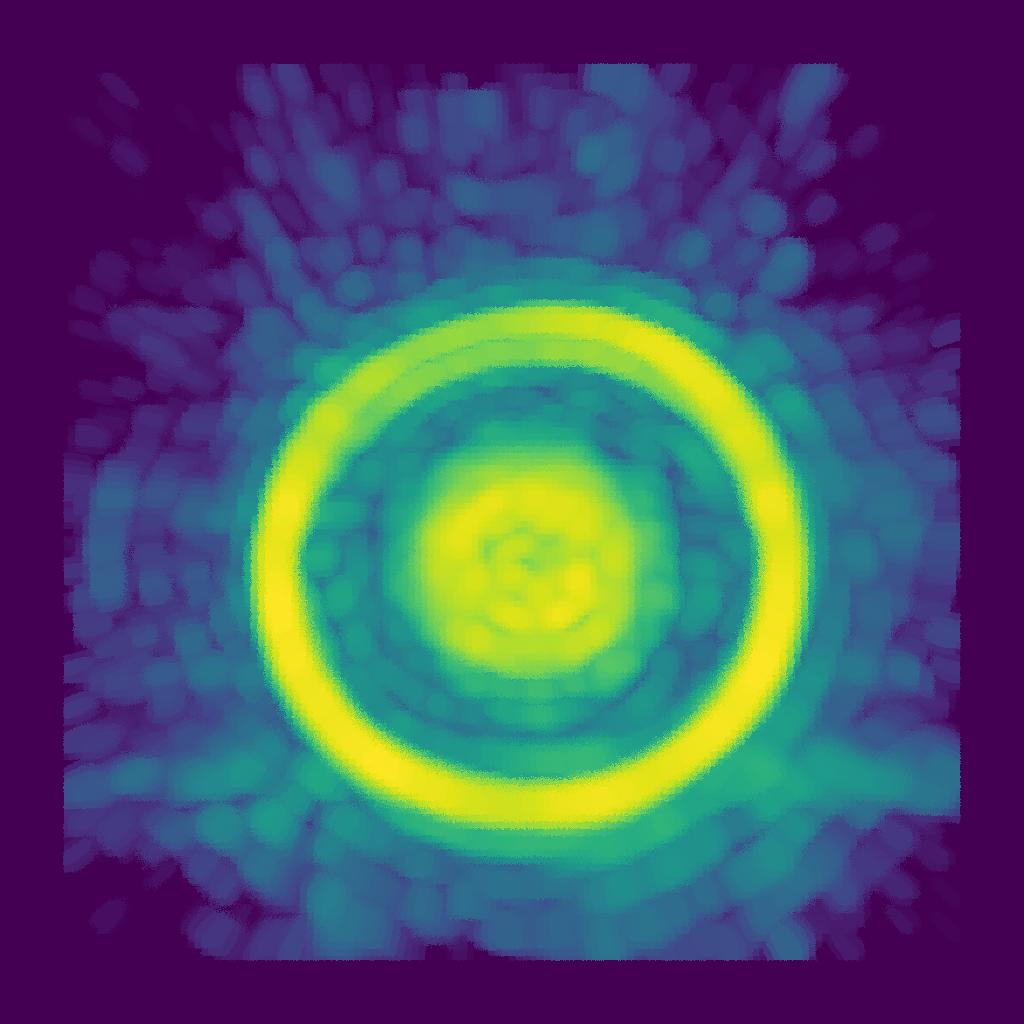} &
			\includegraphics[width=\linewidth]{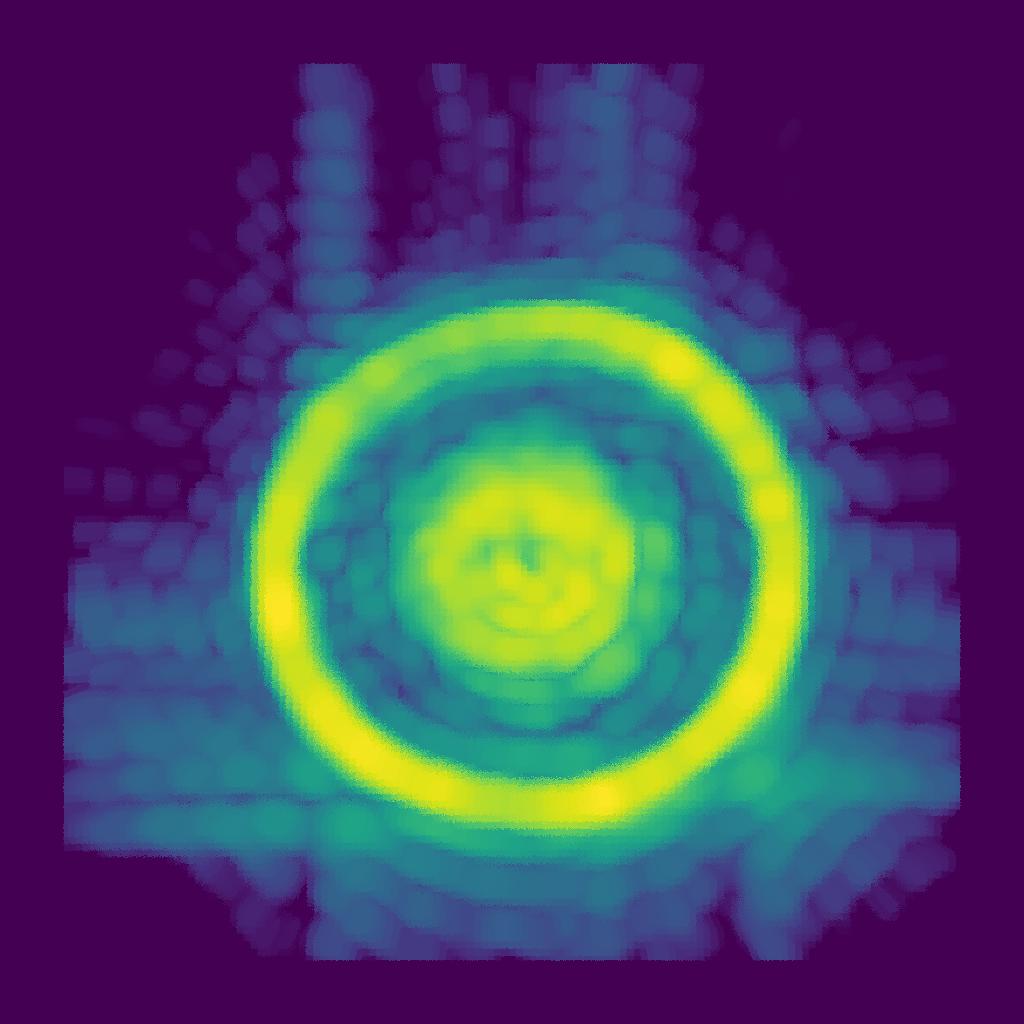} &
			\includegraphics[width=\linewidth]{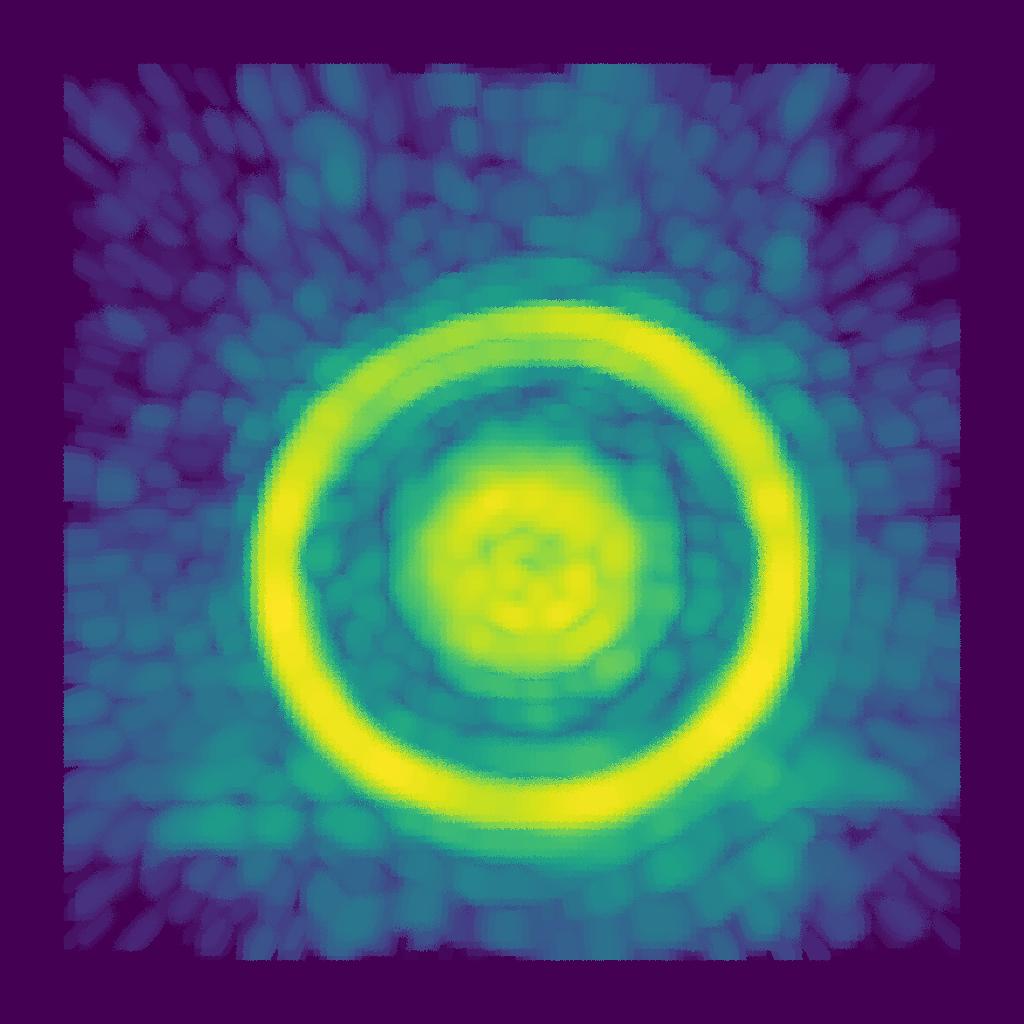} &
			\includegraphics[width=\linewidth]{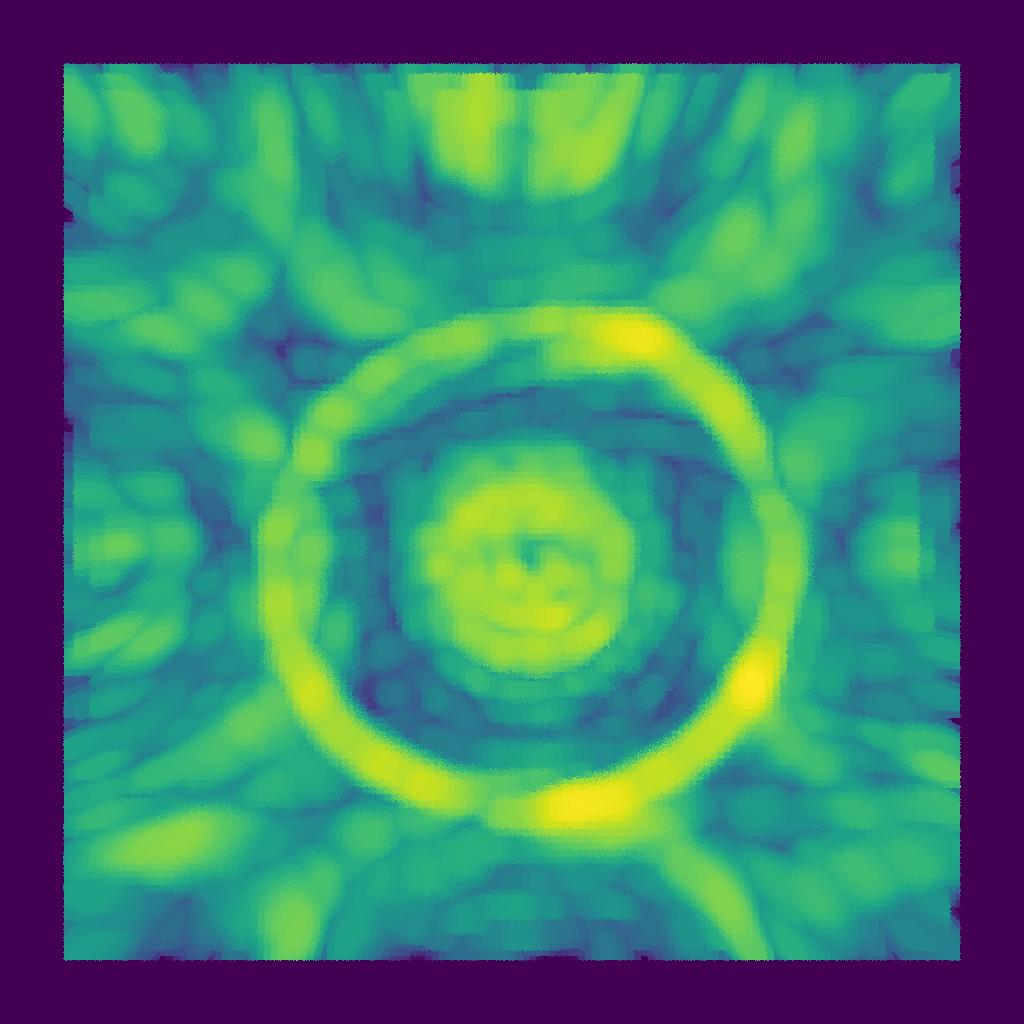} &
			\includegraphics[width=\linewidth]{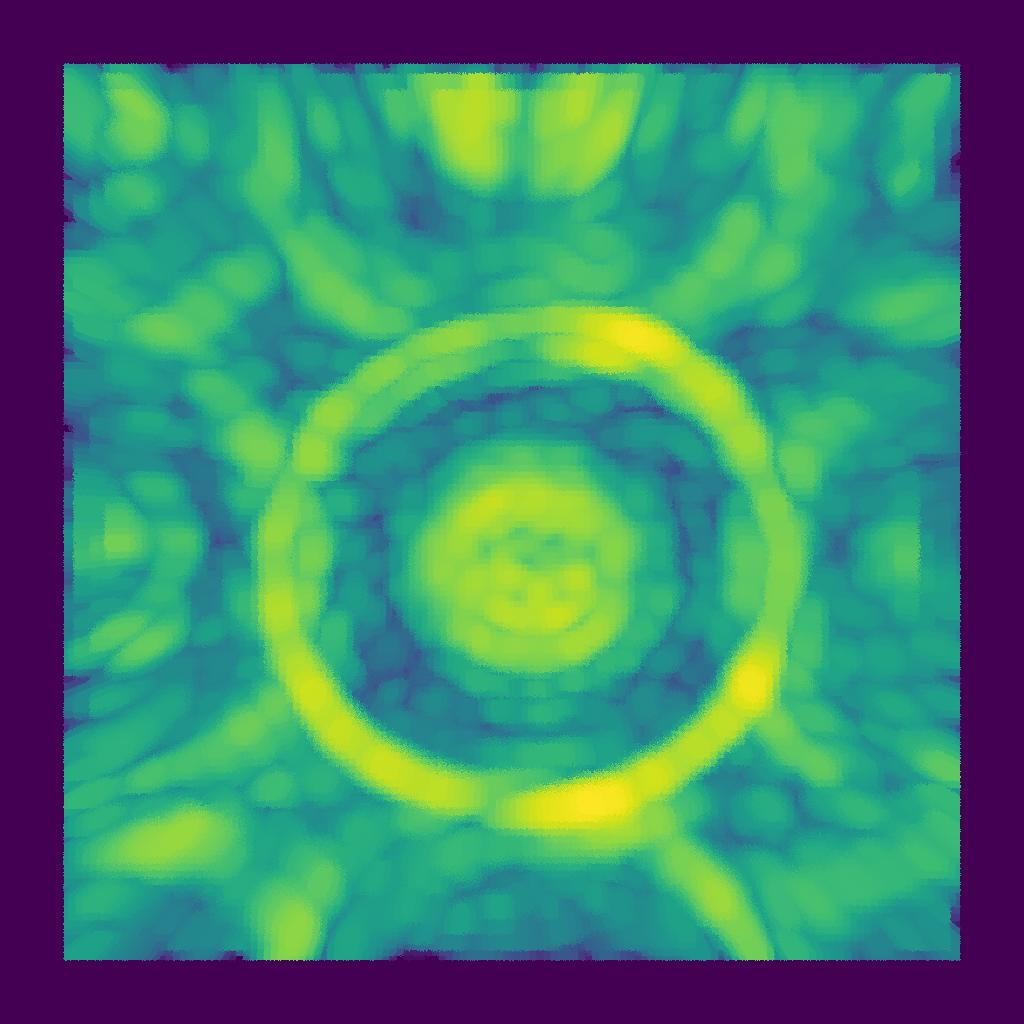} \\
			
			\rotatebox[origin=l]{90}{\small{\hspace{0pt}\obj{Hand Pr.}}} &
			\includegraphics[width=\linewidth]{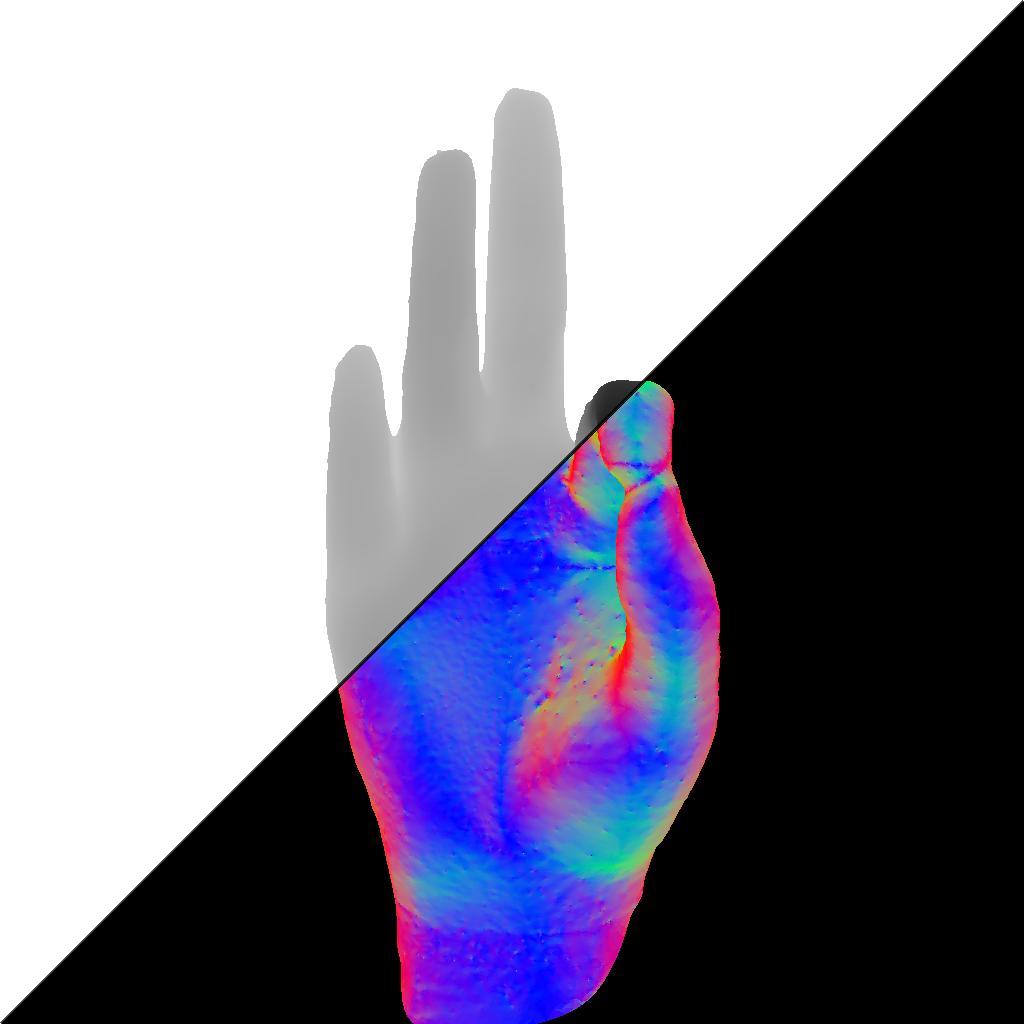} &
			\includegraphics[width=\linewidth]{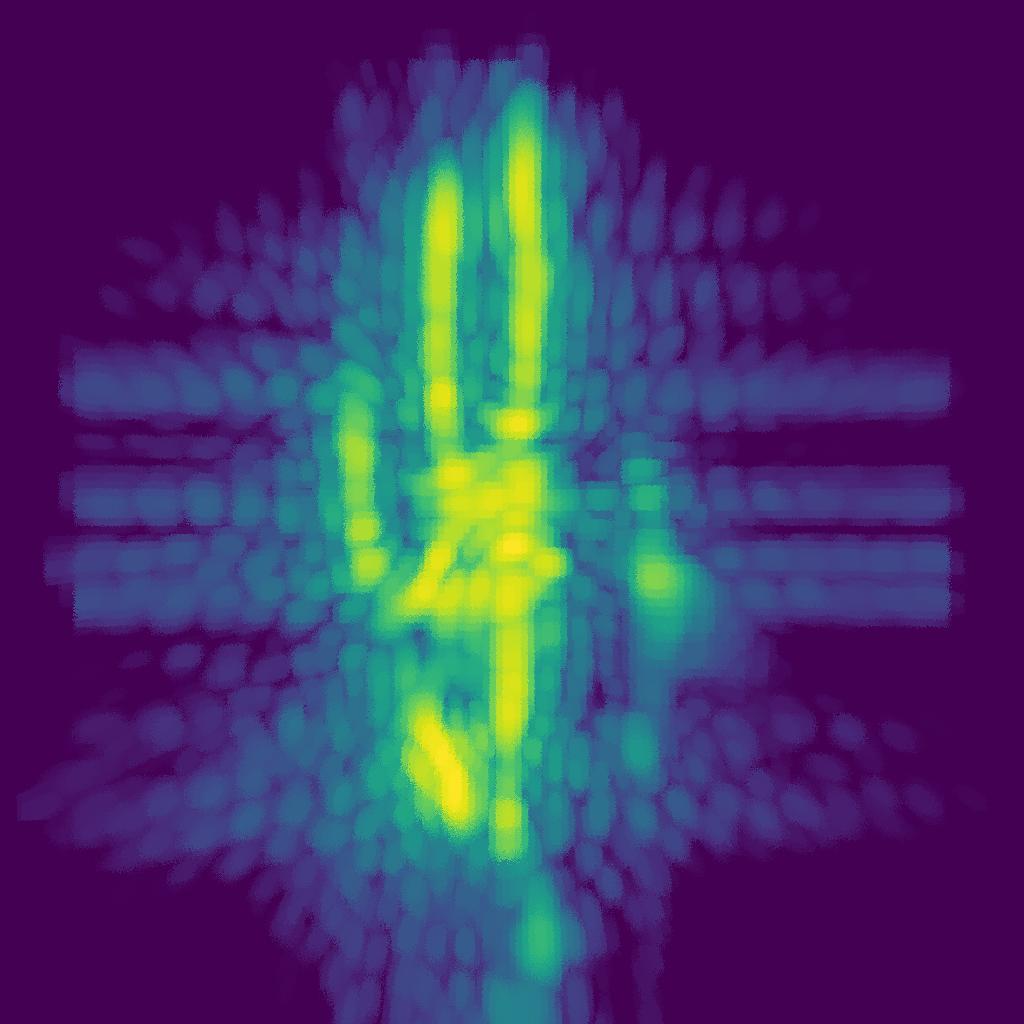} &
			\includegraphics[width=\linewidth]{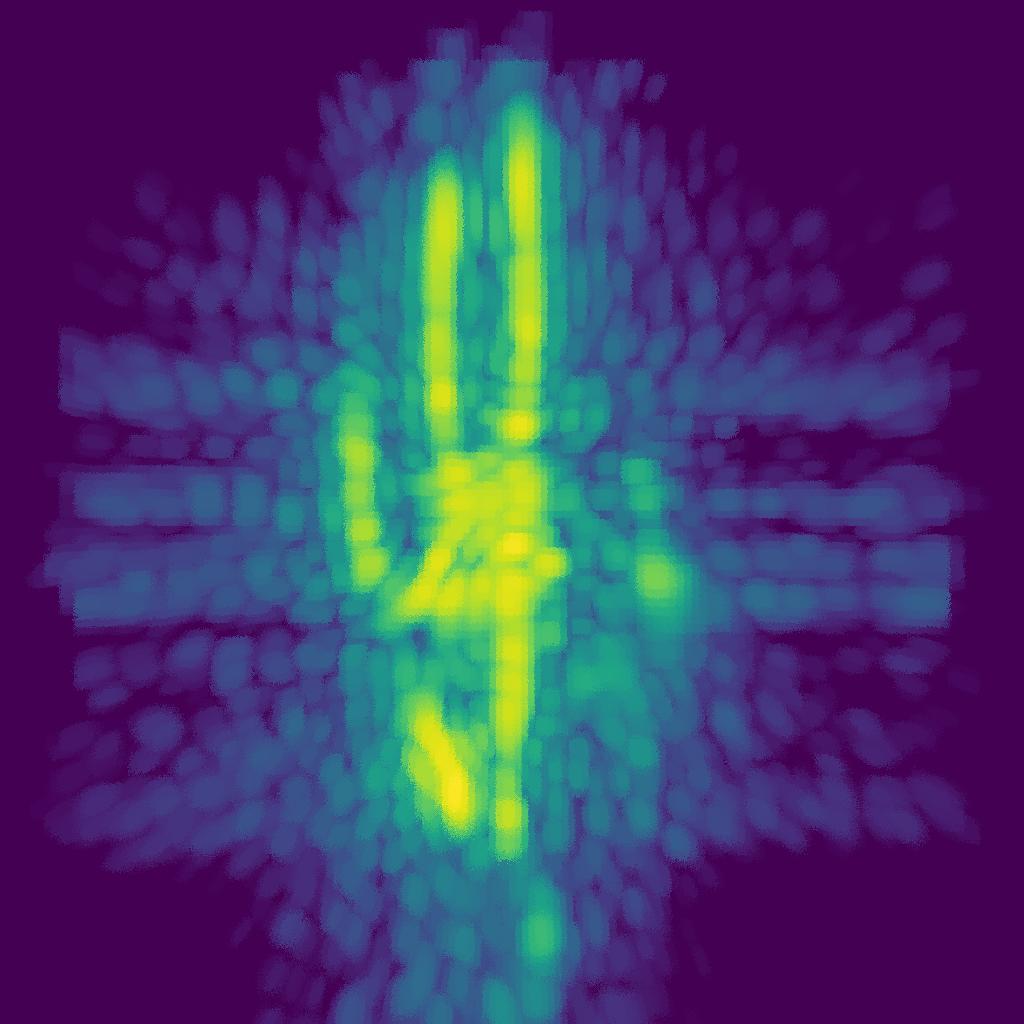} &
			\includegraphics[width=\linewidth]{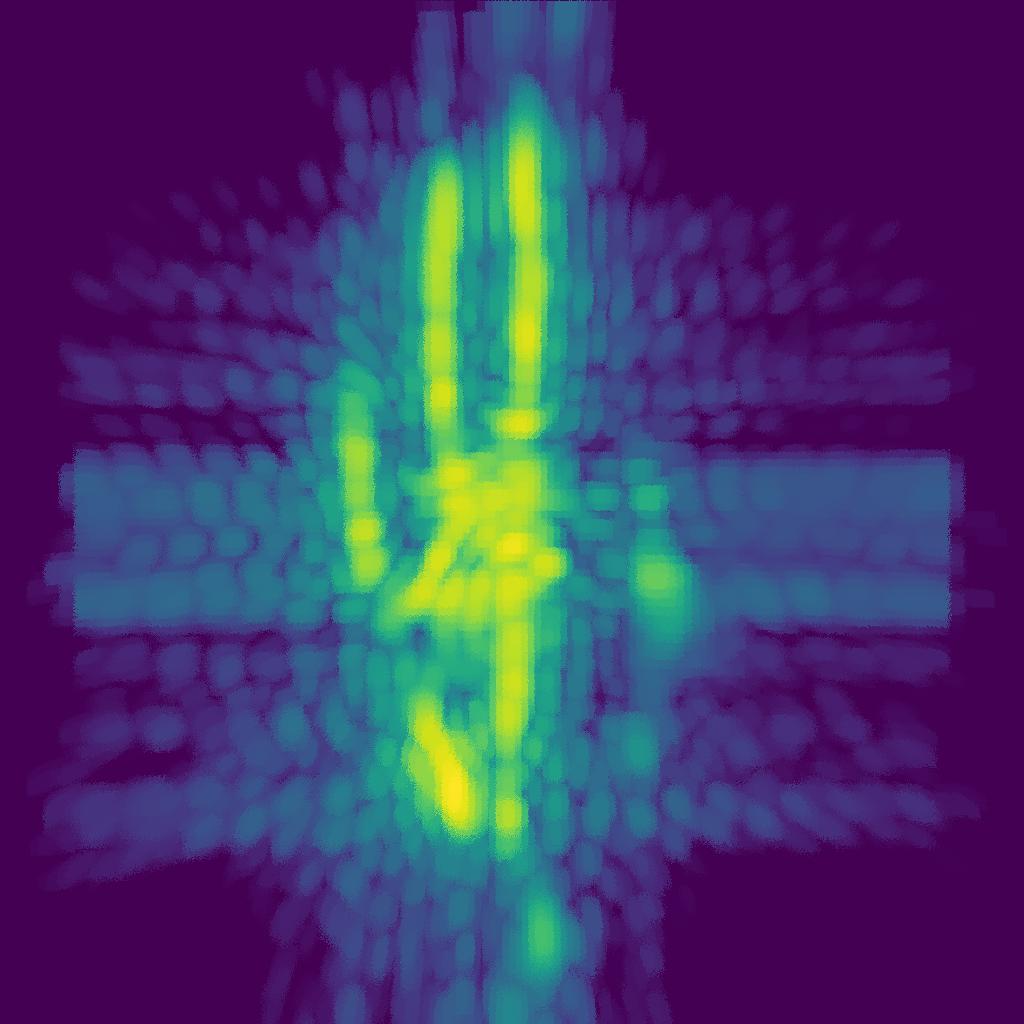} &
			\includegraphics[width=\linewidth]{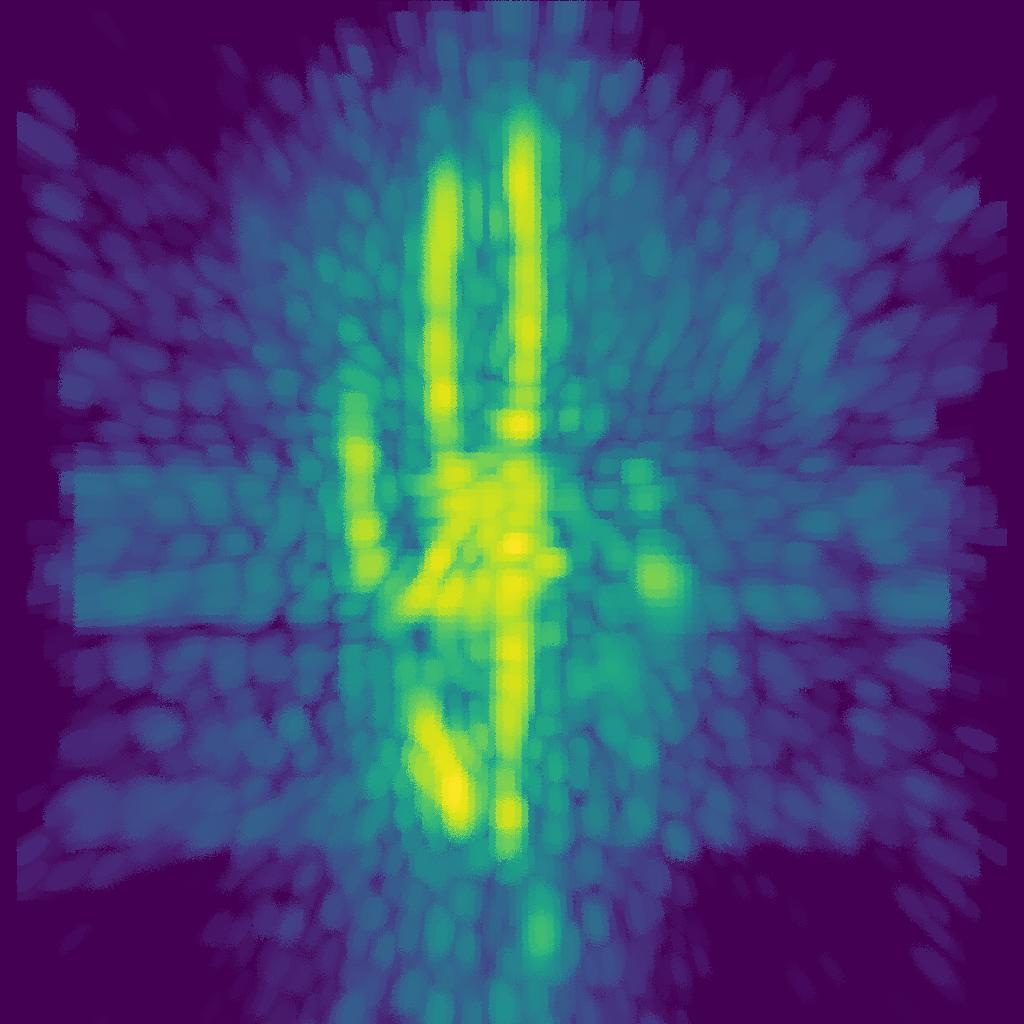} &
			\includegraphics[width=\linewidth]{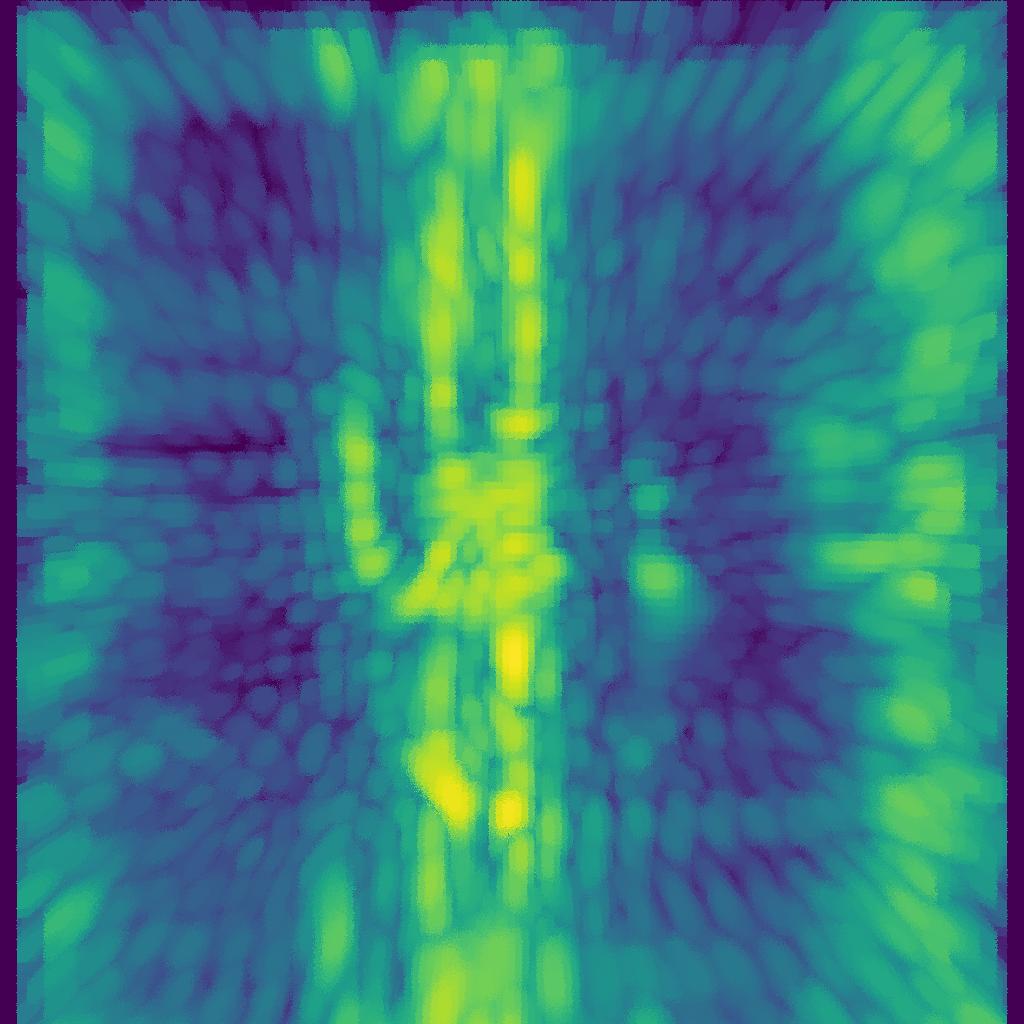} &
			\includegraphics[width=\linewidth]{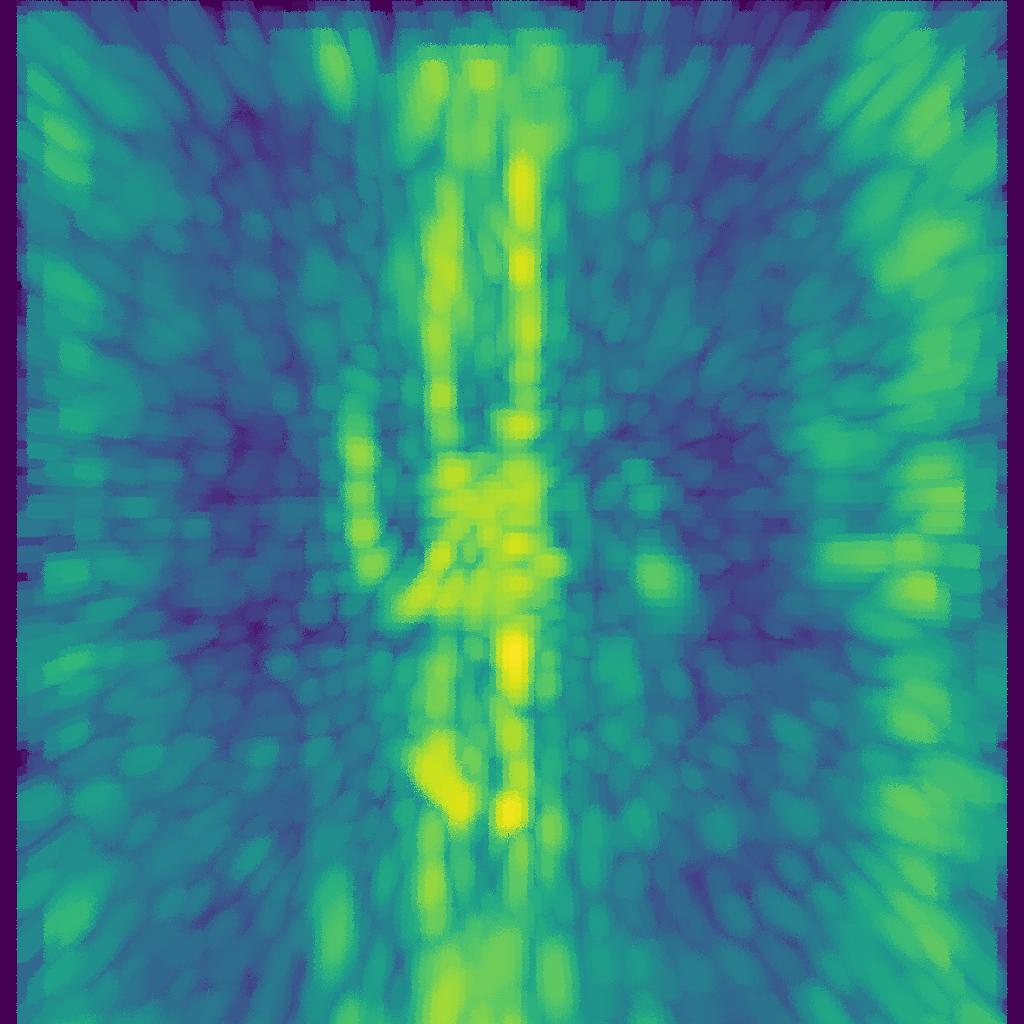} \\
			
			\rotatebox[origin=l]{90}{\small{\hspace{0pt}\obj{S1 Hand}}} &
			\includegraphics[width=\linewidth]{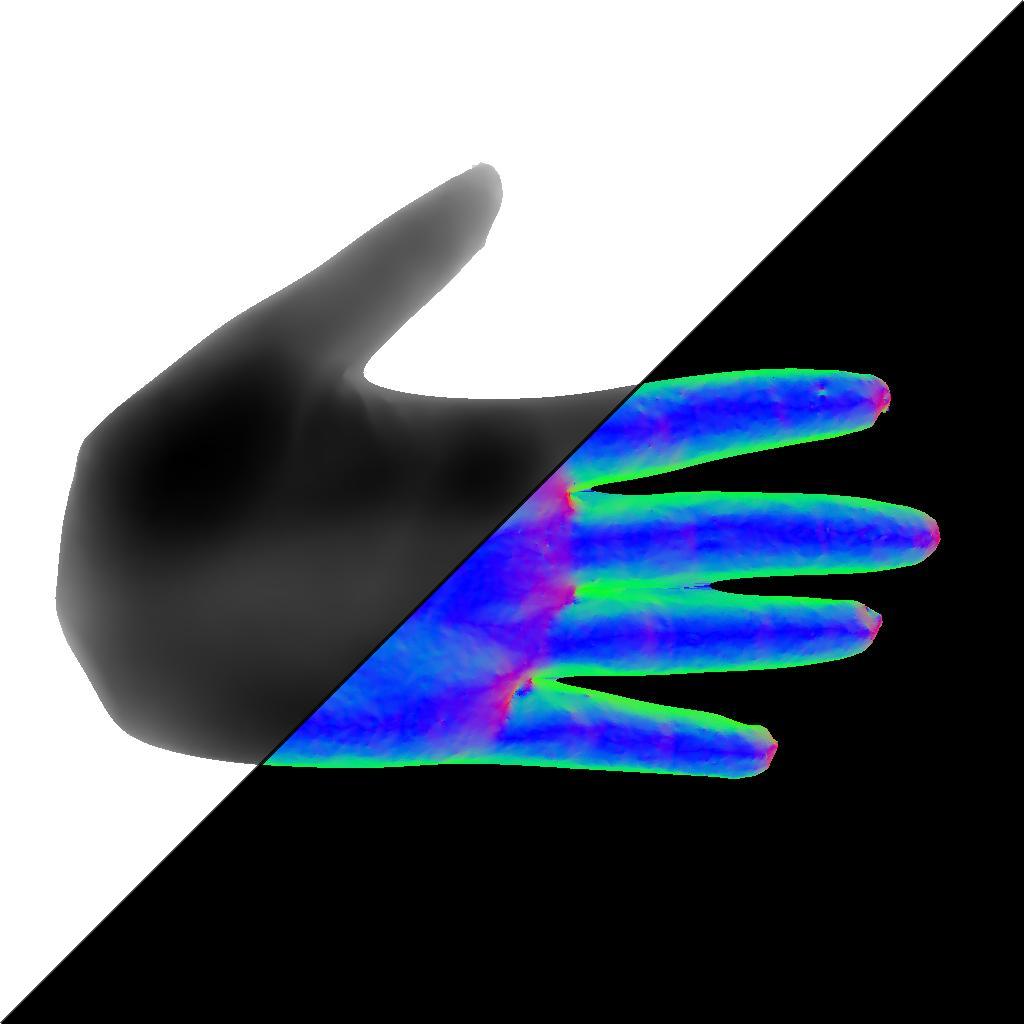} &
			\includegraphics[width=\linewidth]{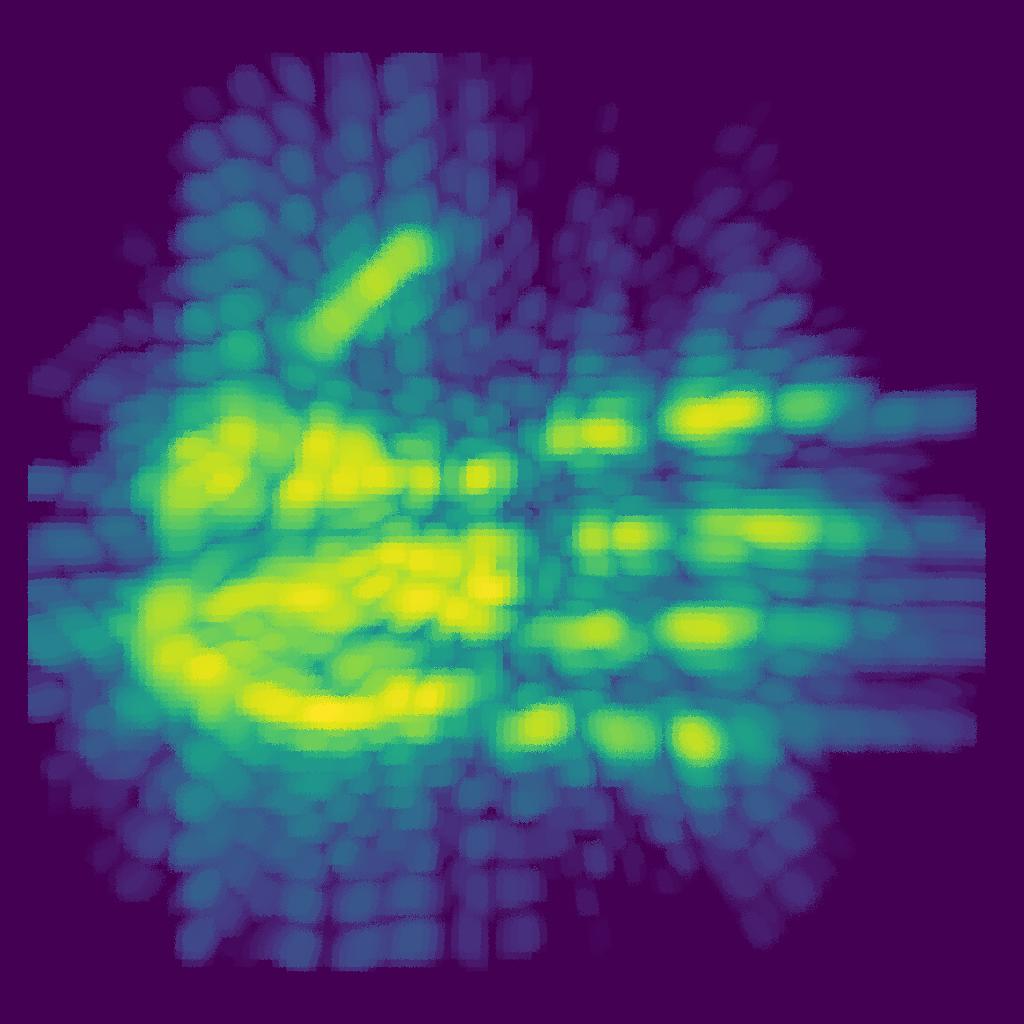} &
			\includegraphics[width=\linewidth]{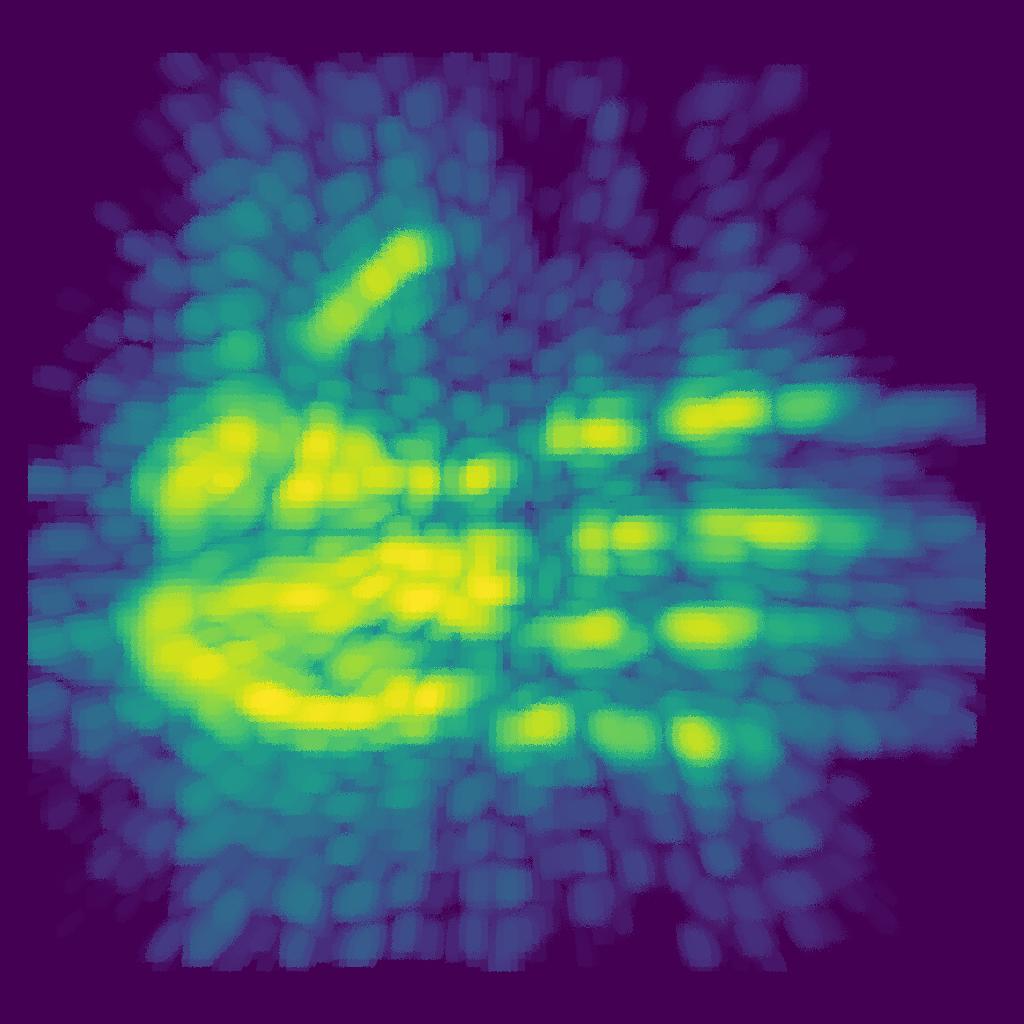} &
			\includegraphics[width=\linewidth]{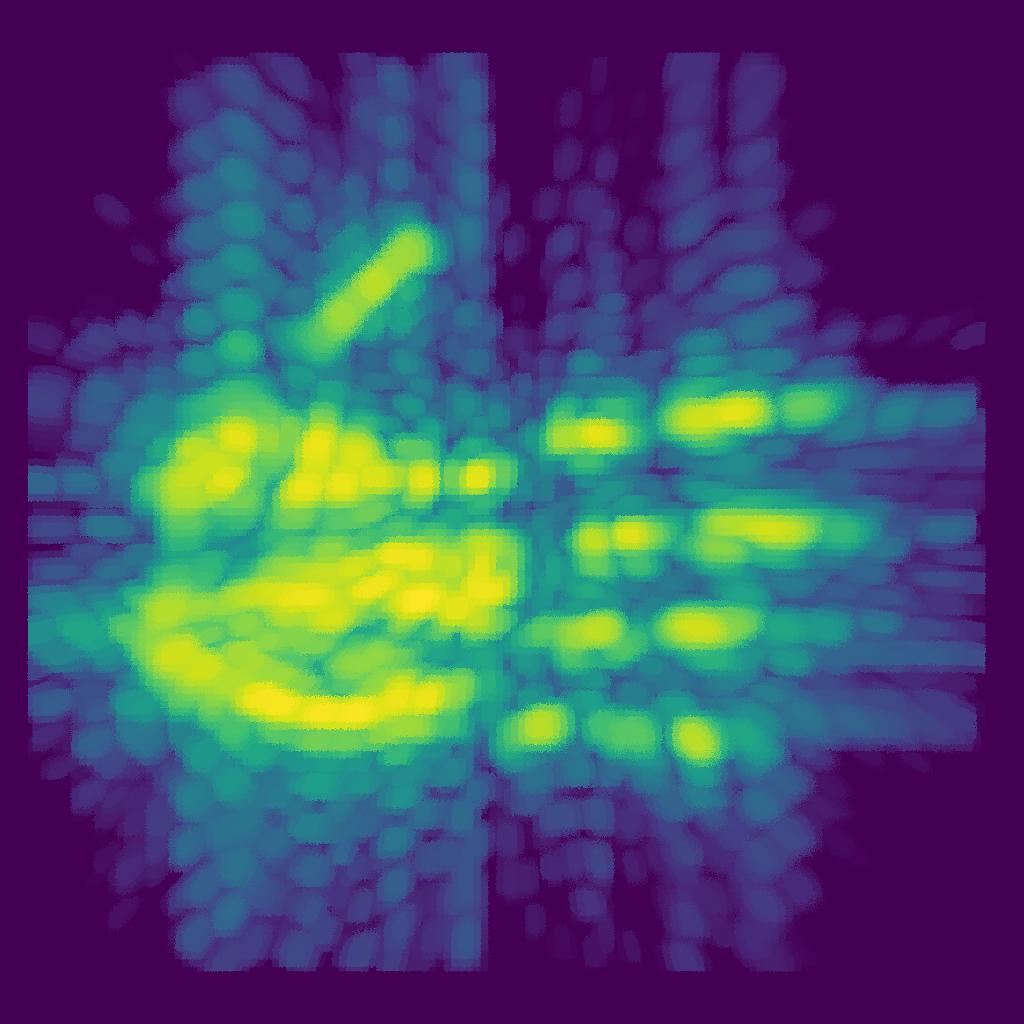} &
			\includegraphics[width=\linewidth]{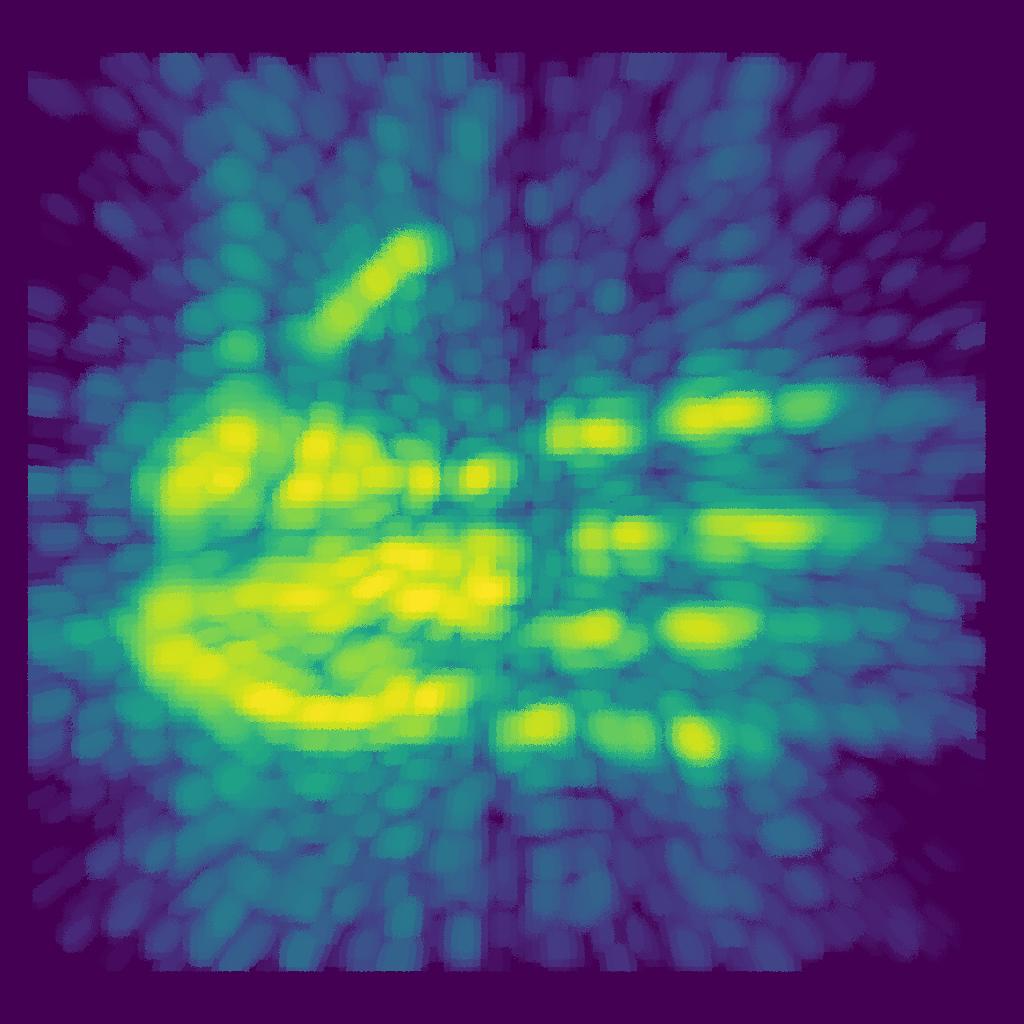} &
			\includegraphics[width=\linewidth]{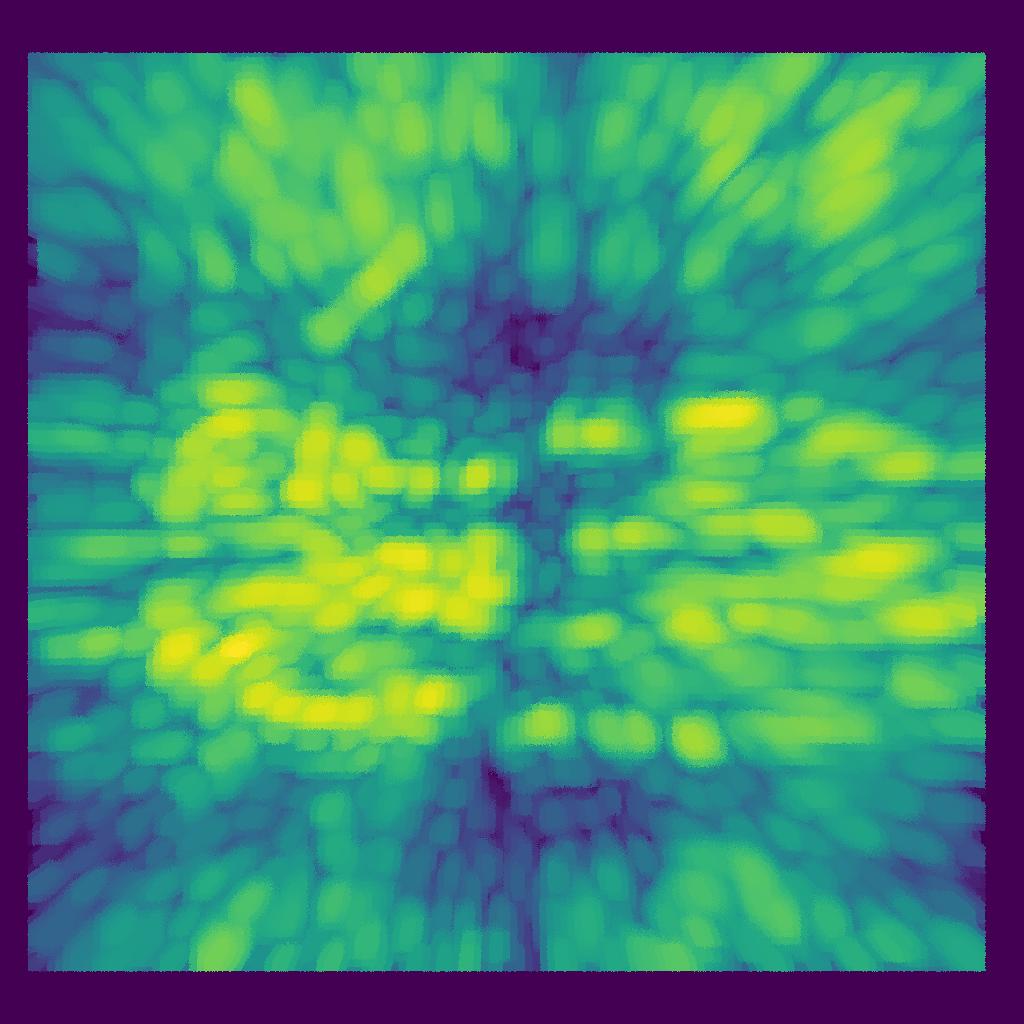} &
			\includegraphics[width=\linewidth]{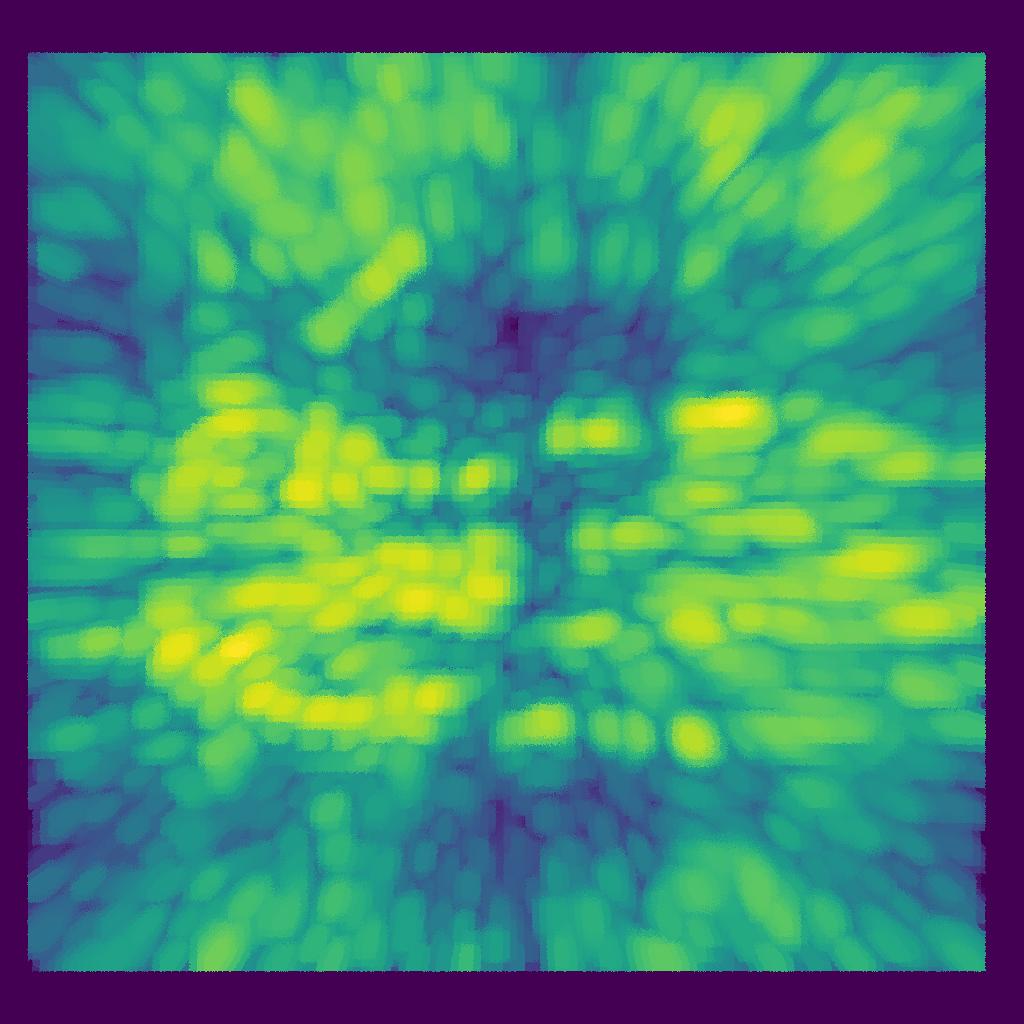} \\
			
			\rotatebox[origin=l]{90}{\small{\hspace{5pt}\obj{Disk}}} &
			\includegraphics[width=\linewidth]{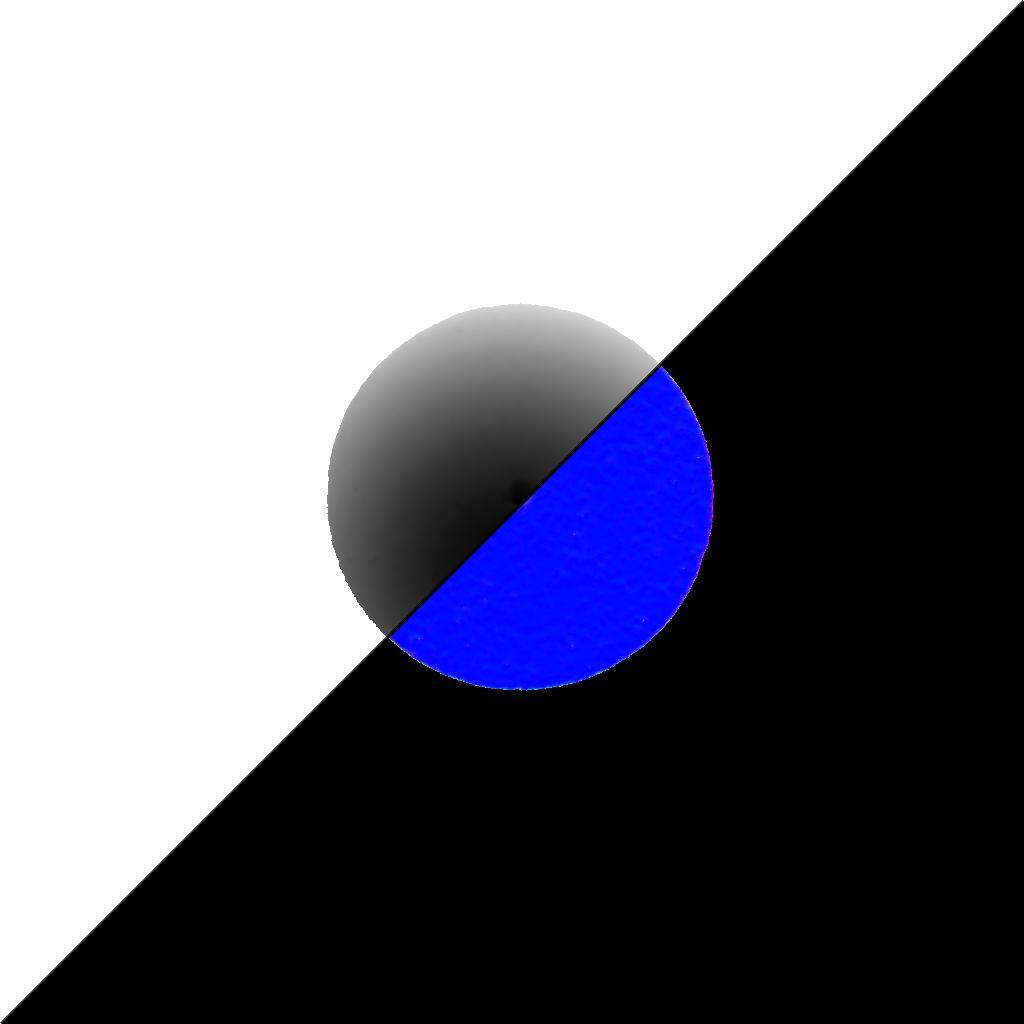} &
			\includegraphics[width=\linewidth]{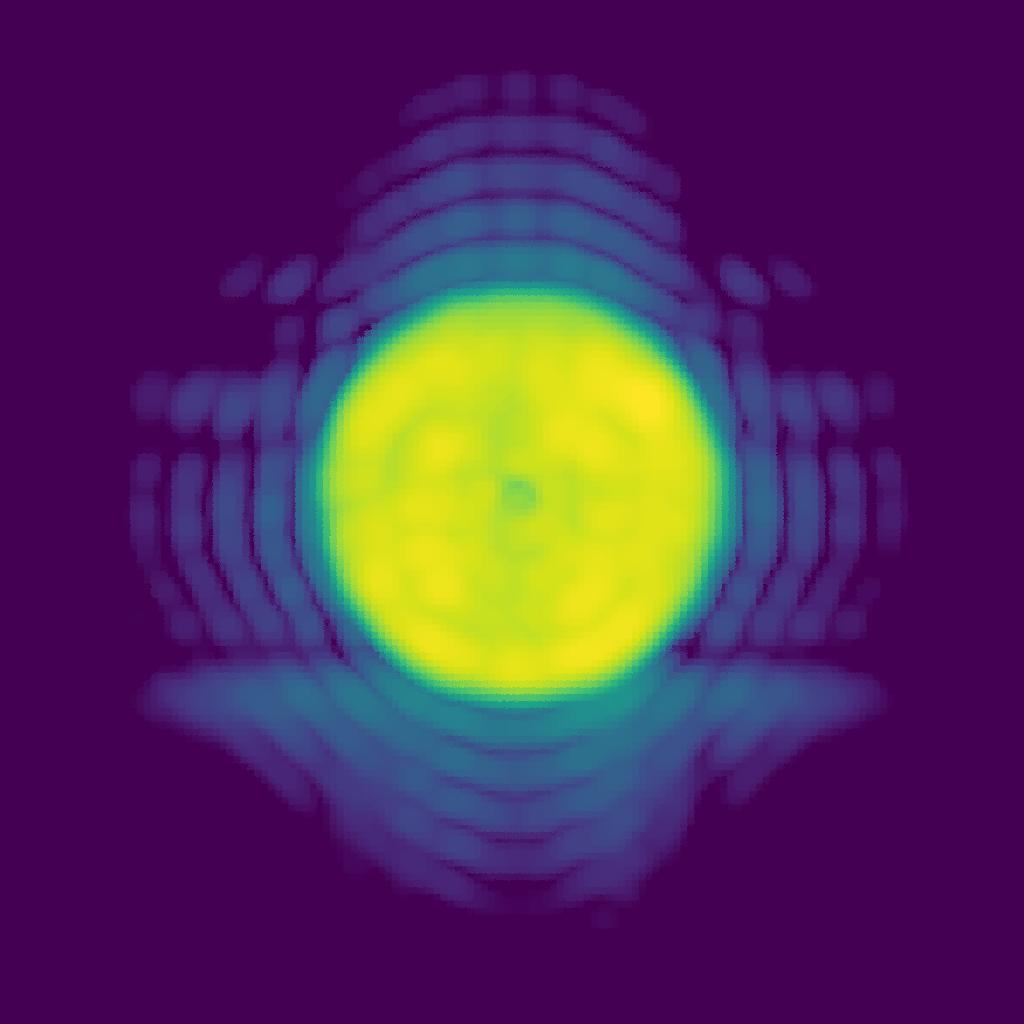} &
			\includegraphics[width=\linewidth]{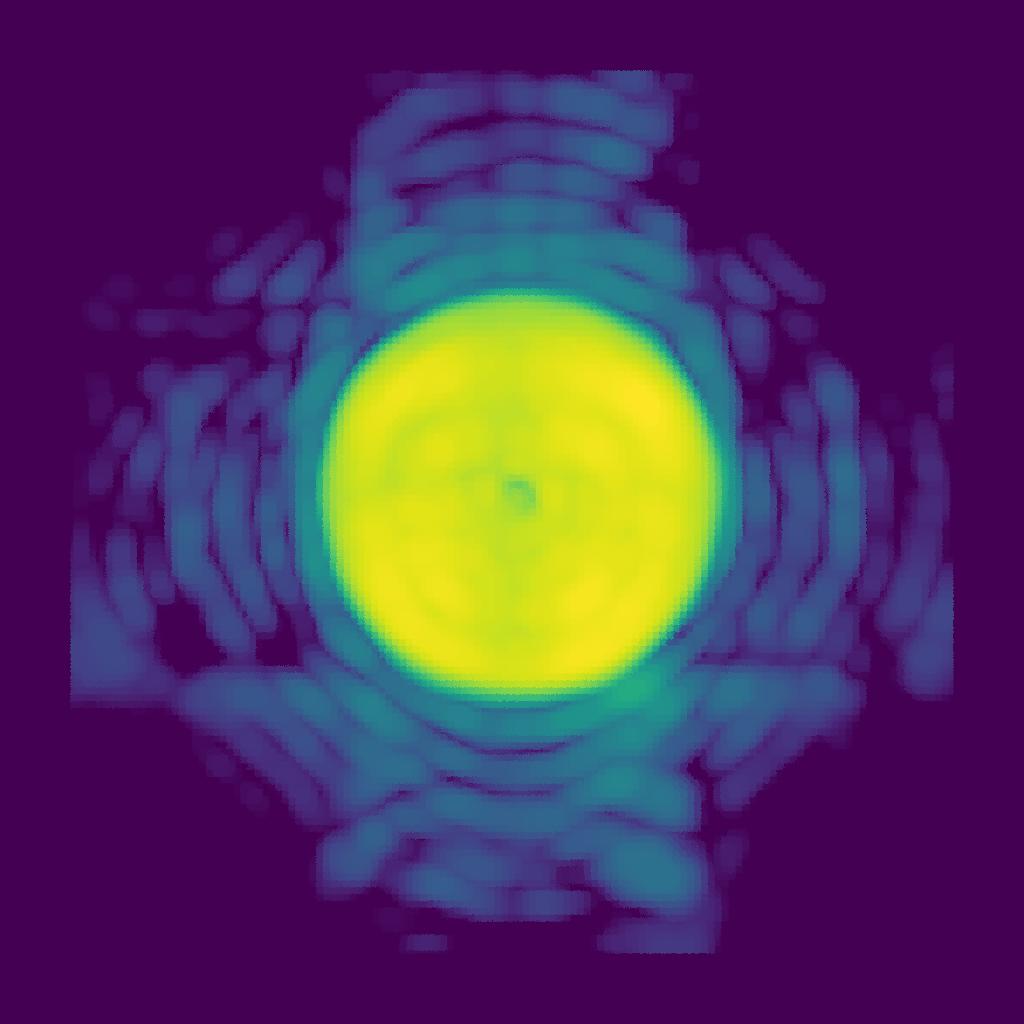} &
			\includegraphics[width=\linewidth]{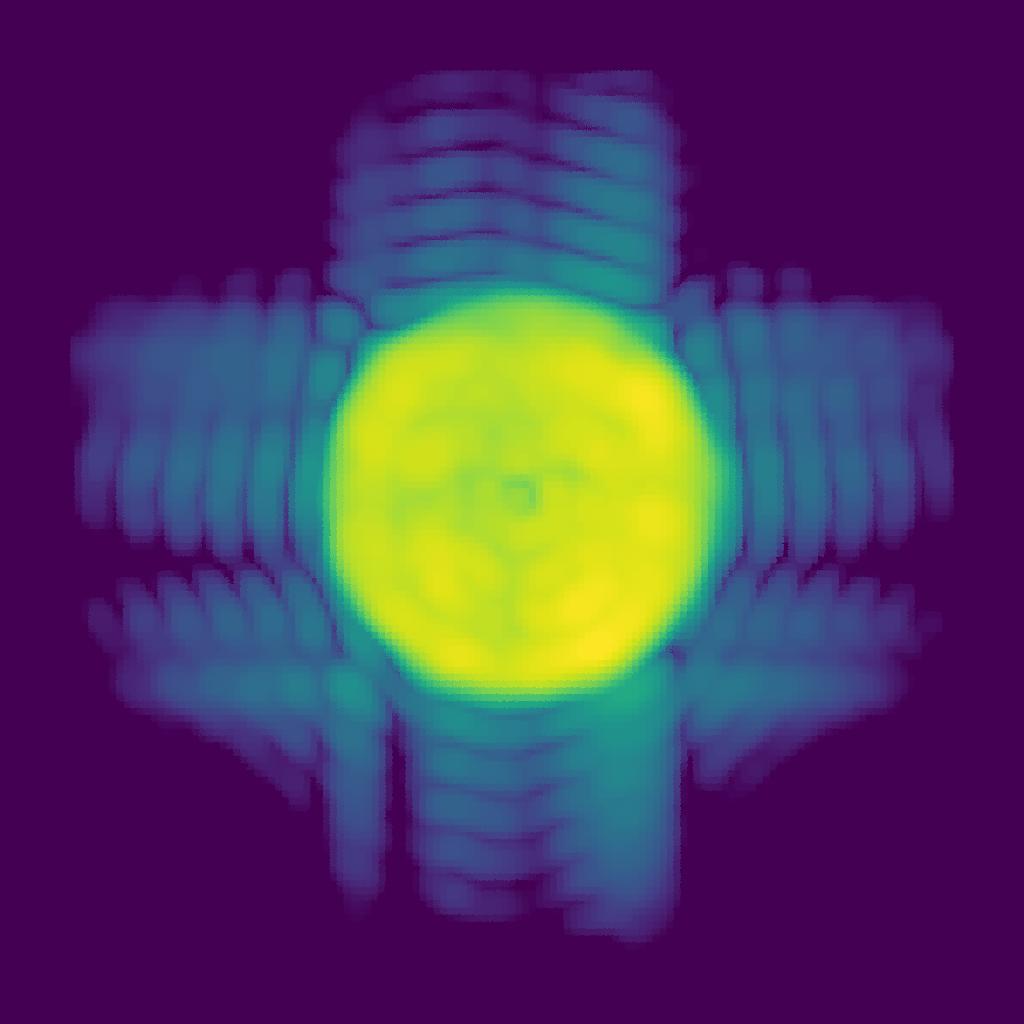} &
			\includegraphics[width=\linewidth]{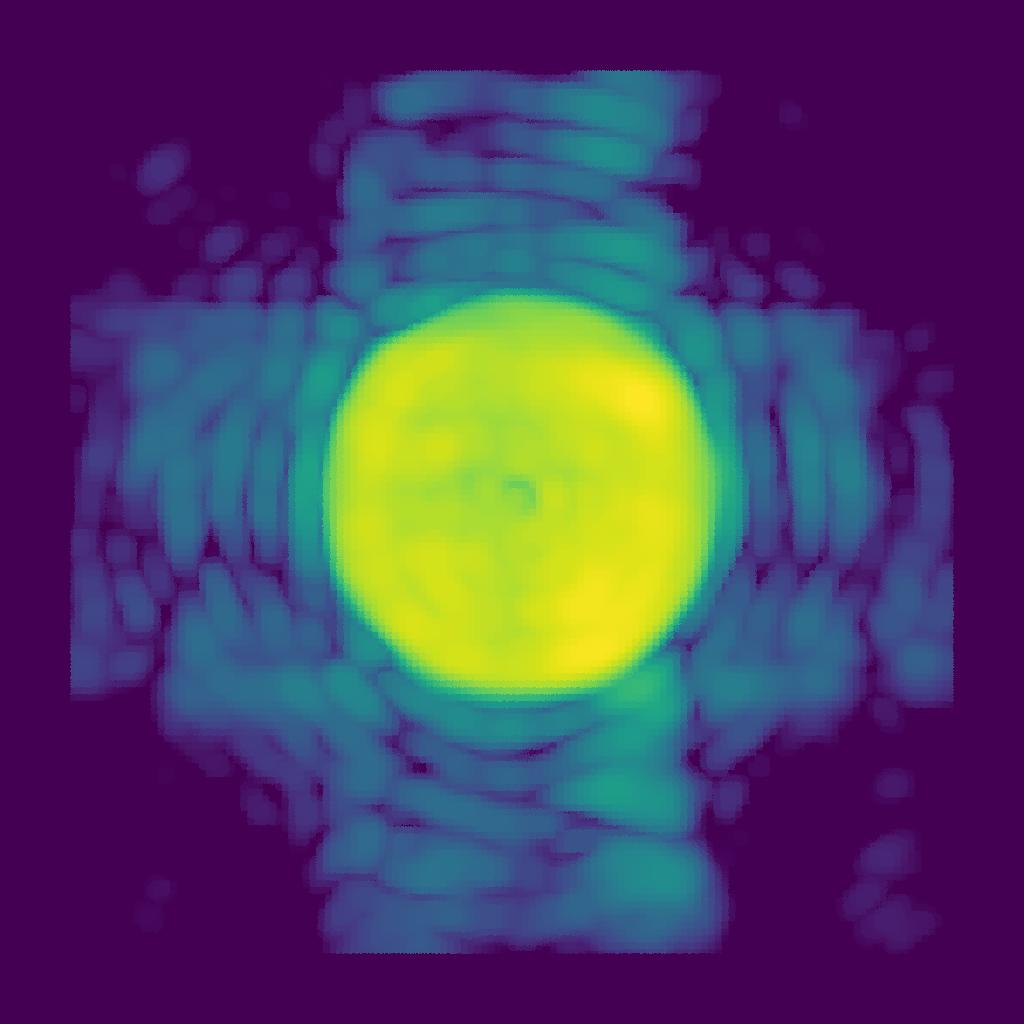} &
			\includegraphics[width=\linewidth]{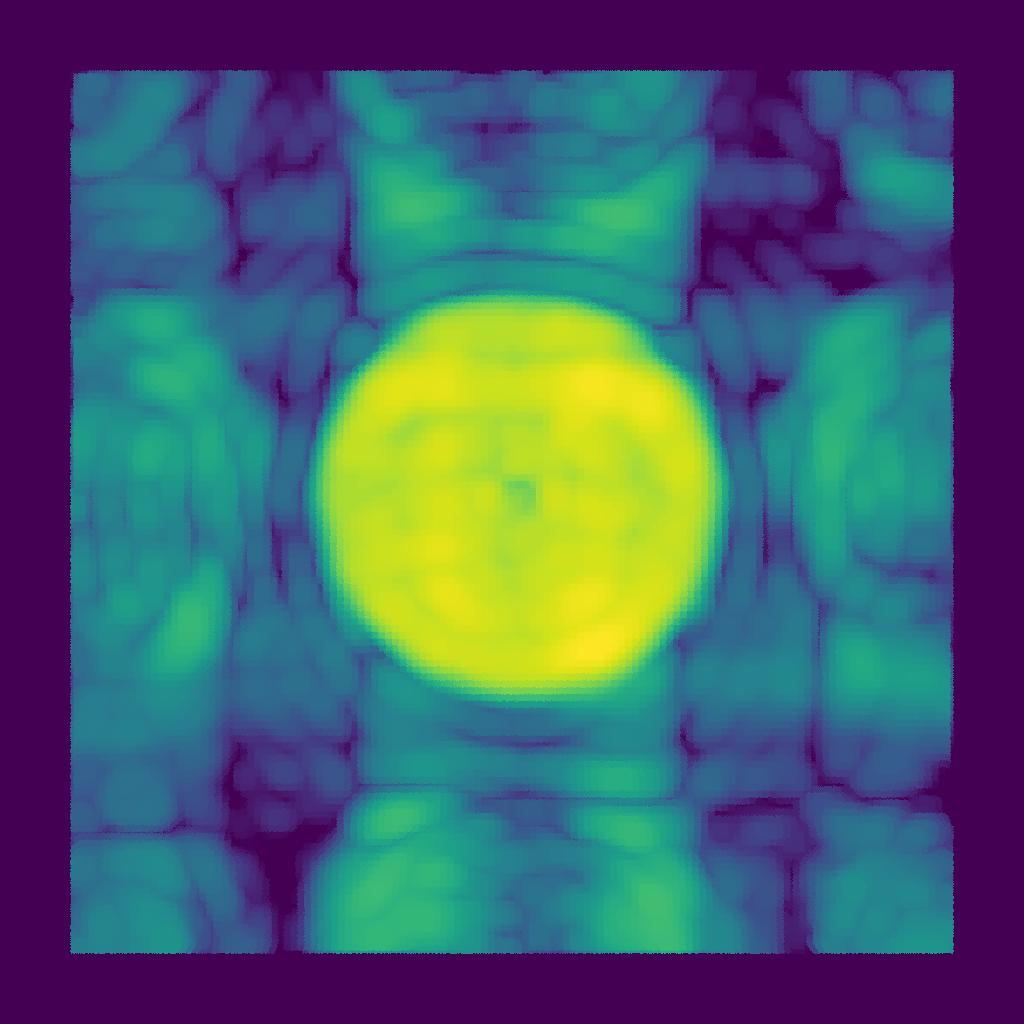} &
			\includegraphics[width=\linewidth]{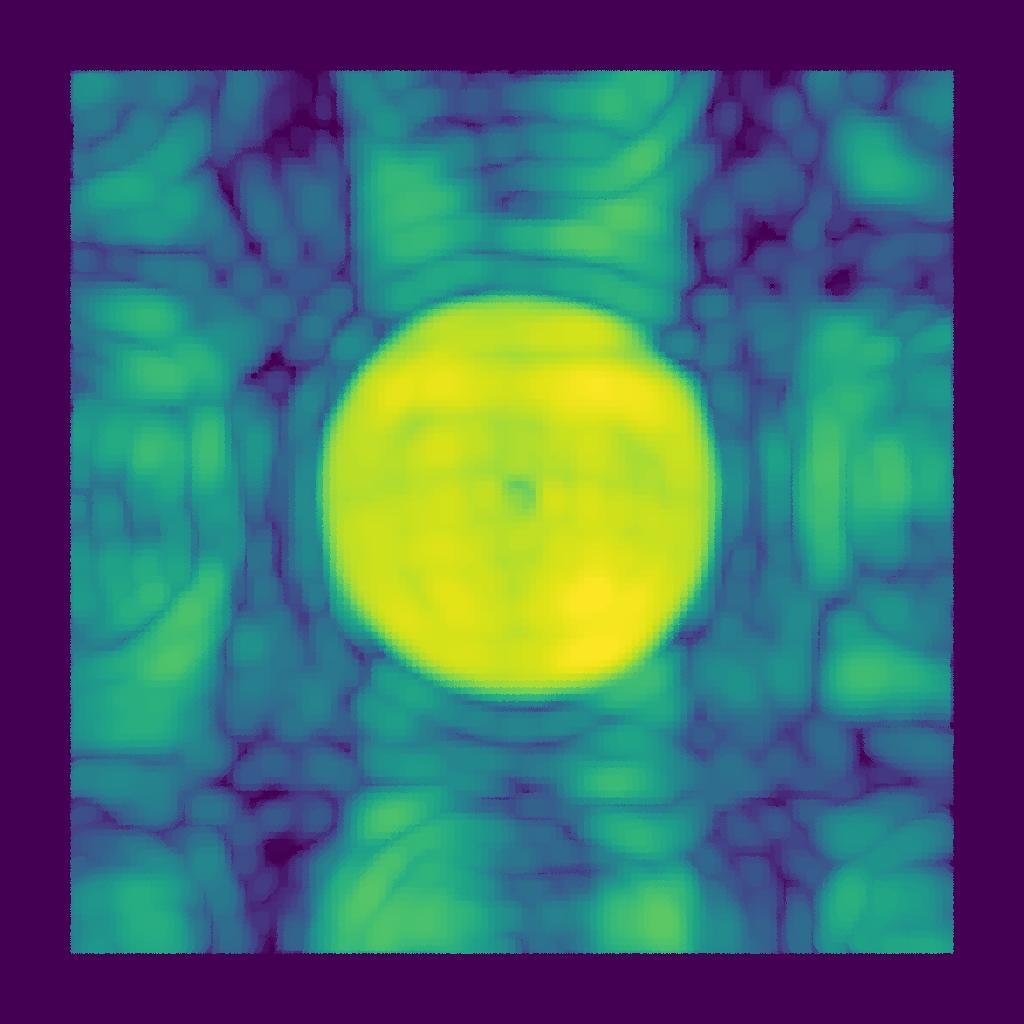} \\
			
			\rotatebox[origin=l]{90}{\small{\hspace{5pt}\obj{Plate}}} &
			\includegraphics[width=\linewidth]{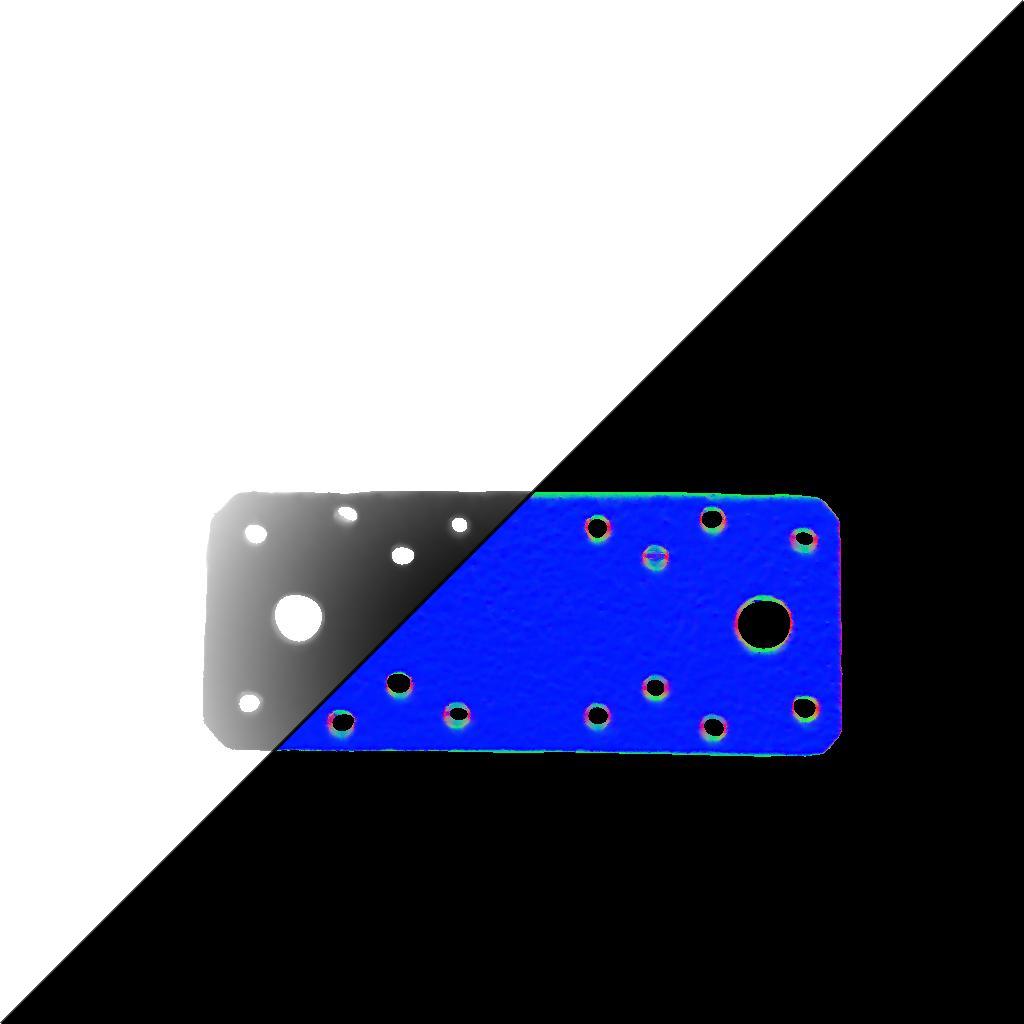} &
			\includegraphics[width=\linewidth]{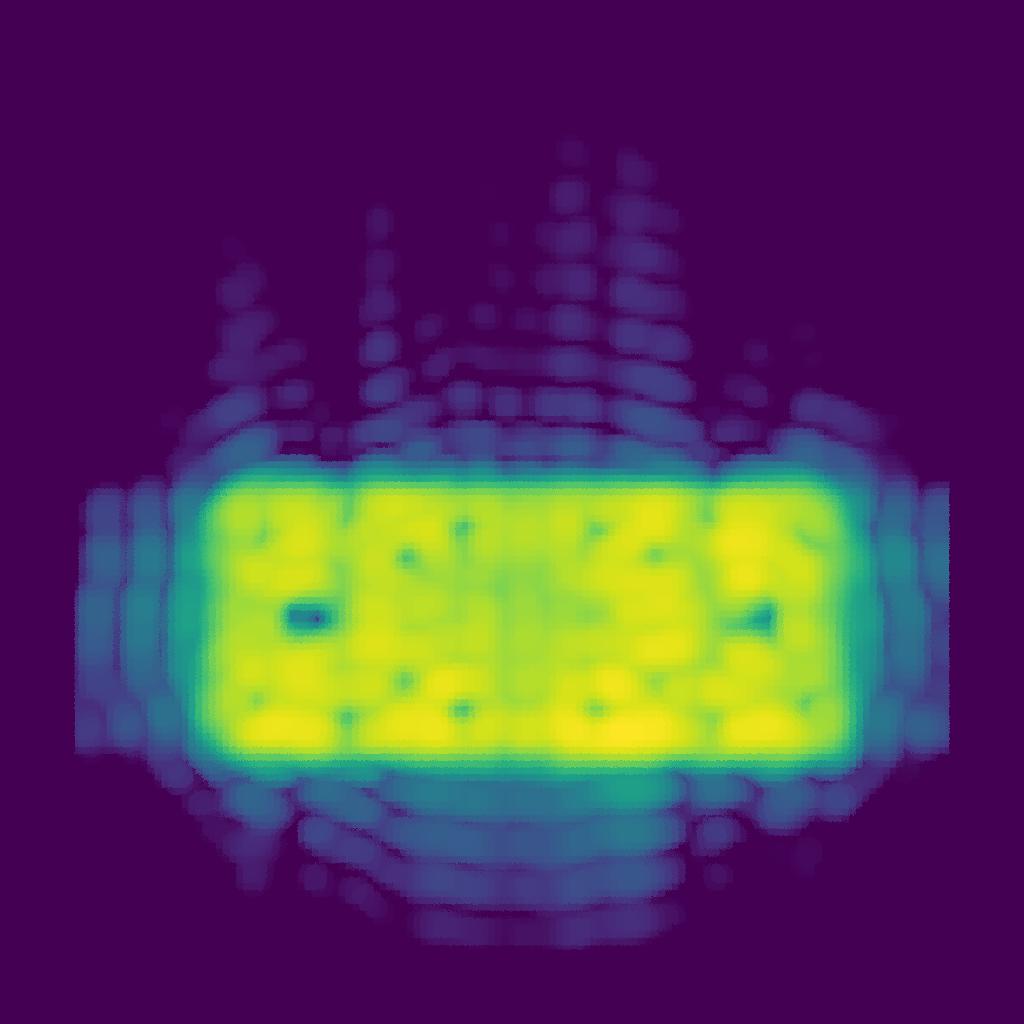} &
			\includegraphics[width=\linewidth]{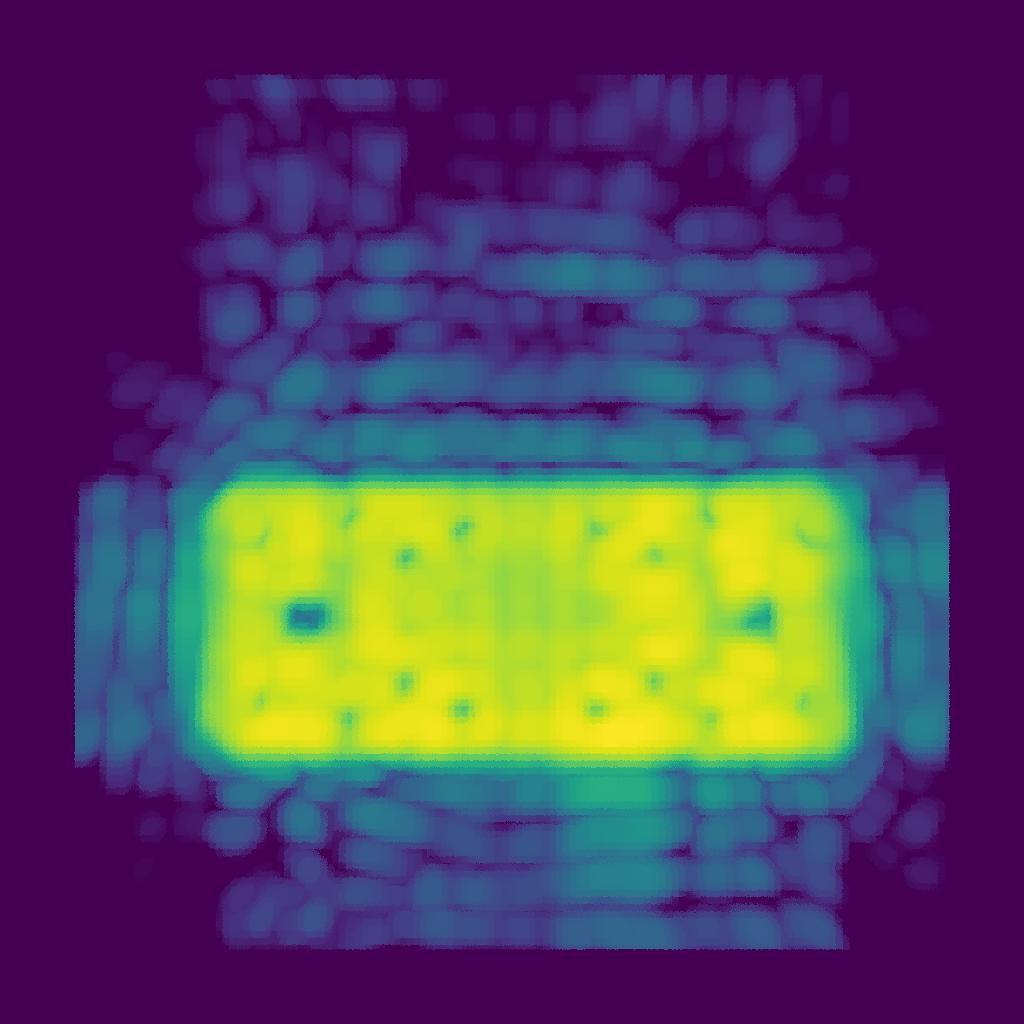} &
			\includegraphics[width=\linewidth]{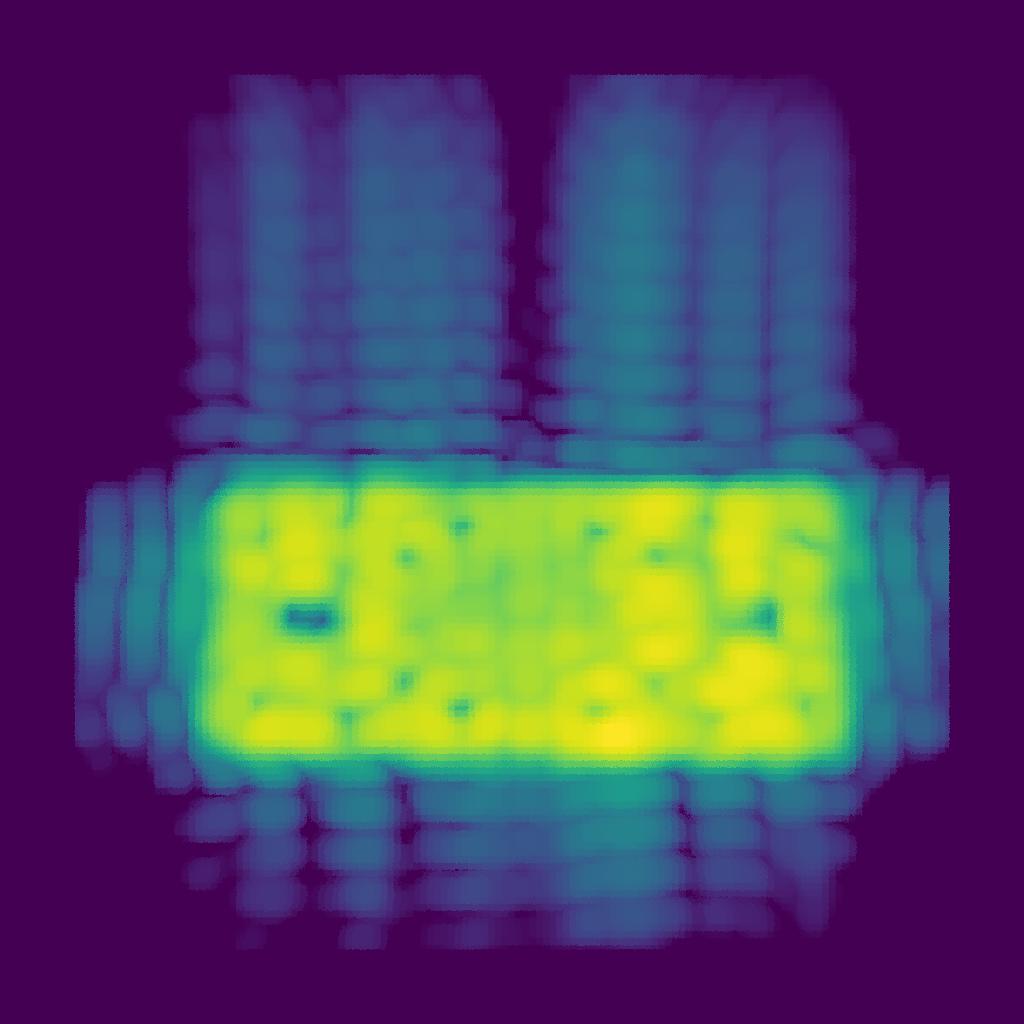} &
			\includegraphics[width=\linewidth]{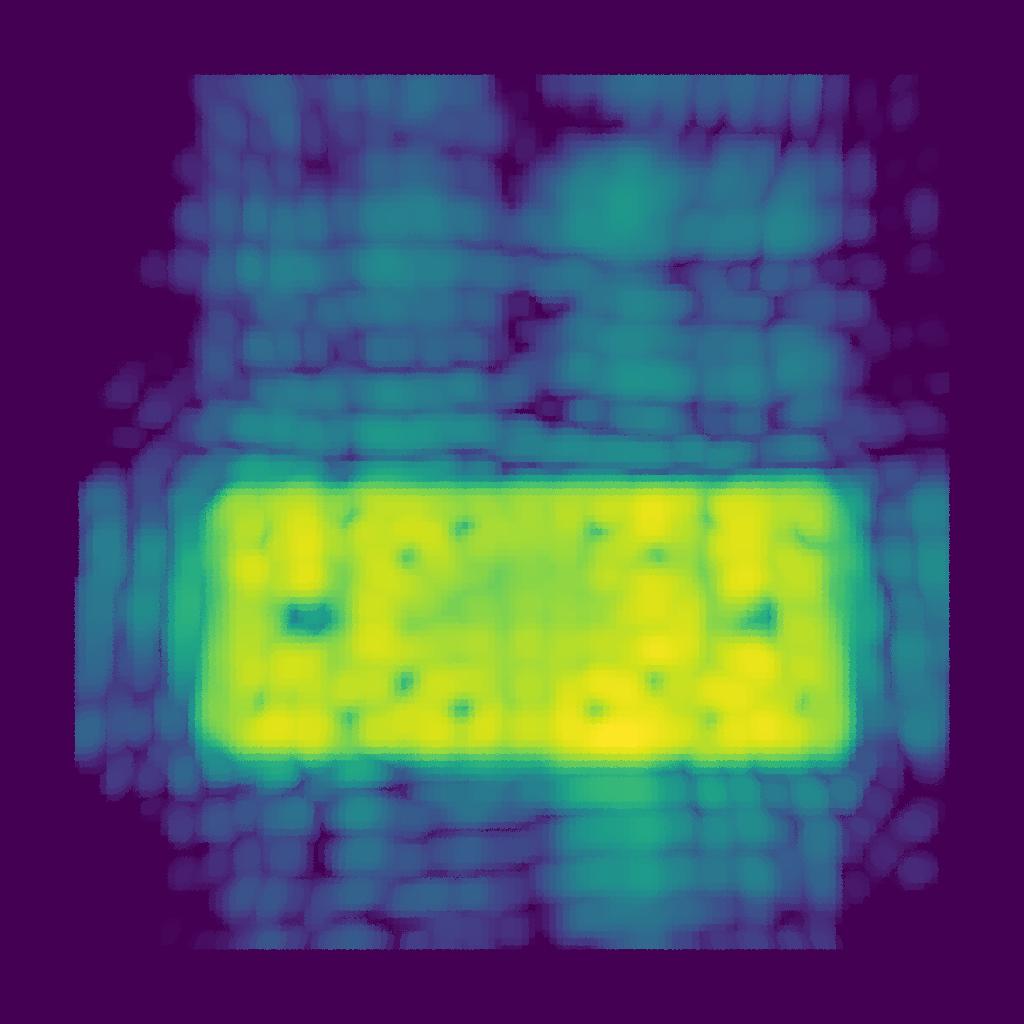} &
			\includegraphics[width=\linewidth]{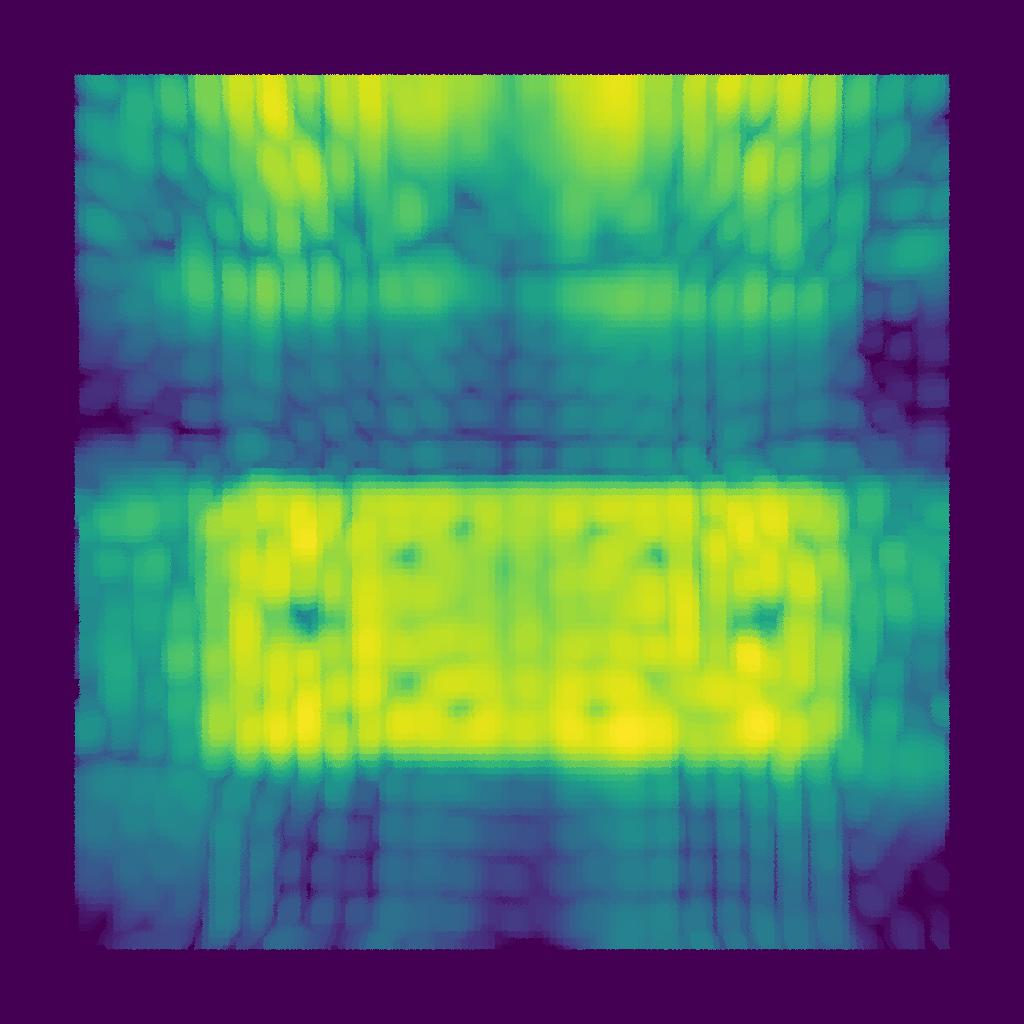} &
			\includegraphics[width=\linewidth]{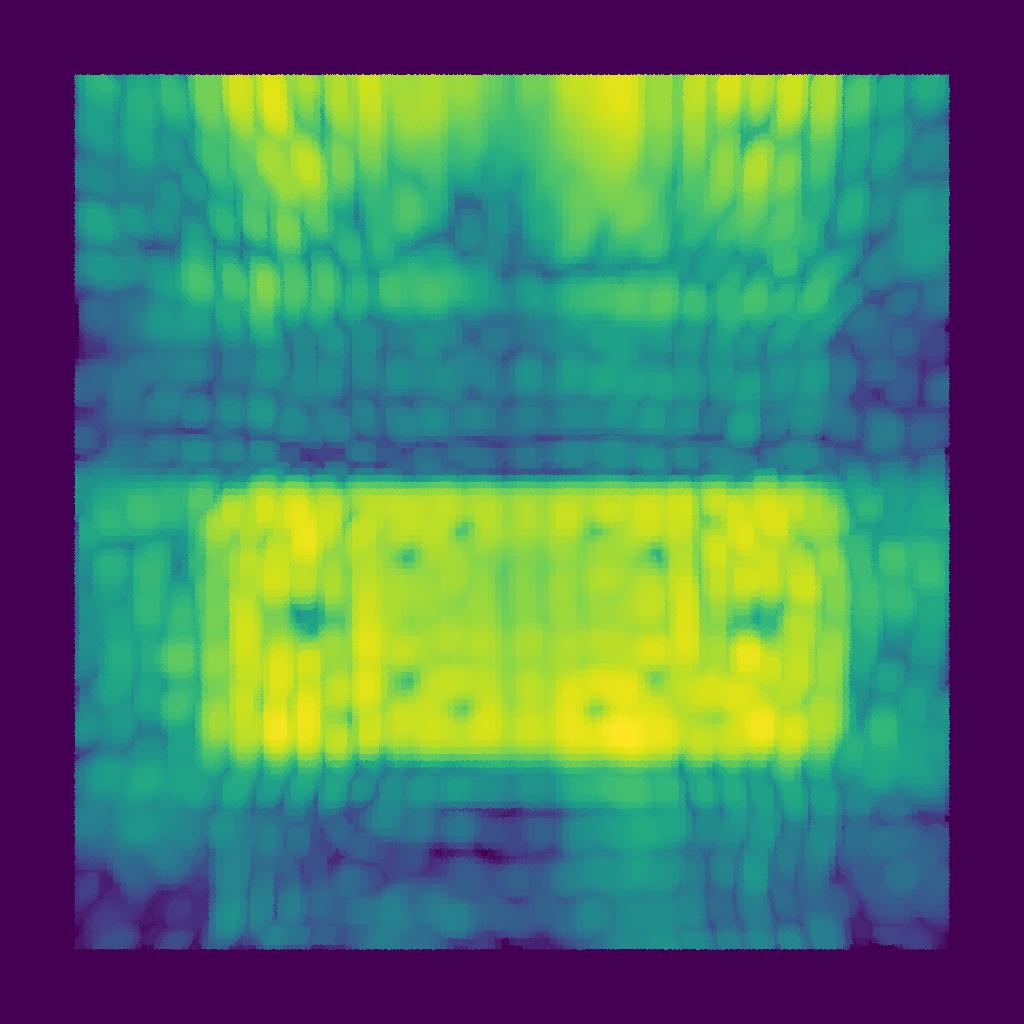} \\
		\end{tabularx}
		\setlength{\tabcolsep}{6pt} %
		\renewcommand{\arraystretch}{1.0} %
		\caption{\label{fig:results_inverse_aperture}%
			Inverse radar rendering results for different antenna configuration. 100\% considers all transmitting and receiving antennas in the MIMO array ($94\times94$), 50\% considers every second respective antenna in the array ($47\times47$), and 25\% only considers every fourth antenna in the array ($24\times24$). Despite obvious artifacts and ambiguities due to a reduced antenna density, the optimization process still performs robustly, and thus generalizes to other MIMO configurations. Objects from top to bottom: \obj{Cardboard}, \obj{Plunger}, \obj{Hand Printed F}, \obj{S1 Hand Open} \obj{Metal Disk (Thin)}, and \obj{V1 Metal Plate}.}
		\Description{The image is a large grid of color-coded, frontal radio-frequency reconstructions arranged in rows (for each object) and columns.
			The columns compare pairs of reconstructions where simulated results are shown next to real measurements, repeated across three levels of antenna configurations, utilizing 100\%, 50\% and 25\% of the antennas from left to right.
			For all of these configurations, the simulation results closely resemble real-world measurements.
			The color of the reconstructed point clouds ranges from dark purple to bright yellow, indicating weaker to stronger signal responses and forming recognizable object shapes.
			As the images progress to the right, the object shapes become blurrier and more distorted, with increased noise and streaking. Despite this, the main outlines of each object remain visible across all columns.}
    \vspace{-3ex} %
	\end{table}
	To demonstrate the versatility of the dataset, in addition to the original experiments conducted by Hofmann et al.~\shortcite{hofmann_2025}, we utilized the high number of antenna signals available in MAROON to examine the impact of a varying antenna configurations and aperture sizes.
	To this end, we used 100\%, 50\%, and 25\% of the antennas, which were simulated by selecting every, every second, or every fourth RX/TX antenna from the raw phasor data in the dataset, respectively.	
	We showcase results for the three different antenna configurations and six different objects in \autoref{fig:results_inverse_aperture}. %
	
	\setcounter{table}{4}
	\renewcommand{\tableautorefname}{Table} 
	\renewcommand{\tablename}{\OldTablename}  %

	\subsection{Multimodal Depth Sensing}

	While the backprojection algorithm is employed in many near-field high-resolution RF ToF applications, its reconstruction time is typically orders of magnitude slower than the sensor's capture rate (see \autoref{table:sensors}).
	As a robust and computationally efficient alternative, Wirth et al.~\shortcite{wirth_2025} introduce a multimodal image reconstruction method that builds upon the previous frequency shift keying (FSK) approach proposed by Bräunig et al.~\shortcite{braeunig_2023_fsk}.
	By utilizing an optical depth camera as a secondary sensor, point-wise depth priors are integrated into the 2FSK signal processing pipeline, allowing just two frequencies to adjust these depth priors towards the target object's actual depth. 
	The depth prior is essential to determine the correct period of the sinusoidal wave signal that is otherwise limited to a small unambiguous range.
	We refer the interested reader to~\cite{wirth_2025} for all technical details about the algorithm.
	The authors evaluated the proposed \textit{MM-2FSK} method using the active stereo depth sensor and MIMO imaging radar from our pre-released dataset. 
	In this section, we extend their work by comparing the method across all optical sensors in our dataset, simulating various capture scenarios influenced by the optical depth sensor.

	\paragraph{Ablation with respect to Optical Depth Sensors} 
	Drawing from insights about sensor-specific characteristics, we examine how different depth imagers affect depth deviations in the MM-2FSK method. 
	We follow the evaluation procedure detailed in~\cite{wirth_2025}, employing the most promising frequency configuration\,---\,specifically, two frequencies at 72 and 82 GHz, resulting in a frequency difference of $\Delta f=10$~GHz.
	
	\begin{figure}[!b]
		\centering
		\includegraphics[width=1.0\linewidth]{images/qualitative_mmfsk_sensors_new.pdf} 
		\caption{\label{fig:ablation_mm2fsk_sensors}%
			Top-down views of the RF ToF reconstructions obtained with the MM-2FSK method, fused with the ground-truth MVS point cloud. \textit{Left} to \textit{right}: we vary the supporting sensor providing the depth prior. \textit{Top} to \textit{bottom}: we display the \obj{Flowerpot (Transparent)}, \obj{V2 Metal Plate}, and \obj{Bunny} objects.\vspace*{-3ex}}	
		\Description{A grid of point-cloud reconstructions, arranged in four columns and three rows.
			Each row corresponds to a different object.
			Across each row, the columns compare the MM-2FSK reconstructions of varying optical depth support by utilizing MVS (ground truth), NIR ToF, active stereo and passive stereo sensors from left to right.
			Small orientation arrows and miniature reference images appear at the left of each row, indicating viewpoint and the ground-truth object appearance.
			The observations and discussion of this Figure are documented in Section 8.2.}
	\end{figure}
	In \autoref{fig:ablation_mm2fsk_sensors}, we display top-down views of the MM-2FSK reconstructions for three objects, overlaid with the ground-truth point cloud while varying the sensor that provides the depth prior. 
	For the \obj{Flowerpot (Transparent)} (\textit{top row}), we notice numerous points reconstructed behind the object for all depth imagers except the ground truth. 
	The transparency causes parts of the background or ground surface to be reconstructed, resulting in a depth prior positioned behind the ground truth.
	With limited depth correction capabilities~\cite{wirth_2025}, the MM-2FSK method can not correct outliers when the optical depth prior lies within a different signal period than the ground truth.
	Similarly, for the \obj{V2 Metal Plate}, large surface areas are reconstructed behind the object for both NIR ToF and active stereo sensors, due to multi-path effects and sensor oversaturation arising from the object's perfect specularity. 
	
	Lastly, there are a few depth outliers behind the \obj{Bunny}, primarily associated with the NIR ToF and active stereo sensors, stemming from incorrect depth priors due to the triangulation of flying pixels, which the MM-2FSK method cannot correct.
	
	Moreover, \autoref{table:ablation_2fsk_sensors} lists the mean depth deviation of all objects in relation to the varying depth priors.
	By assessing the depth deviation always relative to the MVS setup, we expect that its respective depth prior yields the best performance.

	Consistent with earlier evaluations (cf. \autoref{sec:obj_dist_spec_dev}), no single sensor outperforms others across all objects. 
	For 3D errors (\mone~and \mtwo), the active stereo and NIR sensors yield the best performance, while for projective errors (\mthree and \mfour), both passive stereo and active stereo sensors provide the most accurate results.
	\begin{table}[!htbp]
		\caption{\label{table:ablation_2fsk_sensors}%
			Ablation study of the MM-2FSK method with different depth priors, each from another optical depth imager. The mean depth deviation from the ground truth, given in centimeters, is averaged over all objects at 30~cm distance. The \textbf{best} and \textbf{\textit{second best}} results per metric are highlighted. \vspace{-1ex}%
        }
		\begin{tabularx}{\linewidth}{@{}cYYYY@{}} 
			\toprule 
			Depth Prior & { \mone } & { \mtwo }  & { \mthree }  & { \mfour } \\ 
			\midrule
			MVS & \best{0.51} & \best{0.18} & \best{0.19} & \best{0.17} \\
			NIR ToF & 1.19 & \secbest{1.67} & 1.58 & 1.48 \\
			Active Stereo & \secbest{0.82} & 1.74 & 1.36 & \secbest{1.36}  \\
			Passive Stereo & 0.92 & 1.91 & \secbest{1.29} & 1.41 \\
			\bottomrule
		\end{tabularx} 	
        \vspace{-3ex} %
	\end{table}

	\section{Limitations}
	During sensor characterization, we observed that, even though the depth deviation of the MIMO imaging radar is on par with that of optical sensors, the reconstructions exhibit considerably more holes, where no valid depth is estimated.
	While it is intuitive to assume that reconstruction quality is influenced by object material (limiting the returned signal amount\,---\,akin to optically transmissive materials), we observe that in our experiments the object geometry is the primary factor of influence and has a greater impact than in optical sensors; however, disentangling the effects of geometry and material remains challenging, as precise impacts on fine-grained surface details cannot be easily assessed. 
	Without a direct mapping between point targets and RX antennas, depth evaluation concerning these surface-level details is infeasible without backprojection, or any other depth processing algorithm.
	On the other hand, the reconstructed outcomes after signal processing may not align with reality, as these methods typically incorporate a systematic bias by relying on the Born approximation~\cite{sherif_2014}.
	To overcome these limitations, we highlighted one particular work of Hofmann et al.~\shortcite{hofmann_2025}, taking the first step towards automatic material characterization in the radio-frequency domain using inverse rendering.
	We anticipate that further analysis of these material parameters, combined with improvements of radio-frequency simulation frameworks~\cite{schuessler_2021}, will considerably aid in disentangling the potential error sources behind missing reconstructions and enhancing current signal processing methods.
	
	The proposed evaluation framework for sensor characterization is tailored to point cloud comparisons, and is, therefore, independent of the RF signal processing algorithm; however, it requires the spatial co-localization of sensors.
	To achieve the latter, it needs to be verified, whether the respective spatial calibration method may be applicable to other high-resolution radar systems; alternatively, it can be substituted with any other calibration method tailored to the radar system of interest.

	Furthermore, the object reconstructions were evaluated solely for valid locations in the ground truth, excluding artifacts like ghost targets or other forms of noise that may arise from violations of the Born approximation.
	Lastly, we did not capture different orientations of flat objects, which would be an interesting future avenue to explore object orientation in isolation from geometry complexity.

	\section{Conclusion}%
	We presented a novel multimodal dataset, MAROON, that allows us to characterize, for the first time, near-field MIMO imaging radars in direct relation with traditional depth imagers from the optical frequency domain for close-range applications.
	The dataset comprises depth images of a variety of objects, synchronously captured by four mutually calibrated depth imagers and a ground-truth multi-view stereo system. We subsequently analyzed the data within a comprehensive evaluation framework, offering quantitative and qualitative perspectives on each sensor's depth deviation across multiple metric types, objects, and object-to-sensor distances. 
	The findings presented are based on aggregate trends and individual object analyses that contribute to the understanding of the addressed sensor characteristics; however, we believe that our dataset still invites further analysis, exploiting the high diversity of the 45 objects that could not be fully addressed in the scope of this paper.
	
	Moreover, we presented two representative applications, utilizing the collected data.
	First, we built upon previous work to characterize the materials of our captured objects, which is an interesting future direction to disentangle material-specific effects from geometric influences. 
	Second, we conducted extended experiments on a recently proposed multimodal depth estimation approach~\cite{wirth_2025}, using our dataset as a baseline to evaluate its performance. 
	In connection with this work, we examined the impact of different optical sensor modalities to identify suitable depth priors for radar signal processing.
	
	We hope that by highlighting these promising research directions, along with the release of our MAROON dataset, our work will give rise to further studies of multimodal sensor systems in a joint reference frame.

\begin{anonsuppress}
		\section*{Acknowledgement}
		The authors would like to thank the Rohde \& Schwarz
		GmbH \& Co. KG (Munich, Germany) for providing the radar
		imaging devices. 
		This work was funded by the Deutsche Forschungsgemeinschaft (DFG, German Research Foundation) – SFB 1483 – Project-ID 442419336, EmpkinS.
		The authors gratefully acknowledge the scientific support and HPC resources provided by the Erlangen National High Performance Computing Center (NHR@FAU) of the Friedrich-Alexander-Universität Erlangen-Nürnberg (FAU) under the NHR project b175dc. NHR funding is provided by federal and Bavarian state authorities. NHR@FAU hardware is partially funded by the German Research Foundation (DFG) – 440719683.
\end{anonsuppress}

	\clearpage
	\bibliographystyle{ACM-Reference-Format}
	\bibliography{Bibliography}

	\ifseparatesupp
	\else 
	\makeatletter
	\renewcommand{\thesection}{S\arabic{section}}
	\renewcommand{\thefigure}{S\arabic{figure}}
	\renewcommand{\thetable}{S\arabic{table}}
	\renewcommand{\theequation}{S\arabic{equation}}
	\makeatother
	
\ifseparatehead
\twocolumn[\sffamily%
\noindent \parbox{\linewidth}{{\Huge MAROON: A Dataset for the Joint Characterization of Near-Field\\[0.2\baselineskip]High-Resolution Radio-Frequency and Optical Depth Imaging\\[0.2\baselineskip]Techniques}}
\bigskip

\leftline{\LARGE\uppercase{Supplementary Material}}
\bigskip\medskip

\leftline{\LARGE\uppercase{Vanessa Wirth}, \large{Friedrich-Alexander-Universität Erlangen-Nürnberg, Germany}}
\smallskip
\leftline{\LARGE\uppercase{Johanna Bräunig}, \large{Friedrich-Alexander-Universität Erlangen-Nürnberg, Germany}}
\smallskip
\leftline{\LARGE\uppercase{Nikolai Hofmann}, \large{Friedrich-Alexander-Universität Erlangen-Nürnberg, Germany}}
\smallskip
\leftline{\LARGE\uppercase{Martin Vossiek}, \large{Friedrich-Alexander-Universität Erlangen-Nürnberg, Germany}}
\smallskip
\leftline{\LARGE\uppercase{Tim Weyrich*}, \large{Friedrich-Alexander-Universität Erlangen-Nürnberg, Germany and University College London, UK}}
\smallskip
\leftline{\LARGE\uppercase{Marc Stamminger*}, \large{Friedrich-Alexander-Universität Erlangen-Nürnberg, Germany}}

\bigskip\bigskip%
]
\else
\fi

\section{Spatially Resolved Depth Sensing}
\label{sec:spatially_resolved_extended}
It is common practice for stereo sensors to contain two cameras, $C_1$ and $C_2$, of known relative spatial location. In the event of parallel optical axes, this location is defined as the baseline $B$. 
To compute depth, pixels in image of $C_1$ are matched to pixels of $C_2$, forming correspondence pairs.
For every correspondence pair, the depth $d$ is computed from the disparity $D$, which represents the difference between their pixel positions~\cite{optics_giancola_2018}:
\begin{equation}
d = f\frac{B}{D}\;.
\end{equation}
\paragraph{Spatial Resolution} The depth resolution $\delta_z$ of spatially resolved sensors is limited by the disparity resolution $\Delta D$~\cite{zanuttigh_2016}:
\begin{equation}
\delta_z = \frac{z^2}{Bf}\Delta D\;.
\end{equation}
We denote the ground-truth depth as $z$ and the focal length as $f$.
The disparity resolution is dependent on $\delta_x$ and $\delta_y$.
For camera-based systems, $\delta_x$ and $\delta_y$ are typically expressed through the optical transfer function (OTF)~\cite{williams_2002}.

\section{Time-resolved Sensors (Time-of-Flight)}
\label{sec:time_resolved_extended}
Time-of-Flight sensors can be roughly categorized into direct Time-of-Flight (dToF) and indirect Time-of-Flight (iToF) depth sensing methods.
dToF sensors transmit a signal pulse and directly measure the time it takes for the pulse to return. Due to their high cost, however, they are less commonly used in close-range applications.
More cost-efficient than dTof are continuous wave (CW) signal modulations that measure time indirectly (iToF) based on the phase shift $\Delta \varphi$ between the transmitted and received signal ~\cite{zanuttigh_2016}.
The general form of a continuous sinusoidal carrier signal $s_\text{c}$ can be described by two equal formulas of traveling time $t$ and traveling distance $\rho$, respectively:
\begin{align}
s_{\text{c}}(t) &= A \cdot \cos(2\pi tf + \phi_\text{c})  \\
&= A \cdot \cos(\underbrace{2\pi \frac{\rho}{\text{c}}f + \phi_\text{c}}_{\varphi}) = \widehat{s}_\text{c}(\rho)\;.
\label{eq:wave}
\end{align} 
$A$ and $f$ are the known signal amplitude and frequency, respectively and $\text{c}$ is the speed of light, $\varphi$ is the phase and $\phi_c$ is a constant phase offset.
As a transmitted signal $s_\text{t} = s_\text{c}(t_1) = \widehat{s}_\text{c}(\rho_1)$ of known phase and amplitude reflects at a target, the received signal $s_\text{r} = s_\text{c}(t_2) = \widehat{s}_\text{c}(\rho_2)$ has a relative traveling distance of $2\cdot\Delta \rho = (\rho_2 - \rho_1)$ between the transmitter and receiver.
The range, $\Delta \rho$, is related to the relative phase shift $\Delta \varphi$~\cite{zanuttigh_2016} by:
\begin{equation}
\Delta \rho = c \frac{\Delta \varphi}{4 \pi f}\;.
\end{equation}
The general assumption of dToF sensors is that a signal directly reflects at the first target and therefore the range equals half of the traveling distance.
The depth resolution of a ToF sensor is specific to the utilized wavelength and spatial arrangement of transmitters and receivers.

\ifseparatehead
{\let\thefootnote\relax 
	\footnotetext{* Both authors contributed equally to this research.}
}
\else 
\fi

\paragraph{NIR AMCW Time-of-Flight} 
AMCW ToF algorithms usually operate on the SIMO principle, as they do not require as expensive sensor apertures as imaging radars, and often have more receivers and transmitters than can be effectively managed computationally in MIMO depth estimation algorithms~\cite{zanuttigh_2016}.
The range resolution of a NIR AMCW ToF sensor can be expressed as~\cite{kinect_paredes_2023}:
\begin{equation}
\delta_z = \frac{c}{f_{\text{m}}} \sqrt{\frac{P_l + P_a}{P_l} \cdot \frac{I}{k_{o} q_e \rho \Delta t}}\;.
\end{equation}
Environment-specific parameters are the power of ambient light $P_a$, and the reflectivity of the target $\rho$.
Hardware-specific parameters are the modulation frequency $f_{\text{m}}$, the power of the illumination unit $P_l$, the total illumination area $I$, the quantum efficiency $q_e$, the integration time $\Delta t$, and a constant parameter for the optical system, $k_o$.
Due to unknown hardware-specific parameters, we were unable to determine the exact range resolution for NIR AMCW ToF (Azure Kinect) in \ifseparatesupp{Table~1}\else\autoref{table:sensors}\fi\ of the main paper. We refer to~\cite{kinect_paredes_2023} for an experimental approach of determining the effective range and lateral resolution.

\paragraph{MIMO FSCW Time-of-Flight} The spatial resolution of a square-shaped MIMO FSCW imaging sensor can be expressed as~\cite{sherif_2021}:
\begin{align}
\delta_{x,y} &= \frac{c}{4 f_{\text{max}}} \cdot \sqrt{4 \left(\frac{z}{L}\right)^2 + 1} \\
\delta_z &= \frac{0.5 \cdot c}{\Delta f+ \left(1 - \frac{1}{\sqrt{1+0.5(L/z)^2}} \right)\cdot f_{\text{min}}}\;.
\end{align}
We denote the size of the square aperture as $L$.

\section{Sensor Parameters and Settings}
The sensor settings in \ifseparatesupp{Table~1}\else\autoref{table:sensors}\fi\  of the main paper are chosen with respect to a trade-off between fair sensor comparability and practical applicability. 
We uniformly list the frame rate computed from the time takes to \textit{capture} the relevant data of one depth frame.
Note that this may not necessarily include the computation of depth.
For instance, the QAR50 has a capture rate of $\approx$ 70~fps while the backprojection algorithm has an average computing time of 78~s such that the overall frame rate is below 1~fps.
Furthermore, we manually adjusted each optical sensor's exposure time, if possible, to ensure similar lighting conditions.
In summary, we selected the sensor settings that optimize quality while, when feasible, maintaining a comparable frame rate to that of the other sensors. 
Additionally, we adhered to the manufacturer's recommendations for optimal practical use in interactive applications.

\subsection{Radar Field of View}
\label{sec:radar_field_of_view}

\begin{figure}[!htbp]
	\centering
	\includegraphics[width=0.8\linewidth]{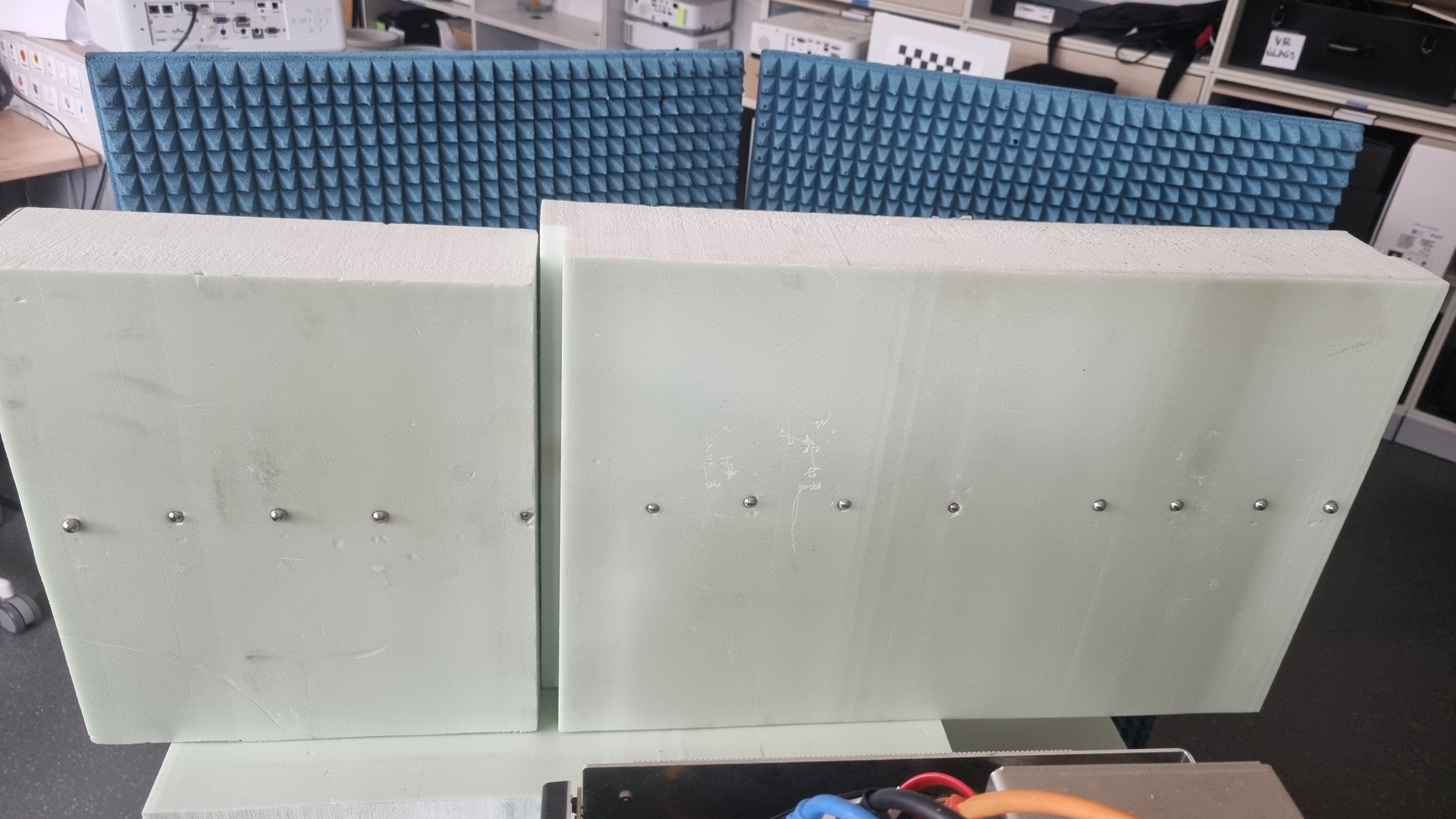}
	\caption{\label{fig:styrofoam_board}%
		Styrofoam board with mounted metal spheres of $\diameter1$~cm.}
	\Description{Two large rectangular boards standing side by side on a desk.
		Each board has a horizontal row of small metal spheres embedded across its front surface at roughly the same height.}
\end{figure}

\begin{figure}[!htbp]
	\centering
	\includegraphics[width=1.0\linewidth]{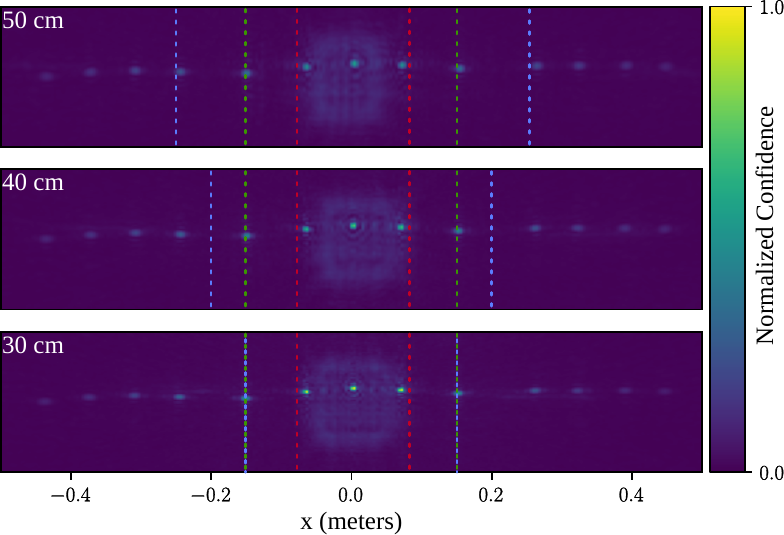}
	\caption{\label{fig:radar_fov}%
		Confidence map of multiple point scatterers, displaced along the x-axis of the antenna aperture for the three object-to-sensor distances of MAROON. The confidence values are normalized across the three reconstructions. The vertical, dashed lines mark the horizontal extent of the {\color[RGB]{193, 0, 37} antenna aperture}, {\color[RGB]{58, 152, 0} reconstruction volume}, and approximated {\color[RGB]{91, 125, 255} $53\times53^\circ$ perspective camera frustum}. }
	\Description{Three horizontal panels in three rows, each representing a different distance labeled at the left: 50 cm, 40 cm, and 30 cm. 
		Along the horizontal axis of each panel, labeled in meters from -0.5 to 0.5, small repeating bright dots appear symmetrically to the left and right of the center, which correspond to the radio-frequency reconstructions of point scatterers.
		The color intensity of these reconstructions varies from dark purple to yellow-green, indicating low to high confidence values, with brighter spots clustered near the center of each panel, at an x-axis value of 0 meters.
		A vertical color bar on the right shows the confidence scale, ranging from 0 at the bottom to 1 at the top.
		As the distance decreases from top to bottom row, the central region of each point scatterer becomes brighter and more pronounced.}
\end{figure}

Optical sensors typically model the field of view using a perspective camera model. 
In theory, MIMO radars %
can also be viewed as an array of small cameras such that the antenna aperture acts as a unified perspective camera, with its field of view defined by the union of all individual antenna frustums.
In practice, modeling the complex antenna radiation pattern as a conventional camera frustum is a crude approximation, as the extents of the visible area are not as straightforward to define as for optical sensors. %

To demonstrate this, we conducted an experiment where we mounted several 1~cm diameter metal spheres on a styrofoam board at a fixed horizontal distance around the aperture origin, as shown in \autoref{fig:styrofoam_board}. 
The aim of this experiment is to explore the maximum visible area by measuring the signal response of each pair of spheres placed on opposite sides of the origin. 
The signal response for each metal sphere is illustrated in \autoref{fig:radar_fov}, as part of the confidence value after spatially resolving the raw signal using backprojection.

For point targets with uniform, view-independent scattering properties, the imaging radar's visible area encompasses all the mounted metal spheres, covering an horizontal area of approximately 90~cm (and potentially even further). 

In contrast, targets with extended surfaces and non-uniform scattering properties are reconstructed within a more limited area that roughly corresponds to the size of the antenna aperture (marked in \textit{\color[RGB]{193, 0, 37} red}). 
Here, this is observed for the styrofoam board, which reflects a minimal amount of the emitted signal and is typically considered nearly invisible. 
In this scenario, where only empty space is reconstructed alongside point targets, the signal response of the styrofoam board behaves similarly to other planar surface targets in MAROON. 

In summary, the visible area of a MIMO radar has similarities with a continuous Gaussian function centered at the aperture origin. 
To determine the effective visible area, we compute the \textit{full width at half maximum} (FWHM); here, it represents the horizontal extent of the reconstruction area where confidence values exceed 50~\% of the maximum. 
For the three object-to-sensor distances of 30~cm, 40~cm, and 50~cm, this extent is approximately between [$-0.15$, $0.15$] meters, aligning closely with the \textit{\color[RGB]{58, 152, 0} green}-marked reconstruction volume used for evaluation. 
The corresponding fields of view of $53^\circ$, $41^\circ$, and $33^\circ$ differ significantly across these distances, making it challenging to find a unified perspective camera frustum.
A suitable perspective camera frustum would also need to fully encompass the $13.8\times13.8$~cm aperture at a distance of 0~cm, i.e., the aperture origin.
Contrary to camera-based systems, where the spatial origin usually lies within the sensor extents, however, this frustum would yield an approximate field of view of $65^\circ$, with the camera origin located $\approx 10.8$~cm behind the aperture.
These observations suggest that an orthographic camera model, as utilized for backprojection, is a more suitable approximation for describing the visible volume of the RF ToF sensor.

However, to maintain consistency with the parameters given for camera-based systems, we assume an approximated $53^\circ$ field of view in \ifseparatesupp{Table 1}\else\autoref{table:sensors}\fi\ of the main paper, which encompasses all of the extents measured with the FWHM and is highlighted in \textit{\color[RGB]{91, 125, 255} blue} in \autoref{fig:radar_fov}.

\section{Dataset Post-processing and Evaluation}
\label{sec:post_processing_evaluation}
Example images of all 45 objects in MAROON can be found in~\autoref{fig:object_images}. We compare the reconstructions produced by the four presented depth imagers with a ground-truth reconstruction in a common metric space and describe the methods used in this process.

\paragraph{Projection into 3D}
We acquire a point cloud of the object's surface utilizing the 2D depth and auxiliary data provided by MAROON. For a given pixel position $(u,v)$ and its corresponding depth $d$ from an optical depth sensor, we first verify its validity using the segmentation map of the same resolution\,---\,a step that has already been performed for radar during depth filtering.
Subsequently we project each valid triple $(u,v,d)$ back into 3D space using the given transformation matrix $\boldsymbol{T}\in \mathbb{R}^{4\times 4}$:
\begin{equation}
\begin{pmatrix}
x \\ y \\ d \\ 1
\end{pmatrix} = 
\underbrace{\left(
	\begin{array}{cccc}
	& & & \\[-1ex]
	\multicolumn{3}{c}{\boldsymbol{I}} & \boldsymbol{t} \\
	& & & \\[-1ex]
	0 & 0 & 0 & 1
	\end{array}\right)^{-1}}_{\boldsymbol{T}^{-1}}
\begin{pmatrix}
u \cdot a \\ v\cdot a \\ d \\ 1
\end{pmatrix}\;.
\label{eq:project}
\end{equation}
For all optical depth imagers, this equation is the inverse of a perspective transformation with intrinsic camera matrix $\boldsymbol{I} \in \mathbb{R}^{3 \times 3}$, pixel offset vector $\boldsymbol{t}\in\mathbb{R}^3 = \boldsymbol{0}$ and $a = d$.
Analogously for radar data, the equation is the inverse of an orthographic transformation with a scale matrix $\boldsymbol{I}$, pixel offset~$\boldsymbol{t}$, and $a = 1$.

\paragraph{Joint Alignment}
To estimate the deviation of a sensor reconstruction $\boldsymbol{R}_s \in \mathbb{R}^{M\times 3}$ from the GT, $\boldsymbol{R}_g \in \mathbb{R}^{N\times 3}$, we use the previously determined spatial calibration parameters ${\boldsymbol{K}_{g \rightarrow s} \in \mathbb{R}^{4 \times 4}}$
to transform $\boldsymbol{R}_g$ from the GT space $g$ into the sensor space $s$:
\begin{equation}
\widetilde{\boldsymbol{R}}_g^s = \widetilde{\boldsymbol{R}}_g  \boldsymbol{K}_{g \rightarrow s}^T\;.
\end{equation}
$\widetilde{\boldsymbol{R}}$ denotes the homogeneous version of $\boldsymbol{R}$. 
We use the notation $\boldsymbol{R}^*$ to indicate a reconstruction that has been transformed to sensor space $*$.

\subsection{Radar Depth Filtering}
\label{sec:radar_depth_filtering}
For our experiments and dataset post-processing, we chose an empirical threshold of $-14$~dB over all objects, which\,---\,to the best of our knowledge\,---\,has proven to yield the best balance of noise pruning while retaining relevant object measurements. In \autoref{fig:signal_threshold}, we show how the signal-to-noise ratio of the radar confidence map, i.e., the pixel-wise values of $\kappa$, behaves over different thresholds for the representative capture of the \textsl{\textsf{{S1 Hand Open}}}.

Additionally, we performed an ablation study to evaluate how different thresholds impact the mean depth deviation across all MAROON objects at a 30 cm object-to-sensor distance. In \autoref{table:ablation_thresholding}, we present results for the four metrics discussed in the main paper. 
We include a weighted mean for each row, giving double weight to \mone{}, as it is the only metric that is sensitive to the completeness of the point cloud.
Notably, we find that the performance concerning \mone{} is inversely related to that of the other metrics, which are more sensitive to signal noise and depth quality. 
The best trade-off between completeness and noise is achieved with a threshold ranging from -14~dB to -17~dB.	

\begin{figure}[!htbp]
	\centering
	\includegraphics[width=1.0\linewidth]{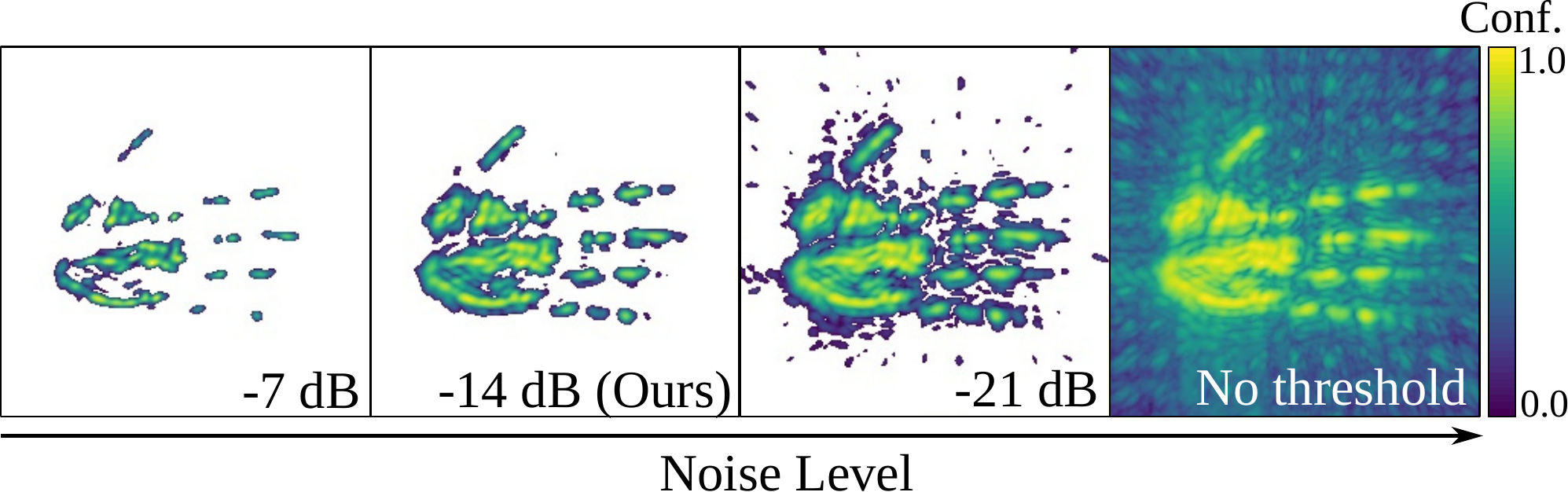}
	\caption{Visualization of the 2D confidence map of the \obj{S1 Hand Open} at various thresholds. The filtered confidence map is subsequently used to extract valid depth information.}
	\Description{Four side-by-side panels showing confidence-colored radio-frequency images.
		From left to right, the text labels beneath each panel show increasing thresholds, with -7~dB, -14~dB, and -21~dB, respectively, ending with a panel that shows the reconstruction with no filtering applied.
		A corresponding arrow from left to right indicates the increasing noise level of reconstructions.}
	\label{fig:signal_threshold}
\end{figure}
\begin{table}[!htbp]
	\caption{\label{table:ablation_thresholding}%
		Ablation study with different signal thresholds used for depth filtering. The depth deviation is expressed in centimeters across the four metrics presented in the main paper, averaged for all objects at a 30 cm object-to-sensor distance. Additionally, we provide a weighted mean for each row, assigning double the weight, $w$, to \mone{}, as it is most sensitive to point cloud completeness. The \best{best} and \textbf{\textit{second best}} results per metric are highlighted.}
	\begin{tabularx}{\linewidth}{@{}cYYYYY@{}} 
		\toprule 
		Threshold & \multirowcell{2}{ \mone \\ ($w=2$)} & \multirowcell{2}{ \mtwo \\ ($w=1$)}  & \multirowcell{2}{ \mthree \\ ($w=1$) }  & \multirowcell{2}{ \mfour \\ ($w=1$) } & Weighted Mean \\ 
		\midrule
		-~\,7~dB & 1.37 & \best{0.67} & \best{0.78} & \best{0.76} & 0.99 \\
		-10~dB & 1.07 & \secbest{0.73} & \secbest{0.82}  & \secbest{0.81} & 0.90 \\
		-14~dB & 0.82 & 0.9 & 0.95 & 0.85 & \best{0.87} \\
		-17~dB & 0.68 & 1.05 & 1.03 & 0.95 & \secbest{0.88} \\
		-21~dB & \secbest{0.51} & 1.54 & 1.16 & 1.07 & 0.96 \\
		--- & \best{0.35} & 5.75 & 1.45 & 1.45 & 1.87\\
		\bottomrule
	\end{tabularx} 
\end{table}

\begin{figure*}[htbp]
	\centering
	\includegraphics[width=\linewidth]{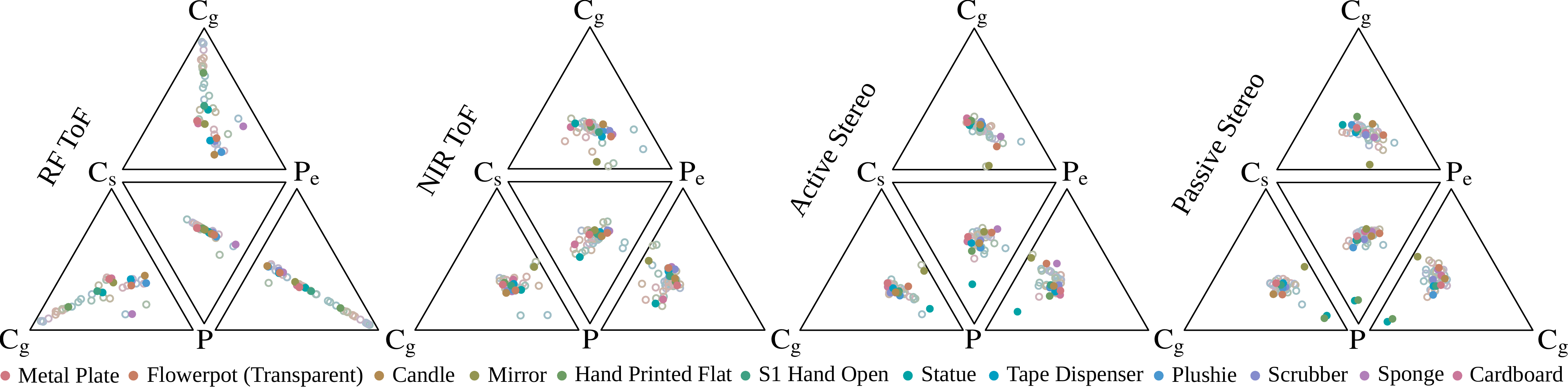}
	\caption{\label{fig:metrics_correlation_types}%
		A complementary view on the depth deviation across different \textit{metric types}.
		For each triplet of metrics $M_i\in\{\text{\mone},\text{\mtwo},\text{\mthree},\text{\mfour}\}$, we convert each mean depth deviation $\mu_i$ to affine coordinates, $\bar\mu_i=\mu_i/\sum_i\mu_i$, that map an object's errors in to a triangle whose corners correspond to metrics $M_i$.
		All 45 MAROON objects are shown as circles, with selected objects from \ifseparatesupp{Table 4}\else\autoref{table:metrics}\fi\ highlighted in solid colors.
		Samples closer to a triangle corner indicate a higher relative depth deviation in the corresponding metric.}
	\Description{Four groups of triangular plots arranged horizontally, each group corresponding to a different sensor from left to right: RF ToF, NIR ToF, active stereo, and passive stereo.
		Within each group, four connected triangles form a larger triangle layout (three at the corners, one at the center). 
		Each small triangle contains many scattered circular points, corresponding to different objects.  A subset of these is highlighted in solid colors and distinguished by a legend running along the bottom of the image.
		Each corner of a small triangle is labeled with a short symbol, corresponding to a metric type. A triangle group forms all possible permutations of these labeled triangle corner triples.
		The observations and trends of this figure are described in Section S5. }
\end{figure*}
\begin{figure*}[htbp]
	\centering
	\includegraphics[width=\linewidth]{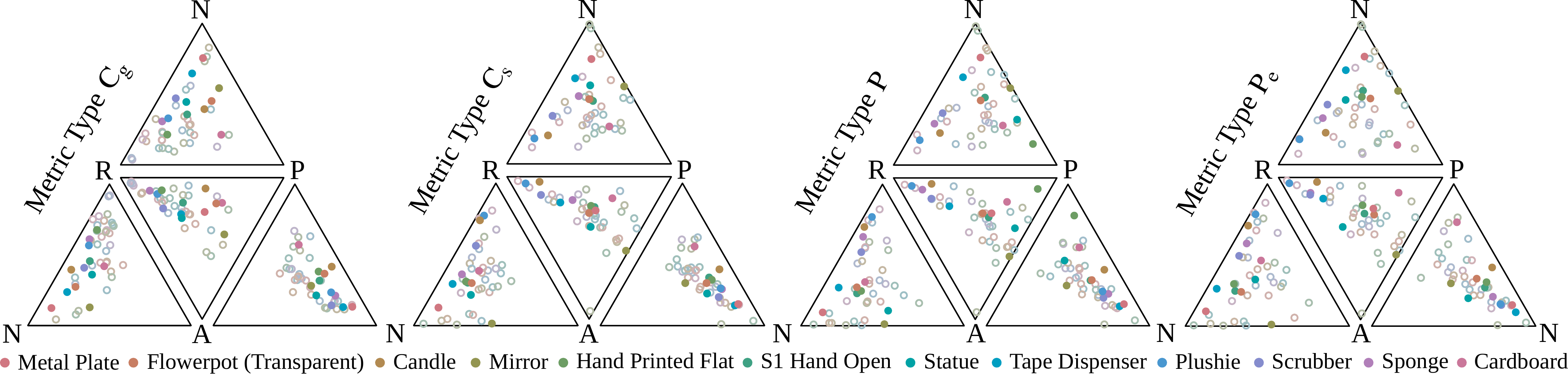}
	\caption{\label{fig:metrics_correlation_sensors}%
		A complementary view on the depth deviation across different \textit{sensors}.  The sensors are denoted as \textbf{R} (RF ToF), \textbf{N} (NIR ToF), \textbf{A} (Active Stereo), and \textbf{P} (Passive Stereo), respectively. 
		Analogously to \autoref{fig:metrics_correlation_types}, we convert the mean depth deviations $\mu_i$ to affine coordinates within triangles corresponding to all possible sensor triples. All 45 MAROON objects are shown as circles, with selected objects from \ifseparatesupp{Table 4}\else\autoref{table:metrics}\fi\ highlighted in solid colors.
		Samples closer to a triangle corner indicate a higher relative depth deviation for the corresponding sensor.}
	\Description{Four groups of triangular plots arranged horizontally, each group corresponding to a different metric from left to right: $C_g$, $C_s$, $P$, $P_e$.
		Within each group, four connected triangles form a larger triangle layout (three at the corners, one at the center). 
		Each small triangle contains many scattered circular points, corresponding to different objects.  A subset of these is highlighted in solid colors and distinguished by a legend running along the bottom of the image.
		Each corner of a small triangle is labeled with a short symbol, corresponding to a sensor type. A triangle group forms all possible permutations of these labeled triangle corner triples.
		The observations and trends of this figure are described in Section S5. }
\end{figure*}

\begin{figure*}[!htbp]
	\centering
	\includegraphics[width=\linewidth]{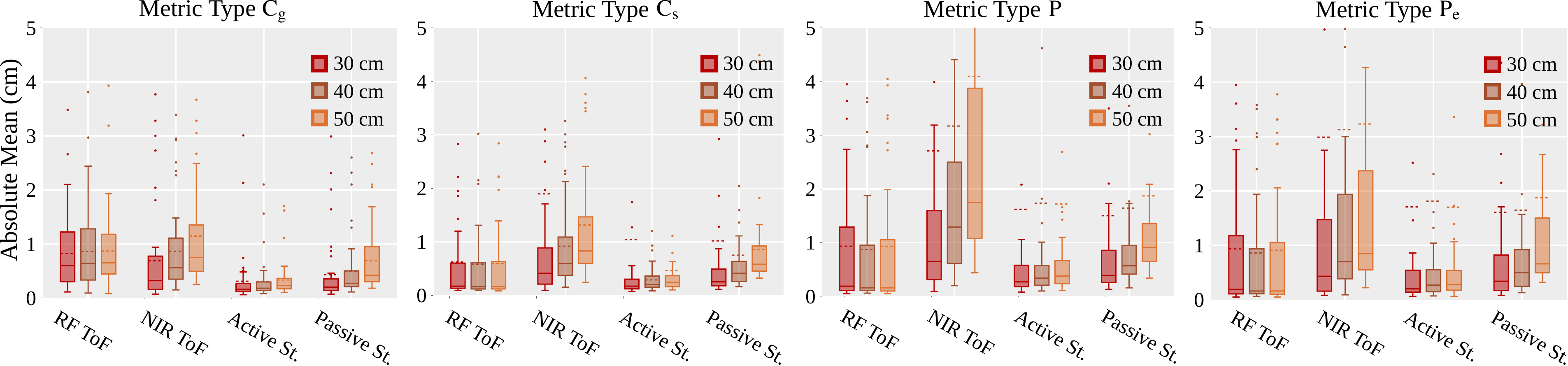}
	\caption{Box plots, visualizing the distribution of the mean error across all objects with respect to different object-to-sensor distances. Solid (---) and dashed (- -) horizontal lines indicate the median and the mean of the distribution, respectively. The results are discussed in \autoref{sec:discussion_depth}.}
	\Description{Four side-by-side box-plot panels, each labeled with a different metric type from left to right: $C_g$, $C_s$, $P$, $P_e$.
		In every panel, the x-axis lists the four depth imagers, while the y-axis shows the absolute mean with respect to the metric, measured in centimeters.
		Each sensor is represented by three grouped box plots in three shades of red and orange, corresponding to the three different sensor-to-object distances.
		The boxes vary in height and spread, indicating differences in error variability between methods and distances.}
	\label{fig:metrics_distance}
\end{figure*}

\subsection{Radar Material Classification}
To investigate the radar signal response and depth deviation with respect to different materials, we divided the 45 objects of MAROON into six classes. These assignments are listed in \autoref{table:material_classes}. The goal of this classification is to highlight material differences on a coarse level, noting the large object variety that still persists within one material class.
Furthermore, we list the objects that are larger than the antenna aperture. It is important to consider these objects when interpreting the depth deviation trends presented in the main paper, as their reconstructions may be incomplete due to the portions that fall outside the antenna aperture.

\subsection{Additional Results}
\label{sec:additional_results}

We provide additional quantitative results for all 45 objects with respect to the depth deviation from the ground truth in~\autoref{table:supp_metrics_1}, ~\autoref{table:supp_metrics_2}, ~\autoref{table:supp_metrics_3}, and ~\autoref{table:supp_metrics_4}.

\section{Extended Discussion of Depth Deviation}
\label{sec:discussion_depth}
First, we present complementary perspectives on the data, where we put the depth deviation of all 45 objects into relation with the different metric types and, subsequently, the different types of depth imagers.
We analyze each representation in turn, highlighting common trends, in combination with previously stated results of \ifseparatesupp{Section~6.2 in the main paper}\else \autoref{sec:results}\fi.

\subsection{General Trends}
\label{sec:discussion_general_trends}

Interpreting the extensive numerical data on depth deviations in \autoref{table:supp_metrics_1},  \autoref{table:supp_metrics_2},  \autoref{table:supp_metrics_3}, and  \autoref{table:supp_metrics_4} can be challenging, so we provide visual, complementary views in this section.
In order to relate different quantities to each other, we use barycentric interpolation based on triples of metric types (\autoref{fig:metrics_correlation_types}) and sensors (\autoref{fig:metrics_correlation_sensors}), respectively.
For each triple $(\mu_a,\mu_b,\mu_c)$, the mean values for depth deviation ($\mu$) of each object are transformed to affine coordinates $(w_a, w_b, w_c)$ by using the formula $w_{\{a,b,c\}} = \mu_{\{a,b,c\}} / (\mu_a+\mu_b+\mu_c)$.
Circle locations closer to a triangle corner indicate higher relative depth deviation. Moreover, as the triples are drawn from a set of four, the triangles are arranged in the shape of an unfolded tetrahedron, highlighting that each triangle's contents can be seen as a projection of barycentric coordinates within a (3D) tetrahedron $w_{\{a,b,c,d\}} = w_{\{a,b,c,d\}} / (w_a + \cdots + w_d)$.

\paragraph{Interpretation of Metrics} In \autoref{fig:metrics_correlation_types}, we provide a qualitative comparison of each sensor's depth deviation from GT with respect to the four presented metrics.
In dense reconstructions, as is typical for optical depth sensors, metrics based on nearest neighbors (here, \mone{} and \mtwo{}) are bound to be lower than those based on projection (\mthree{} and \mfour{}); they also tend to be more resilient against noise.
For RF reconstructions, however, that are prone to sparse depth maps, Chamfer distances often create false matches; accordingly, \mone{} dominates for RF ToF compared to the other metrics. For further discussion regarding the sparsity of RF reconstructions, see \ifseparatesupp{Section~7.3}\else \autoref{sec:discussion_radar_signal}\fi~in the main paper.

For optical sensors, a marginal trend towards the corners of \mtwo{} and \mthree{}, away from the silhouette-resilient \mfour{} and \mone{}, cf.
\ifseparatesupp{Table~3 in the main paper}\else  \autoref{table:metric_meaning}\fi, indicates the presence of noise at object silhouettes.

\begin{figure*}[!htbp]
	\centering
	\includegraphics[width=1.0\linewidth]{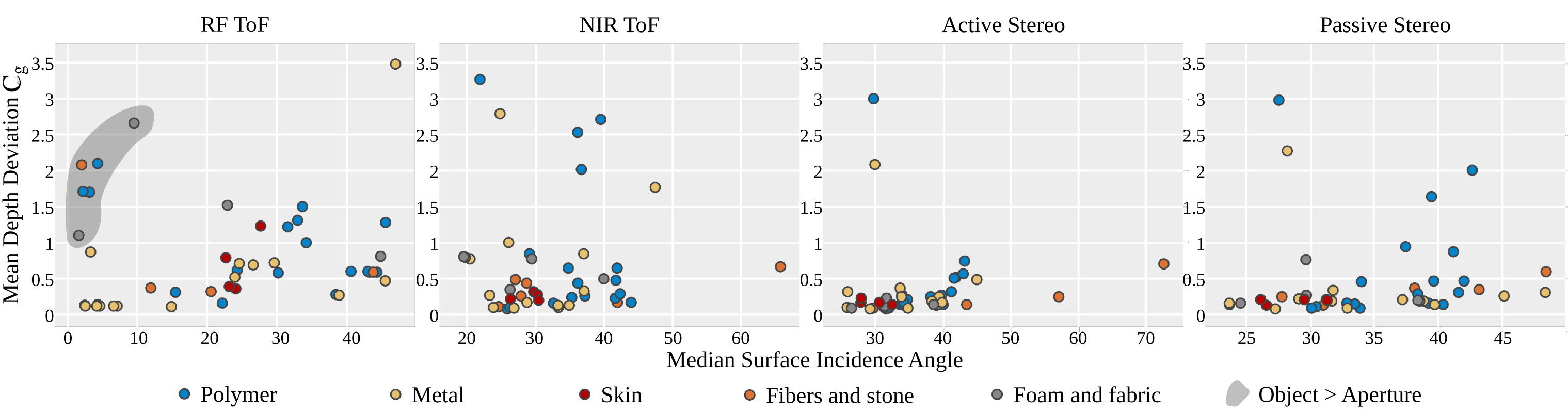}
	\caption{
		Extended experiments of~\ifseparatesupp Figure 5\else \autoref{fig:radar_intensity}\fi~for all four depth imagers, where object geometry (median surface incidence angle) is put in relation to mean depth deviation. Measurements, where large objects appear outside the radar's antenna aperture, are highlighted in gray regions, as they exhibit higher depth deviations compared to the ground-truth reconstructions, which may extend beyond this aperture; this is attributed to the comparably small field of view and the surface reflection characteristics with respect to radio waves (see \autoref{sec:radar_field_of_view}).}
	\Description{Four side-by-side plots from left to right, corresponding to the sensors: RF ToF, NIR ToF, active stereo, and passive stereo. 
		All plots show colored dots that correspond to individual object measurements, where each color represents a different object material such as metal, polymer, fibers and stone, foam and fabric, and skin.
		The plots share a common y-axis for depth deviation w.r.t. $C_g$, and a common x-axis with respect to geometry, as described in the caption.
		The trends of these plots are discussed in Section S6. Gray shaded clusters highlight groups of points that behave differently from the rest, indicating cases where the size of the objects exceed the extents of the antenna aperture.}
	\label{fig:median_angle_per_sensor}
\end{figure*}

\begin{figure*}[!htbp]
	\centering
	\includegraphics[width=0.9\linewidth]{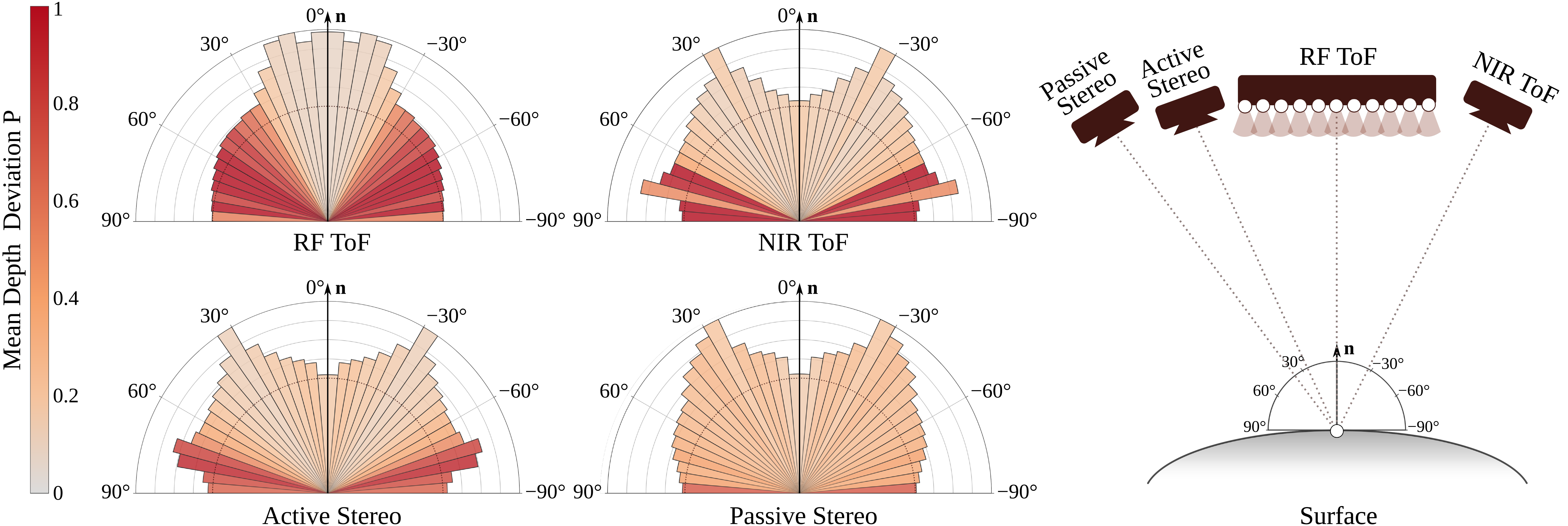}
	\caption{Depth deviation with respect to metric \mthree{} per $5^\circ$-binned surface incidence angle, normalized to $[0,1]$ across the four presented depth imagers. The length of each bin indicates the relative data distribution, with the minimum bin length represented as a dotted hemisphere contour. The corresponding sensor setup is shown on the \textit{right}, providing an intuitive understanding of the data distribution in relation to each sensor's viewing direction and the majority of ground-truth surface normals pointing into the direction of $\boldsymbol{n}$.}
	\Description{The image contains four hemispherical polar charts labeled RF ToF, NIR ToF, Active Stereo, and Passive Stereo, where colored surface angle bins vary in intensity from light (low depth deviation) to dark red (high depth deviation). Darker colors appear more strongly at larger angles away from the center of the hemisphere, where surface angles are close to $0^\circ$. 
		A greater length of each angle bin indicates a higher occurence in the data; for RF ToF, the highest occurence lies around the hemisphere center, whereas for optical imagers the angle bins with the greatest length lie at approximately at $30^\circ$.
		On the right of the plot, a simple 2D sketch of the sensor setup shows four sensors positioned above a curved surface, each pointing toward it from different directions, illustrating how the sensor's viewing angles relate to the surface orientation. While the surface is approximately parallel to the RF ToF sensor, all optical sensors have a viewing angle of $30^\circ$, which coincides with the distribution of angle bins with the greatest length. The figure is discussed in S6.}
	\label{fig:error_per_angle}
\end{figure*}

\paragraph{Relative Depth Deviation across Sensors} In \autoref{fig:metrics_correlation_sensors}, we observe a considerable spread of depth deviations across different sensors. As noted in \ifseparatesupp{Section~6.2 of the main paper}\else \autoref{sec:results}\fi, the relative depth deviation between sensors ranges from 1.9 to 3.4~mm, that is, the variation in the depicted normalized error occurs within a comparatively small range of absolute errors.
As a general trend, the two stereo sensors have the lowest depth deviation (see triangles N--R--A and N--R--P), with a moderate edge for active stereo, particularly for metrics $P$ and $P_\text{e}$, see the $A$--$P$ axis, where passive stereo‘s
performance degrades.
This is consistent with the results of \ifseparatesupp{Section~6.2.1}\else \autoref{sec:obj_dist_spec_dev}\fi~in the main paper, where we find that the active stereo has the highest number of best results.
However, given that relative depth deviations between sensors  differ by only a few millimeters, we conclude that the seemingly improved depth quality of active stereo sensors is of minimal significance.
Moreover, an examination of the
scatter plot
reveals a marginal weight trend towards the corners of the two ToF sensors.
Reconstruction errors between the two ToF sensors are highly object-dependent, and many objects, including many of the highlighted ones, demonstrate multi-path effects due to perfect signal reflections, retro-reflectivity, or partial signal transmission. Further details on this will be discussed in the next section.

\paragraph{Depth Deviation over Distance} In \autoref{fig:metrics_distance}, we observe that the depth deviation of the passive stereo and the NIR ToF sensor is considerably more distance-specific compared to active stereo and RF ToF.
For passive stereo, this may be due to a decrease in effective spatial resolution, where $\delta_z$ directly depends on $\delta_{x,y}$, see \autoref{sec:spatially_resolved_extended}. As the distance between the object and the sensor increases, $\delta_{x,y}$ decreases, resulting in a loss of high-frequency color details while the object appears smaller in the image.
Compared to passive stereo, the active stereo sensor has a comparably higher effective resolution, assuming that the resolution of both sensors differs in accordance with $\delta_{xyz}$ in \ifseparatesupp{Table~1}\else\autoref{table:sensors}\fi~of the main paper.
The unique active NIR pattern may also be less sensitive to decreases in spatial resolution, maintaining the quality of correspondence matches.
The trend for the NIR ToF sensor aligns with findings from Bamji et al.~\shortcite{kinect_bamji_2018} for 30--50~cm distances. However, we argue that absolute errors for a target with 20\% reflectivity do not fully represent all objects in our experiments. We suggest that, in addition to the expected decrease in spatial resolution, a greater depth deviation with increasing distance arises from the signal-to-noise ratio with respect to environmental light, which typically decreases over distance due to the inverse-square law.

Notably, RF ToF does not exhibit a distance-de\-pen\-dent depth deviation, unlike its optical counterpart. This seems to contradict \ifseparatesupp{Table~1}\else \autoref{table:sensors}\fi~ of the main paper, where $\delta_z$ for RF ToF degrades more rapidly with depth, compared to optical sensors; however, the theoretical decay with depth stems from the worsening separability of neighboring point targets, which is not a pertinent scenario in our database, as the recorded targets primarily have smooth surfaces at locations where valid reconstructions are measured.
Moreover, our setup minimizes mmWave interactions with external objects, limiting noise primarily to the object itself. As a result, the signal-to-noise ratio of RF ToF is considerably less sensitive to changes in object-to-sensor distance compared to NIR ToF, assuming no interference from external sources.

\section{Discussion of the Influence of Geometry on Reconstruction Completeness and Depth Deviation}
\label{sec:discussion_geometry_extended}
In \ifseparatesupp{Section~7.3}\else\autoref{sec:discussion_radar_signal}\fi\ of the main paper, we discussed the influence of object geometry on the RF ToF sensor.
Here, we extend our experiments to all four depth imagers.

\paragraph{Influence on Reconstruction Completeness} We visualize the extended results for all depth imagers in \autoref{fig:median_angle_per_sensor}, where we show the mean depth deviation with respect to \mone{} in conjunction with the median surface incidence angle.
We generally observe lower errors for optical sensors compared to RF ToF, indicating that optical depth measurements tend to be more complete.

During dataset capture, we aligned the object surfaces with the RF ToF sensor aperture. 
Due to spatial constraints, all optical sensors were placed next to the antenna aperture, resulting in view directions that do not directly align with the majority of surface normals. 
This is further illustrated in \autoref{fig:error_per_angle} on the \textit{right}, which depicts the sensor placement.
Consequently, the object measurements from optical sensors begin at a median surface incidence angle of approximately 20-30$^\circ$. 
The absence of a significant proportion of smaller angles in the data complicates the identification of notable trends.

\paragraph{Influence on Depth Deviation} We conduct an additional experiment measuring per-angle depth deviation with respect to metric \mthree{}, which is generally more sensitive to depth quality and noise than \mone{}.
For each object, we compute the surface incidence angle and depth deviation for each point-wise measurement using the corresponding ground-truth normal. 
We then aggregate the point-wise measurements across all objects and cluster them into angle bins of 5$^\circ$. 
After calculating the mean depth deviation for each angle bin, we normalize the results to $[0,1]$ across all four sensors. 

The findings are depicted in \autoref{fig:error_per_angle}, where the length of each bar represents the relative quantity of per-point measurements for each angle bin, offering insight into data distribution. 
We illustrate the sensor setup on the \textit{right} to clarify each sensor’s placement and viewing direction, from which we derive the surface incidence angle.

As previously noted for the median angle measurements, the main lobe of per-angle measurements is concentrated around $30^\circ$ for optical sensors and $0^\circ$ for the RF ToF sensor, due to the respective sensor placements. 
In general, we observe a more rapid decline in depth quality for active sensors (NIR ToF, active stereo, RF ToF) compared to passive stereo, which underscores their dependency on well-illuminated (or well-radiated) areas. 
Additionally, the depth quality of the RF ToF sensor significantly decreases at angles greater than $30^\circ$, rendering it more susceptible to object geometry than optical sensors, where we experience a decline in depth quality at angles greater than $60^\circ$.

\section{Extended Discussion of MIMO Radar: Signal Response and Depth Deviation}
\label{sec:discussion_signal_response_extended}

In this section, we extend the experiments of \ifseparatesupp{Section~7.3}\else \autoref{sec:discussion_radar_signal}\fi\ of the main paper, where we
presented both, the signal magnitude and the depth deviation concerning object material, geometry, and size. To demonstrate that the signal magnitude is not only influenced by either object geometry or by object size\,---\,which would be possible due to the high diversity of captured objects that prevents us from isolating one variable while keeping the others constant\,---\,we visualize the signal magnitude in \autoref{fig:radar_intensity_2} (\textit{top}), in relation to object geometry (median surface incidence angle) and size (relative surface area).

\begin{figure}[!tbp]
	\centering
	\includegraphics[width=0.70\linewidth]{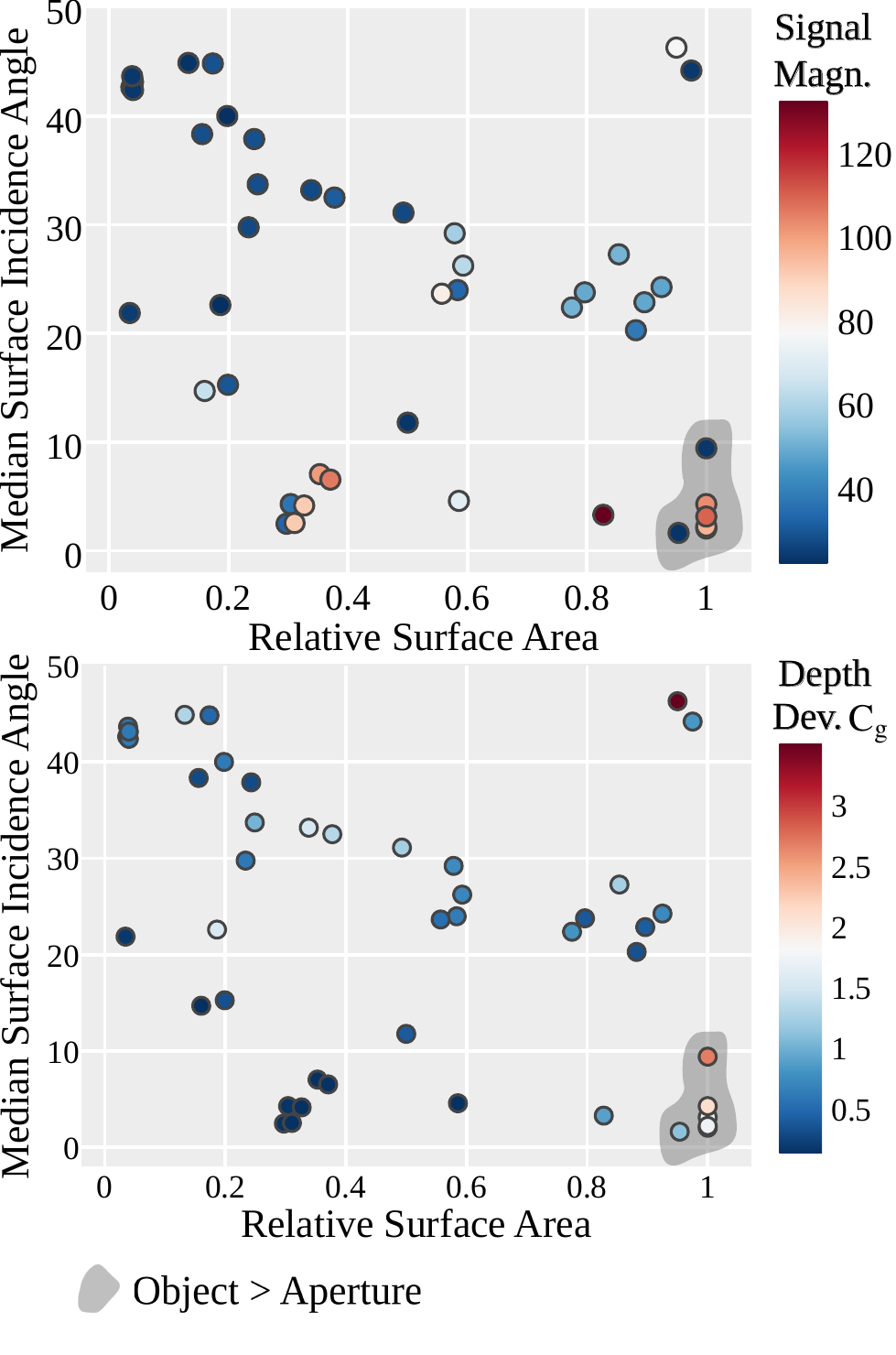}
	\caption{The mean signal magnitude (\textit{top}) and mean depth deviation (\textit{bottom}), put in relation to object geometry (median surface incidence angle) and object size (relative surface area). Large objects outside the radar's antenna aperture exhibit higher depth deviations with respect to the ground-truth reconstructions that possibly extend beyond this {aperture;} this is attributed to the comparably small field of view and the surface reflection characteristics with respect to radio waves (see \autoref{sec:radar_field_of_view}).}
	\Description{The figure contains two vertically stacked scatter plots, visualizing colored dots that correspond to individual object measurements, where color refers to lower (blue) and higher (red) values of signal magnitude (top plot) and depth deviation with respect to $C_g$ (bottom plot).
		Both plots share a common x-axis for relative surface area and a common y-axis that refers to surface angles, as described in the caption.  Gray shaded clusters highlight groups of points that behave differently from the rest, indicating cases where the size of the objects exceed the extents of the antenna aperture. The trends of this figure are discussed in S7.}
	\label{fig:radar_intensity_2}
\end{figure}

\begin{figure}[!tbp]
	\centering
	\includegraphics[width=1.0\linewidth]{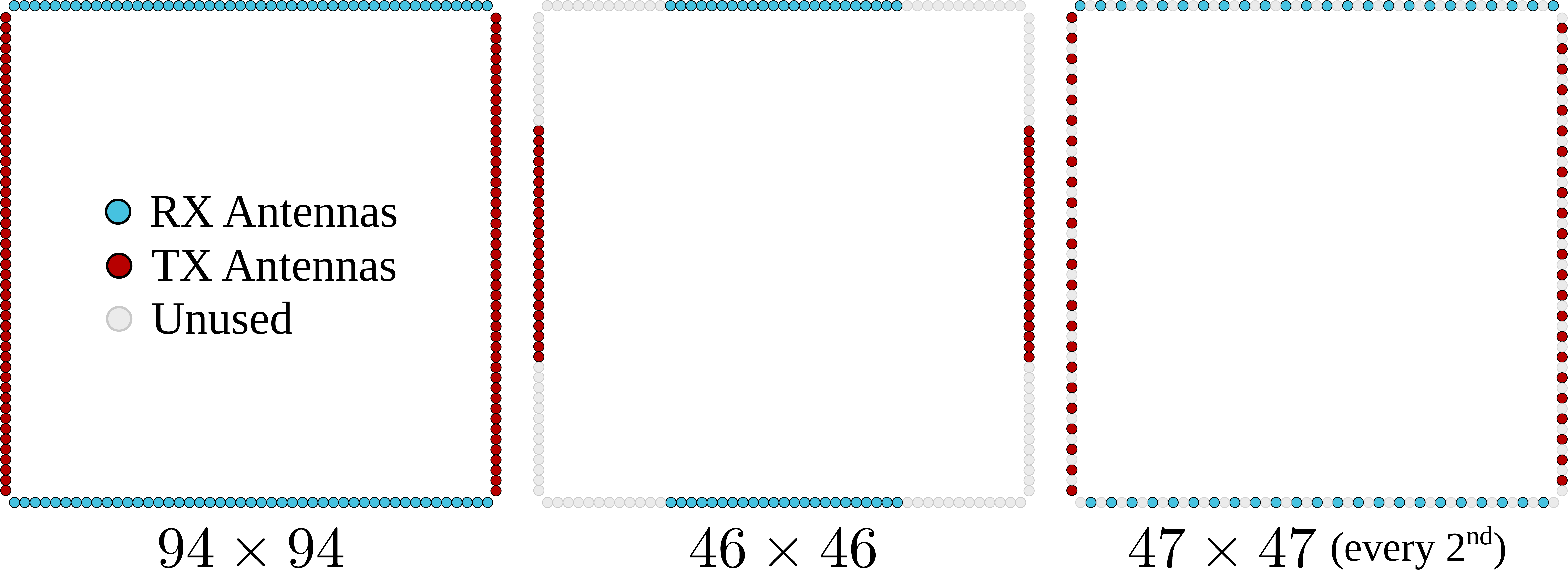}
	\caption{We simulate more consumer-friendly antenna apertures and compare them to the full $94\times94$ antenna array. First, we reduce the aperture size by selecting a spatially centered antenna subset within the MIMO array. Second, we reduce the antenna density, using only every second antenna from the array, while preserving an aperture size comparable to that of the original array. Additionally, we maintain a similar number of antennas between both consumer-friendly variants to ensure a fair comparison.}
	\Description{The image shows three square outlines made of small dots, comparing different antenna layouts side by side. Blue dots represent receiving antennas, red dots represent transmitting antennas, and light gray dots mark unused antennas. The left square outline is densely filled with blue antennas on the horizontal axes and red antennas on the vertical axes. In the middle square outline, only the center area of each axis is colored in red and blue, respectively. In the right outline, the red and blue dots dots are spaced farther apart, with a grey dot in-between each color pair.}
	\label{fig:antenna_architectures}
\end{figure}
\begin{figure}[!tbp]
	\centering
	\includegraphics[width=1.0\linewidth]{images/qualitative_antenna_configuration.pdf}
	\caption{Qualitative evaluation of different antenna configurations for the \obj{Hand Printed Flat} object at 30~cm object-to-sensor distance, using only 50~\% and 25~\% of the original number of antennas.}
	\Description{The image shows multiple panels comparing visual reconstructions of a hand shape. A reconstruction with the full antenna array is depicted on the left, where reconstruction quality is closest to the ground-truth mesh that is displayed next to it.
		On the right, four reconstructions in a $2\times 2$ square layout are depicted. The two rows list the reconstructions that contain a similar number of antennas, where the first row utilizes 50\% and the second row 25\% of the original antenna array.
		The columns differentiate between antenna configurations that are reduced in antenna density and those with a reduced aperture size.
		The four cells from left to right and from top to bottom refer to the following configurations: $47 \times 47$ (every 2nd antenna), $46 \times 46$, $24\times 24$ (every 4th antenna), and $22 \times 22$.
		With fewer antennas, the hand reconstructions become sparser and noisier. 
		Further observations and trends are described in Section S7.1.}
	\label{fig:qualitative_antenna_configuration}
\end{figure}

A general trend on the $x$-axis shows an increase in the signal magnitude from left to right, particularly for objects with a surface incidence angle greater than $10^\circ$.
The majority of objects below this $10^\circ$ angle on the $y$-axis exhibit significantly higher signal magnitudes, regardless of their relative surface area, hence indicating the influence of object geometry.
On the \textit{bottom} of \autoref{fig:radar_intensity_2}, we visualize the mean depth deviation in relation to object geometry and size. While a similar trend is observed for object geometry on the $y$-axis\,---\,where objects of lower surface incidence angle exhibit smaller errors\,---\,no overall trend appears on the $x$-axis, suggesting depth deviation is more influenced by object geometry instead of size.

\subsection{Ablation Study with Reduced Antenna Architectures}
\label{sec:ablation_antenna}

To explore alternative consumer-friendly RF ToF devices, we experiment with different architectures, varying the aperture size and antenna density by selecting only a subset of antennas from the raw phasor measurements in MAROON. 
Two of the selected antenna architectures are shown in \autoref{fig:antenna_architectures}, both with a comparable number of antennas.

We visually compare the reconstructions of four antenna configurations in \autoref{fig:qualitative_antenna_configuration} for the \obj{Hand Printed Flat} at 30~cm object-to-sensor distance. 
Reducing the aperture size (\textit{right column}) results in a loss of continuous object geometry, causing the object to visually resemble a collection of point targets. 
Compared to architectures with larger apertures (\textit{left column}), less of the surface details are preserved. 
Conversely, a decreasing antenna density introduces more localization ambiguities, leading to multiple reconstructions of the object, as seen with the replicated hands in the \textit{lower left} reconstruction.

Further quantitative results are presented in \autoref{table:ablation_antenna_architecture}, where we measure the mean depth deviation in centimeters across all objects at a distance of 30~cm while varying the antenna architecture. 
When comparing configurations with similar numbers of antennas, larger aperture sizes exhibit lower depth deviation than higher-density configurations, as indicated by \mthree{} and \mfour{}; this supports prior qualitative observations that surface quality declines more rapidly with reduced aperture size. 
In contrast, decreasing the antenna density leads to a rapid rise in noise, likely due to the previously mentioned localization ambiguities, which tend to appear in areas where holes typically arise, as \mone{} behaves inversely proportional to \mtwo{}.

\begin{table}[htb]
	\caption{\label{table:ablation_antenna_architecture}%
		Ablation study on different antenna configurations. The depth deviation is expressed in centimeters across the four metrics presented in the main paper, averaged for all objects at a 30 cm object-to-sensor distance. The \best{best} and \textbf{\textit{second best}} results per metric are highlighted.}
	\begin{tabularx}{\linewidth}{@{}cYYYY@{}} 
		\toprule 
		
		Antenna Config. & { \mone } & { \mtwo }  & { \mthree }  & { \mfour } \\ 
		\midrule
		$46\times46$ & 1.01 & \best{1.09} & \secbest{1.09} & \secbest{1.08} \\
		$22\times22$ & 0.82 & 1.59 & 1.30 & 1.24 \\
		\midrule
		$47\times47$ {\small(every $2^{\text{nd}}$)} & \secbest{0.73} & \secbest{1.2} & \best{1.02} & \best{0.95} \\
		$24\times24$ {\small(every $4^{\text{th}}$)} & \best{0.59} & 5.97 & 1.41 & 1.35 \\
		\midrule
		$94\times94$ & 0.82 & 0.90 & 0.94 & 0.85 \\
		\bottomrule
	\end{tabularx} 	
\end{table}

\begin{table}[H]
	\caption{\label{table:ablation_frequencies}%
		Ablation study on various frequency configurations defined by the range of the modulation frequency $f_m$, frequency difference $\Delta f$, and the corresponding number of frequency steps $N_f$. 
		Frequency units are given in GHz.
		The \textit{last row} depicts the full-bandwidth frequency configuration from the main paper.
		The depth deviation is expressed in centimeters, averaged for all objects at a 30 cm object-to-sensor distance. The \best{best} and \textbf{\textit{second best}} results per metric are highlighted.}
	\begin{tabularx}{\linewidth}{@{}lYYYY@{}} 
		\toprule 
		Frequency Configuration (in GHz) & { \mone } & { \mtwo }  & { \mthree }  & { \mfour } \\
		\midrule 
		$f_m \in [72.00, 81.92]$, $\Delta f=0.16$, $N_f=64$ & \secbest{0.81} & \secbest{0.91} & \secbest{0.95} & \secbest{0.86} \\
		$f_m \in [72.00, 81.76]$, $\Delta f=0.31$, $N_f=32$ & 0.90 & 1.56 & 1.24 & 1.15 \\
		$f_m \in [72.00, 81.45]$, $\Delta f=0.59$, $N_f=16$ & 0.90 & 1.56 & 1.24 & 1.16 \\
		$f_m \in [72.00, 80.82]$, $\Delta f=1.10$, $N_f=8$ & 0.86 & 3.58 & 2.49 & 2.20 \\
		$f_m \in [72.00, 79.56]$, $\Delta f=1.89$, $N_f=4$ & 0.91 & 4.82 & 4.2 & 4.04 \\
		\midrule
		$f_m \in [76.96, 82.00]$, $\Delta f=0.078$, $N_f=64$ & \best{0.76} & 1.19 & 1.21 & 1.11 \\
		$f_m \in [79.48, 82.00]$, $\Delta f=0.078$, $N_f=32$ & \best{0.76} & 1.98 & 1.85 & 1.67 \\
		$f_m \in [80.74, 82.00]$, $\Delta f=0.078$, $N_f=16$ & 0.77 & 3.33 & 2.87 & 2.71 \\
		$f_m \in [81.37, 82.00]$, $\Delta f=0.078$, $N_f=8$ & 0.91 & 5.15 & 4.14 & 3.97 \\
		$f_m \in [81.69, 82.00]$, $\Delta f=0.078$, $N_f=4$ & 0.98 & 6.22 & 4.82 & 4.71 \\
		\midrule
		$f_m \in [72.00, 82.00]$, $\Delta f=0.078$, $N_f=128$ & 0.82 & \best{0.90} & \best{0.94} & \best{0.85} \\
		\bottomrule
	\end{tabularx} 
\end{table}

\subsection{Ablation Study with Reduced Frequencies}
\label{sec:ablation_frequencies}

In this experiment, we adjust the frequency configuration of the RF ToF sensor to simulate various sensors.
For this, we compute frequency subsets of the raw phasor data provided in the dataset.
Following a similar approach to the antenna aperture ablations described in ~\autoref{sec:ablation_antenna}, we implement two key variations: first, by operating with a smaller bandwidth, and second, by varying the frequency differences through subsampling every second, fourth, eighth frequency, and so on.

The results are shown in~\autoref{table:ablation_frequencies}, with mean depth deviations quantified in centimeters across all metrics and objects at a distance of 30~cm while varying the frequency configuration. 
Compared to the full frequency spectrum used in the main paper (\textit{last row}), the 64-frequency-stepped configuration performs on par, indicating that a configuration with half the frequencies may serve as a viable alternative maintaining the same accuracy. 

Additionally, we observe a general trend of increasing depth deviation with a decreasing number of frequencies. 
Among the variations, adjusting the frequency difference rather than the bandwidth yields better overall results and may present an interesting sensor configuration that could enhance the RF ToF sensor's capture rate\,---\,as fewer frequencies require less capture time.

\begin{table*}[htbp]
	\centering
	\caption{\label{table:material_classes}%
		Assignment from objects to material classes with optional description about the assignment process.}	
	\begin{tabular}{@{}llr@{}} 
		\toprule
		Class & Objects & Additional Description \\
		\midrule
		\multirow{9}*{Metal} & V1 Metal Plate, V2 Metal Plate & \\
		& Metal Disk (Thin), Metal Disk (Thick) & \\
		& Hand Printed: Flat, B, F, U & Coated with metal lacquer\\
		& Brazen Rosette & \\
		& Corner Reflector & \\
		& Cardboard Box & Coated with metal lacquer\\
		& Mirror & Metal surface beneath partially transmissive glass \\
		& Metal Angle & \\
		& Statue & \\
		\midrule
		\multirow{6}*{Fibers and stone} & Wood Plane & \\
		& Cardboard & \\
		& Book & Primarily made out of paper \\
		& Concrete Stone & \\
		& Wood Ball & \\
		& Bunny Box & Large wooden box, comparably small plastic bunny \\
		\midrule
		\multirow{13}*{Polymer} & Plumber  &\\
		& Silicone Cup & \\
		& Christmas Ball: V1, V2, V3 & \\
		& Candle & \\
		& Bottle & \\
		& Sandpaper (k120), Sandpaper (k80) & \\
		& Flowerpot (Transparent), Flowerpot (Brown) & \\
		& Polystyrene Plate & \\
		& Water Cube & Water wrapped in a plastic cube \\
		& Scrubber & \\
		& Pool Ball & \\ 
		& Bunny & \\
		& Tape Dispenser & \\
		\midrule
		\multirow{2}*{Skin} & S1 Hand Open, S1 Hand Open (Rev.) & \\
		& S2 Hand Open, S2 Hand Open (Rev.) & \\
		\midrule
		\multirowcell{4}{Foam and fabric \\ (Primarily transmissive)} & Rubber Foam Plane & \\
		& Sponge & \\
		& Plushie & \\
		& Foam Plane & \\
		\midrule
		\multirow{5}*{Objects outside FOV} & Polystyrene Plate & \\
		& Sandpaper (k120), Sandpaper (k80) & \\
		& Wood Plane &  \\
		& Foam Plane & \\
		& Rubber Foam Plane & \\
		\bottomrule	
	\end{tabular}
\end{table*}

\begin{figure*}[htbp]
	\centering
	\includegraphics[width=0.82\linewidth]{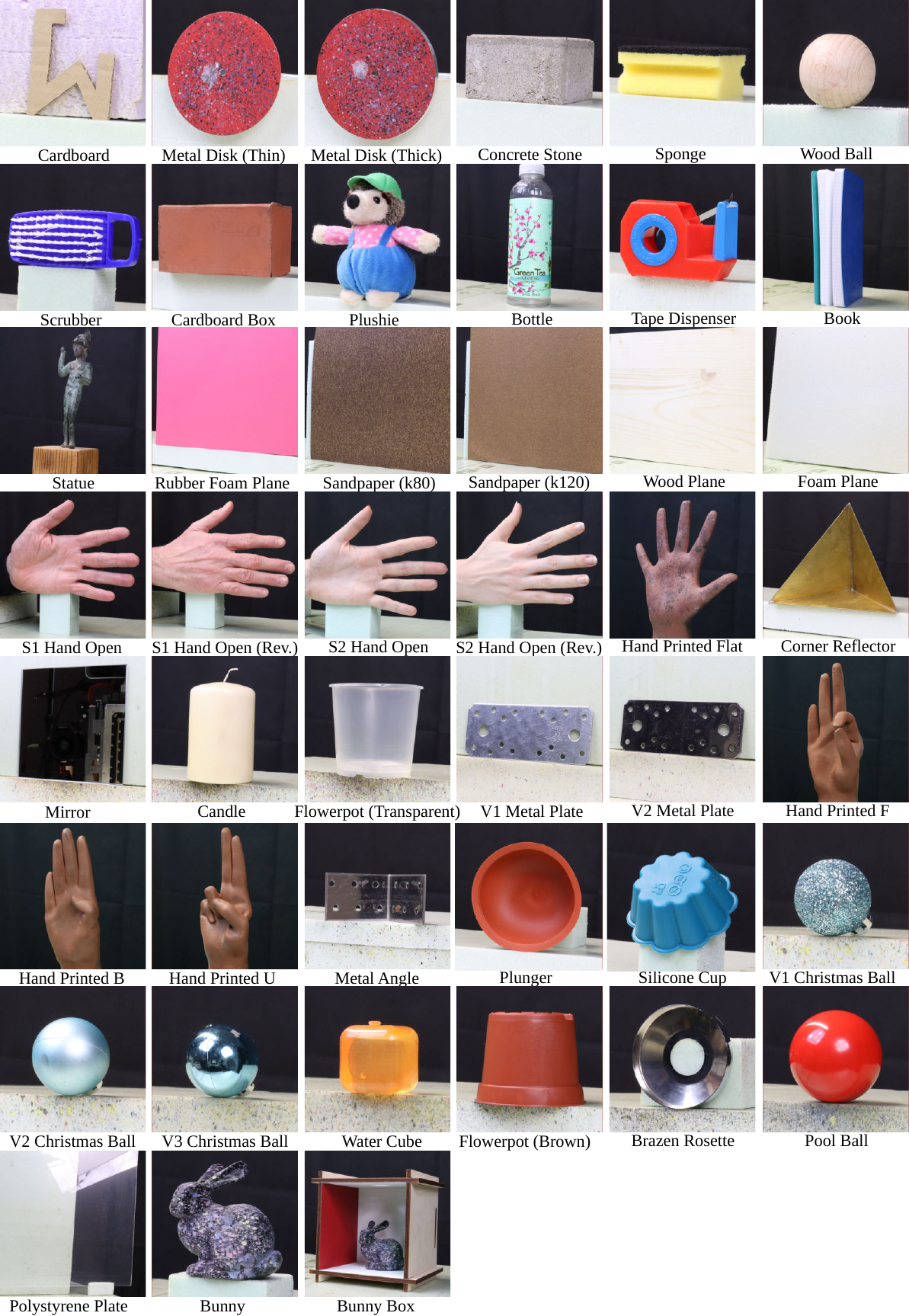}
	\caption{Example images of all objects in MAROON.}
	\label{fig:object_images}
\end{figure*}

\begin{table*}[]
	\caption{\label{table:supp_metrics_1}%
		We measure the depth deviation with respect to \protect\mone{}\protect, \protect\mtwo{}\protect, \protect\mthree{}\protect, \protect\mfour{}\protect~and an additional signed version of \protect\mthree{}\protect,\protect\mfour{}\protect, which is denoted as \protect\mthree{}\protect*,\protect\mfour{}\protect*. All metrics are listed in the form $(\mu \pm \sigma)$, consisting of the mean $\mu$ and standard deviation $\sigma$ in centimeters, computed over the entire metric domain, respectively. The best results among all sensors of one metric type are highlighted in \bf{bold} and the worst results are \underline{underlined}.}
	\vspace*{-0.25 cm} 
	\begin{tabularx}{\textwidth}{@{}lcYYYYYY@{}} 
		\toprule
		& \multirowcell{2}{Metric \\ Type} & \multirowcell{2}{ Cardboard }  & \multirowcell{2}{ Metal Disk \\ (Thin) }  & \multirowcell{2}{ Metal Disk \\ (Thick) }  & \multirowcell{2}{ Concrete Stone }  & \multirowcell{2}{ Sponge }  & \multirowcell{2}{ Wood Ball }  \\ 
		&&&&&&& \\ 
		\hline 
		RF ToF & \multirowcell{4}{\mone{}} 
		&  0.13 ($\pm$  0.06) &  \underline{0.12} ($\pm$  \underline{0.04}) &  \underline{0.12} ($\pm$  \textbf{0.05}) &  \underline{0.14} ($\pm$  \underline{0.07}) &  \underline{1.52} ($\pm$  \underline{0.97}) &  \underline{0.59} ($\pm$  \underline{0.40})   \\ 
		NIR ToF & &  0.10 ($\pm$  0.06) &  0.09 ($\pm$  \textbf{0.03}) &  0.08 ($\pm$  \textbf{0.05}) &  0.09 ($\pm$  \textbf{0.04}) &  0.79 ($\pm$  0.45) &  0.17 ($\pm$  0.12)   \\ 
		Active Stereo & &  \textbf{0.08} ($\pm$  \textbf{0.05}) &  \textbf{0.06} ($\pm$  \underline{0.04}) &  \textbf{0.07} ($\pm$  \textbf{0.05}) &  \textbf{0.08} ($\pm$  \textbf{0.04}) &  \textbf{0.18} ($\pm$  0.17) &  \textbf{0.13} ($\pm$  \textbf{0.06})   \\ 
		Passive Stereo & &  \underline{0.24} ($\pm$  \underline{0.11}) &  0.07 ($\pm$  \underline{0.04}) &  0.08 ($\pm$  \underline{0.07}) &  0.09 ($\pm$  0.06) &  0.26 ($\pm$  \textbf{0.15}) &  0.34 ($\pm$  0.16)   \\ 
		\hline 
		RF ToF & \multirowcell{4}{\mtwo{}} 
		&  0.15 ($\pm$  0.08) &  \underline{0.13} ($\pm$  \textbf{0.05}) &  0.13 ($\pm$  \textbf{0.06}) &  \underline{0.15} ($\pm$  \underline{0.08}) &  0.54 ($\pm$  0.40) &  \textbf{0.12} ($\pm$  \textbf{0.05})   \\ 
		NIR ToF & &  0.16 ($\pm$  \underline{0.18}) &  0.12 ($\pm$  0.13) &  \underline{0.14} ($\pm$  0.20) &  0.10 ($\pm$  0.06) &  \underline{0.79} ($\pm$  \underline{0.42}) &  0.21 ($\pm$  0.19)   \\ 
		Active Stereo & &  \textbf{0.08} ($\pm$  \textbf{0.05}) &  \textbf{0.07} ($\pm$  \textbf{0.05}) &  \textbf{0.08} ($\pm$  0.07) &  \textbf{0.08} ($\pm$  \textbf{0.04}) &  \textbf{0.17} ($\pm$  \textbf{0.15}) &  0.13 ($\pm$  0.07)   \\ 
		Passive Stereo & &  \underline{0.30} ($\pm$  0.13) &  0.12 ($\pm$  \underline{0.17}) &  \underline{0.14} ($\pm$  \underline{0.33}) &  0.12 ($\pm$  \underline{0.08}) &  0.33 ($\pm$  0.18) &  \underline{0.40} ($\pm$  \underline{0.20})   \\ 
		\hline 
		RF ToF & \multirowcell{4}{\mthree{}} 
		&  0.13 ($\pm$  0.14) &  0.10 ($\pm$  \textbf{0.11}) &  0.11 ($\pm$  \textbf{0.14}) &  0.13 ($\pm$  0.15) &  \underline{2.73} ($\pm$  \underline{2.47}) &  \textbf{0.10} ($\pm$  \textbf{0.09})   \\ 
		NIR ToF & &  0.19 ($\pm$  \underline{0.25}) &  0.13 ($\pm$  0.19) &  0.16 ($\pm$  0.28) &  \textbf{0.10} ($\pm$  \textbf{0.10}) &  1.32 ($\pm$  0.58) &  0.32 ($\pm$  \underline{0.44})   \\ 
		Active Stereo & &  \textbf{0.10} ($\pm$  \textbf{0.12}) &  \textbf{0.08} ($\pm$  0.25) &  \textbf{0.10} ($\pm$  0.17) &  \textbf{0.10} ($\pm$  0.12) &  \textbf{0.29} ($\pm$  \textbf{0.42}) &  0.20 ($\pm$  0.17)   \\ 
		Passive Stereo & &  \underline{0.36} ($\pm$  0.17) &  \underline{0.14} ($\pm$  \underline{0.32}) &  \underline{0.18} ($\pm$  \underline{0.51}) &  \underline{0.16} ($\pm$  \underline{0.17}) &  0.47 ($\pm$  0.51) &  \underline{0.65} ($\pm$  0.33)   \\ 
		\hline 
		RF ToF & \multirowcell{4}{\mfour{}} 
		&  0.12 ($\pm$  0.13) &  0.10 ($\pm$  0.08) &  \underline{0.10} ($\pm$  0.09) &  \underline{0.13} ($\pm$  \underline{0.15}) &  \underline{2.93} ($\pm$  \underline{2.60}) &  \textbf{0.10} ($\pm$  0.09)   \\ 
		NIR ToF & &  0.08 ($\pm$  0.10) &  \underline{0.12} ($\pm$  \textbf{0.06}) &  0.09 ($\pm$  \textbf{0.06}) &  \textbf{0.09} ($\pm$  \textbf{0.07}) &  1.69 ($\pm$  \textbf{0.23}) &  \textbf{0.10} ($\pm$  0.10)   \\ 
		Active Stereo & &  \textbf{0.06} ($\pm$  \textbf{0.08}) &  \textbf{0.06} ($\pm$  \textbf{0.06}) &  \textbf{0.06} ($\pm$  \textbf{0.06}) &  \textbf{0.09} ($\pm$  0.10) &  \textbf{0.42} ($\pm$  0.50) &  0.16 ($\pm$  \textbf{0.07})   \\ 
		Passive Stereo & &  \underline{0.38} ($\pm$  \underline{0.14}) &  0.08 ($\pm$  \underline{0.11}) &  0.08 ($\pm$  \underline{0.10}) &  0.10 ($\pm$  0.10) &  0.55 ($\pm$  0.56) &  \underline{0.69} ($\pm$  \underline{0.26})   \\ 
		\hline 
		RF ToF & \multirowcell{4}{\mthree{}*} 
		&  +0.08 ($\pm$  0.14) &  -0.06 ($\pm$  \textbf{0.11}) &  -0.05 ($\pm$  \textbf{0.14}) &  -0.06 ($\pm$  0.15) &  \underline{+2.25} ($\pm$  \underline{2.47}) &  \textbf{-0.07} ($\pm$  \textbf{0.09})   \\ 
		NIR ToF & &  +0.15 ($\pm$  \underline{0.25}) &  -0.04 ($\pm$  0.19) &  \textbf{-0.01} ($\pm$  0.28) &  -0.08 ($\pm$  \textbf{0.10}) &  +1.30 ($\pm$  0.58) &  +0.09 ($\pm$  \underline{0.44})   \\ 
		Active Stereo & &  \textbf{+0.04} ($\pm$  \textbf{0.12}) &  \textbf{-0.01} ($\pm$  0.25) &  -0.02 ($\pm$  0.17) &  \textbf{-0.04} ($\pm$  0.12) &  \textbf{-0.15} ($\pm$  \textbf{0.42}) &  +0.17 ($\pm$  0.17)   \\ 
		Passive Stereo & &  \underline{+0.36} ($\pm$  0.17) &  \underline{+0.07} ($\pm$  \underline{0.32}) &  \underline{+0.10} ($\pm$  \underline{0.51}) &  \underline{+0.15} ($\pm$  \underline{0.17}) &  +0.22 ($\pm$  0.51) &  \underline{+0.64} ($\pm$  0.33)   \\ 
		\hline 
		RF ToF & \multirowcell{4}{\mfour{}*} 
		&  +0.08 ($\pm$  0.13) &  -0.09 ($\pm$  0.08) &  \underline{-0.08} ($\pm$  0.09) &  -0.07 ($\pm$  \underline{0.15}) &  \underline{+2.40} ($\pm$  \underline{2.60}) &  -0.07 ($\pm$  0.09)   \\ 
		NIR ToF & &  \textbf{-0.00} ($\pm$  0.10) &  \underline{-0.12} ($\pm$  \textbf{0.06}) &  \underline{-0.08} ($\pm$  \textbf{0.06}) &  \underline{-0.08} ($\pm$  \textbf{0.07}) &  +1.69 ($\pm$  \textbf{0.23}) &  \textbf{-0.05} ($\pm$  0.10)   \\ 
		Active Stereo & &  \textbf{+0.00} ($\pm$  \textbf{0.08}) &  -0.04 ($\pm$  \textbf{0.06}) &  -0.05 ($\pm$  \textbf{0.06}) &  \textbf{-0.05} ($\pm$  0.10) &  -0.37 ($\pm$  0.50) &  +0.16 ($\pm$  \textbf{0.07})   \\ 
		Passive Stereo & &  \underline{+0.38} ($\pm$  \underline{0.14}) &  \textbf{+0.01} ($\pm$  \underline{0.11}) &  \textbf{+0.02} ($\pm$  \underline{0.10}) &  \underline{+0.08} ($\pm$  0.10) &  \textbf{-0.34} ($\pm$  0.56) &  \underline{+0.69} ($\pm$  \underline{0.26})   \\ 
		\bottomrule
	\end{tabularx} 
	\newline 
	\newline 
	\begin{tabularx}{\textwidth}{@{}lcYYYYYY@{}} 
		\toprule
		& \multirowcell{2}{Metric \\ Type} & \multirowcell{2}{ Scrubber }  & \multirowcell{2}{ Cardboard Box }  & \multirowcell{2}{ Plushie }  & \multirowcell{2}{ Bottle }  & \multirowcell{2}{ Tape Dispenser }  & \multirowcell{2}{ Book }  \\ 
		&&&&&&& \\ 
		\hline 
		\hline 
		RF ToF & \multirowcell{4}{\mone{}} 
		&  0.58 ($\pm$  0.29) &  0.12 ($\pm$  0.08) &  \underline{0.81} ($\pm$  \underline{0.47}) &  \underline{1.22} ($\pm$  \underline{1.00}) &  0.31 ($\pm$  0.23) &  0.37 ($\pm$  0.56)   \\ 
		NIR ToF & &  \underline{0.64} ($\pm$  \underline{0.34}) &  \textbf{0.11} ($\pm$  \textbf{0.04}) &  0.49 ($\pm$  0.21) &  0.44 ($\pm$  0.51) &  \underline{0.84} ($\pm$  \underline{0.30}) &  \underline{0.59} ($\pm$  \underline{0.65})   \\ 
		Active Stereo & &  0.20 ($\pm$  0.16) &  0.12 ($\pm$  0.07) &  \textbf{0.13} ($\pm$  \textbf{0.14}) &  \textbf{0.16} ($\pm$  \textbf{0.21}) &  0.16 ($\pm$  0.14) &  \textbf{0.13} ($\pm$  \textbf{0.11})   \\ 
		Passive Stereo & &  \textbf{0.14} ($\pm$  \textbf{0.11}) &  \underline{0.29} ($\pm$  \underline{0.20}) &  0.19 ($\pm$  0.21) &  0.28 ($\pm$  0.36) &  \textbf{0.15} ($\pm$  \textbf{0.11}) &  0.15 ($\pm$  0.13)   \\ 
		\hline 
		RF ToF & \multirowcell{4}{\mtwo{}} 
		&  \underline{0.97} ($\pm$  \underline{0.61}) &  \textbf{0.11} ($\pm$  \textbf{0.05}) &  \underline{2.21} ($\pm$  \underline{2.05}) &  0.49 ($\pm$  0.99) &  0.55 ($\pm$  \underline{0.77}) &  \textbf{0.14} ($\pm$  \textbf{0.08})   \\ 
		NIR ToF & &  0.59 ($\pm$  0.30) &  0.14 ($\pm$  0.16) &  0.53 ($\pm$  0.28) &  \underline{0.80} ($\pm$  \underline{1.32}) &  \underline{1.16} ($\pm$  0.60) &  0.20 ($\pm$  0.17)   \\ 
		Active Stereo & &  \textbf{0.17} ($\pm$  \textbf{0.11}) &  0.15 ($\pm$  0.13) &  \textbf{0.12} ($\pm$  0.38) &  \textbf{0.20} ($\pm$  \textbf{0.53}) &  \textbf{0.17} ($\pm$  \textbf{0.17}) &  0.18 ($\pm$  0.15)   \\ 
		Passive Stereo & &  0.19 ($\pm$  0.13) &  \underline{0.35} ($\pm$  \underline{0.24}) &  0.23 ($\pm$  \textbf{0.27}) &  0.41 ($\pm$  0.62) &  0.22 ($\pm$  0.20) &  \underline{0.24} ($\pm$  \underline{0.19})   \\ 
		\hline 
		RF ToF & \multirowcell{4}{\mthree{}} 
		&  \underline{1.28} ($\pm$  \underline{0.76}) &  0.38 ($\pm$  \underline{1.52}) &  \underline{3.64} ($\pm$  \underline{3.23}) &  0.64 ($\pm$  1.63) &  0.67 ($\pm$  \underline{1.08}) &  \textbf{0.13} ($\pm$  0.31)   \\ 
		NIR ToF & &  0.91 ($\pm$  0.47) &  \textbf{0.14} ($\pm$  \textbf{0.19}) &  0.85 ($\pm$  \textbf{0.52}) &  \underline{1.32} ($\pm$  2.00) &  \underline{1.57} ($\pm$  0.70) &  \underline{1.15} ($\pm$  \underline{1.42})   \\ 
		Active Stereo & &  \textbf{0.27} ($\pm$  \textbf{0.34}) &  0.19 ($\pm$  0.22) &  \textbf{0.24} ($\pm$  1.25) &  \textbf{0.49} ($\pm$  \underline{2.49}) &  \textbf{0.24} ($\pm$  \textbf{0.35}) &  0.19 ($\pm$  \textbf{0.25})   \\ 
		Passive Stereo & &  0.29 ($\pm$  0.38) &  \underline{0.47} ($\pm$  0.55) &  0.35 ($\pm$  0.54) &  0.62 ($\pm$  \textbf{0.97}) &  0.29 ($\pm$  \textbf{0.35}) &  0.47 ($\pm$  1.21)   \\ 
		\hline 
		RF ToF & \multirowcell{4}{\mfour{}} 
		&  \underline{1.35} ($\pm$  \underline{0.66}) &  0.25 ($\pm$  \underline{1.19}) &  \underline{3.61} ($\pm$  \underline{3.23}) &  0.63 ($\pm$  1.63) &  0.66 ($\pm$  \underline{1.07}) &  \textbf{0.13} ($\pm$  0.30)   \\ 
		NIR ToF & &  1.16 ($\pm$  \textbf{0.27}) &  \textbf{0.13} ($\pm$  \textbf{0.07}) &  0.82 ($\pm$  0.38) &  \underline{1.16} ($\pm$  \underline{2.16}) &  \underline{1.70} ($\pm$  0.89) &  \underline{1.34} ($\pm$  \underline{1.54})   \\ 
		Active Stereo & &  \textbf{0.22} ($\pm$  \textbf{0.27}) &  0.18 ($\pm$  0.12) &  \textbf{0.16} ($\pm$  0.30) &  \textbf{0.33} ($\pm$  1.13) &  \textbf{0.15} ($\pm$  \textbf{0.21}) &  0.17 ($\pm$  \textbf{0.23})   \\ 
		Passive Stereo & &  0.24 ($\pm$  0.29) &  \underline{0.42} ($\pm$  0.28) &  0.18 ($\pm$  \textbf{0.26}) &  0.55 ($\pm$  \textbf{0.86}) &  0.20 ($\pm$  0.22) &  0.33 ($\pm$  0.95)   \\ 
		\hline 
		RF ToF & \multirowcell{4}{\mthree{}*} 
		&  \underline{+1.26} ($\pm$  \underline{0.76}) &  +0.28 ($\pm$  \underline{1.52}) &  \underline{+3.63} ($\pm$  \underline{3.23}) &  +0.52 ($\pm$  1.63) &  +0.56 ($\pm$  \underline{1.08}) &  -0.10 ($\pm$  0.31)   \\ 
		NIR ToF & &  +0.89 ($\pm$  0.47) &  \textbf{-0.09} ($\pm$  \textbf{0.19}) &  +0.80 ($\pm$  \textbf{0.52}) &  \underline{+1.30} ($\pm$  2.00) &  \underline{+1.57} ($\pm$  0.70) &  \underline{+1.13} ($\pm$  \underline{1.42})   \\ 
		Active Stereo & &  \textbf{-0.01} ($\pm$  \textbf{0.34}) &  -0.10 ($\pm$  0.22) &  \textbf{-0.04} ($\pm$  1.25) &  \textbf{+0.18} ($\pm$  \underline{2.49}) &  \textbf{+0.01} ($\pm$  \textbf{0.35}) &  \textbf{-0.07} ($\pm$  \textbf{0.25})   \\ 
		Passive Stereo & &  \textbf{-0.01} ($\pm$  0.38) &  \underline{+0.42} ($\pm$  0.55) &  +0.17 ($\pm$  0.54) &  +0.51 ($\pm$  \textbf{0.97}) &  +0.19 ($\pm$  \textbf{0.35}) &  +0.39 ($\pm$  1.21)   \\ 
		\hline 
		RF ToF & \multirowcell{4}{\mfour{}*} 
		&  \underline{+1.34} ($\pm$  \underline{0.66}) &  +0.14 ($\pm$  \underline{1.19}) &  \underline{+3.59} ($\pm$  \underline{3.23}) &  +0.52 ($\pm$  1.63) &  +0.57 ($\pm$  \underline{1.07}) &  -0.10 ($\pm$  0.30)   \\ 
		NIR ToF & &  +1.16 ($\pm$  \textbf{0.27}) &  \textbf{-0.13} ($\pm$  \textbf{0.07}) &  +0.82 ($\pm$  0.38) &  \underline{+1.15} ($\pm$  \underline{2.16}) &  \underline{+1.70} ($\pm$  0.89) &  \underline{+1.33} ($\pm$  \underline{1.54})   \\ 
		Active Stereo & &  \textbf{-0.02} ($\pm$  \textbf{0.27}) &  -0.16 ($\pm$  0.12) &  -0.06 ($\pm$  0.30) &  \textbf{+0.21} ($\pm$  1.13) &  \textbf{+0.03} ($\pm$  \textbf{0.21}) &  \textbf{-0.02} ($\pm$  \textbf{0.23})   \\ 
		Passive Stereo & &  -0.07 ($\pm$  0.29) &  \underline{+0.42} ($\pm$  0.28) &  \textbf{+0.02} ($\pm$  \textbf{0.26}) &  +0.52 ($\pm$  \textbf{0.86}) &  +0.08 ($\pm$  0.22) &  +0.27 ($\pm$  0.95)   \\ 
		\bottomrule
	\end{tabularx} 
\end{table*}
\begin{table*}[]
	\caption{\label{table:supp_metrics_2}%
		We measure the depth deviation with respect to \protect\mone{}\protect, \protect\mtwo{}\protect, \protect\mthree{}\protect, \protect\mfour{}\protect~and an additional signed version of \protect\mthree{}\protect,\protect\mfour{}\protect, which is denoted as \protect\mthree{}\protect*,\protect\mfour{}\protect*. All metrics are listed in the form $(\mu \pm \sigma)$, consisting of the mean $\mu$ and standard deviation $\sigma$ in centimeters, computed over the entire metric domain, respectively. The best results among all sensors of one metric type are highlighted in \bf{bold} and the worst results are \underline{underlined}.}
	\vspace*{-0.25 cm}
	\begin{tabularx}{\textwidth}{@{}lcYYYYYY@{}} 
		\toprule 
		& \multirowcell{2}{Metric \\ Type} & \multirowcell{2}{ Statue }  & \multirowcell{2}{ Rubber Foam \\ Plane }  & \multirowcell{2}{ Sandpaper \\ (k80) }  & \multirowcell{2}{ Sandpaper \\ (k120) }  & \multirowcell{2}{ Wood Plane }  & \multirowcell{2}{ Foam Plane }  \\ 
		&&&&&&& \\ 
		\hline 
		\hline 
		RF ToF & \multirowcell{4}{\mone{}} 
		&  0.27 ($\pm$  0.25) &  \underline{1.10} ($\pm$  \underline{1.21}) &  \underline{1.70} ($\pm$  \underline{2.07}) &  \underline{1.71} ($\pm$  \underline{2.23}) &  \underline{2.08} ($\pm$  \underline{2.31}) &  \underline{2.66} ($\pm$  \underline{1.26})   \\ 
		NIR ToF & &  \underline{0.32} ($\pm$  \underline{0.28}) &  0.34 ($\pm$  0.07) &  0.09 ($\pm$  \textbf{0.04}) &  \textbf{0.07} ($\pm$  \textbf{0.03}) &  0.48 ($\pm$  0.15) &  0.80 ($\pm$  0.14)   \\ 
		Active Stereo & &  0.16 ($\pm$  \textbf{0.13}) &  \textbf{0.11} ($\pm$  \textbf{0.05}) &  \textbf{0.07} ($\pm$  \textbf{0.04}) &  0.08 ($\pm$  0.04) &  0.13 ($\pm$  \textbf{0.06}) &  \textbf{0.08} ($\pm$  \textbf{0.06})   \\ 
		Passive Stereo & &  \textbf{0.13} ($\pm$  \textbf{0.13}) &  0.77 ($\pm$  0.67) &  0.08 ($\pm$  \textbf{0.04}) &  0.10 ($\pm$  0.06) &  \textbf{0.12} ($\pm$  0.10) &  0.15 ($\pm$  0.09)   \\ 
		\hline 
		RF ToF & \multirowcell{4}{\mtwo{}} 
		&  \textbf{0.17} ($\pm$  \textbf{0.11}) &  0.34 ($\pm$  \underline{0.72}) &  \underline{0.12} ($\pm$  \textbf{0.05}) &  \underline{0.12} ($\pm$  \textbf{0.05}) &  0.20 ($\pm$  0.12) &  \underline{2.83} ($\pm$  \underline{0.89})   \\ 
		NIR ToF & &  \underline{0.43} ($\pm$  0.42) &  0.38 ($\pm$  0.11) &  0.11 ($\pm$  \underline{0.12}) &  \textbf{0.09} ($\pm$  \underline{0.12}) &  \underline{0.52} ($\pm$  \underline{0.16}) &  0.84 ($\pm$  0.15)   \\ 
		Active Stereo & &  0.19 ($\pm$  \underline{0.76}) &  \textbf{0.13} ($\pm$  \textbf{0.08}) &  \textbf{0.08} ($\pm$  \textbf{0.05}) &  \textbf{0.09} ($\pm$  \textbf{0.05}) &  0.15 ($\pm$  \textbf{0.08}) &  \textbf{0.09} ($\pm$  0.15)   \\ 
		Passive Stereo & &  0.18 ($\pm$  0.69) &  \underline{0.83} ($\pm$  0.71) &  \underline{0.12} ($\pm$  0.09) &  0.11 ($\pm$  0.08) &  \textbf{0.13} ($\pm$  0.11) &  0.16 ($\pm$  \textbf{0.12})   \\ 
		\hline 
		RF ToF & \multirowcell{4}{\mthree{}} 
		&  \textbf{0.20} ($\pm$  \textbf{0.26}) &  \underline{1.62} ($\pm$  \underline{2.21}) &  \textbf{0.09} ($\pm$  0.11) &  \textbf{0.08} ($\pm$  0.10) &  0.18 ($\pm$  0.23) &  \underline{3.95} ($\pm$  \underline{4.22})   \\ 
		NIR ToF & &  0.77 ($\pm$  3.13) &  0.44 ($\pm$  \textbf{0.19}) &  0.10 ($\pm$  0.14) &  0.09 ($\pm$  \underline{0.14}) &  \underline{0.65} ($\pm$  \underline{0.28}) &  1.13 ($\pm$  3.58)   \\ 
		Active Stereo & &  0.90 ($\pm$  5.17) &  \textbf{0.16} ($\pm$  0.24) &  \textbf{0.09} ($\pm$  \textbf{0.09}) &  0.10 ($\pm$  \textbf{0.09}) &  0.20 ($\pm$  \textbf{0.12}) &  0.21 ($\pm$  2.11)   \\ 
		Passive Stereo & &  \underline{1.43} ($\pm$  \underline{7.09}) &  0.91 ($\pm$  1.11) &  \underline{0.13} ($\pm$  \underline{0.17}) &  \underline{0.13} ($\pm$  \underline{0.14}) &  \textbf{0.14} ($\pm$  0.18) &  \textbf{0.19} ($\pm$  \textbf{0.44})   \\ 
		\hline 
		RF ToF & \multirowcell{4}{\mfour{}} 
		&  0.20 ($\pm$  \underline{0.27}) &  \underline{1.62} ($\pm$  \underline{2.21}) &  0.09 ($\pm$  0.11) &  \textbf{0.08} ($\pm$  0.10) &  0.18 ($\pm$  0.23) &  \underline{3.95} ($\pm$  \underline{4.22})   \\ 
		NIR ToF & &  \underline{0.25} ($\pm$  \textbf{0.13}) &  0.43 ($\pm$  0.11) &  \textbf{0.08} ($\pm$  0.10) &  \textbf{0.08} ($\pm$  0.10) &  \underline{0.64} ($\pm$  \underline{0.27}) &  0.91 ($\pm$  0.12)   \\ 
		Active Stereo & &  0.16 ($\pm$  0.21) &  \textbf{0.16} ($\pm$  \textbf{0.09}) &  0.09 ($\pm$  \textbf{0.08}) &  0.10 ($\pm$  \textbf{0.09}) &  0.20 ($\pm$  \textbf{0.10}) &  \textbf{0.09} ($\pm$  \textbf{0.10})   \\ 
		Passive Stereo & &  \textbf{0.10} ($\pm$  \textbf{0.13}) &  0.94 ($\pm$  1.13) &  \underline{0.12} ($\pm$  \underline{0.15}) &  \underline{0.11} ($\pm$  \underline{0.12}) &  \textbf{0.12} ($\pm$  0.13) &  0.17 ($\pm$  0.20)   \\ 
		\hline 
		RF ToF & \multirowcell{4}{\mthree{}*} 
		&  \textbf{-0.04} ($\pm$  \textbf{0.26}) &  \underline{-1.35} ($\pm$  \underline{2.21}) &  -0.04 ($\pm$  0.11) &  -0.04 ($\pm$  0.10) &  \textbf{+0.03} ($\pm$  0.23) &  -0.66 ($\pm$  \underline{4.22})   \\ 
		NIR ToF & &  -0.23 ($\pm$  3.13) &  +0.43 ($\pm$  \textbf{0.19}) &  -0.03 ($\pm$  0.14) &  \textbf{+0.02} ($\pm$  \underline{0.14}) &  \underline{+0.65} ($\pm$  \underline{0.28}) &  \underline{+1.13} ($\pm$  3.58)   \\ 
		Active Stereo & &  +0.71 ($\pm$  5.17) &  \textbf{-0.15} ($\pm$  0.24) &  \underline{-0.06} ($\pm$  \textbf{0.09}) &  \underline{-0.08} ($\pm$  \textbf{0.09}) &  -0.19 ($\pm$  \textbf{0.12}) &  \textbf{+0.08} ($\pm$  2.11)   \\ 
		Passive Stereo & &  \underline{+1.34} ($\pm$  \underline{7.09}) &  +0.34 ($\pm$  1.11) &  \textbf{+0.00} ($\pm$  \underline{0.17}) &  \underline{-0.08} ($\pm$  \underline{0.14}) &  +0.08 ($\pm$  0.18) &  \textbf{+0.08} ($\pm$  \textbf{0.44})   \\ 
		\hline 
		RF ToF & \multirowcell{4}{\mfour{}*} 
		&  -0.04 ($\pm$  \underline{0.27}) &  \underline{-1.35} ($\pm$  \underline{2.21}) &  -0.04 ($\pm$  0.11) &  -0.04 ($\pm$  0.10) &  \textbf{+0.03} ($\pm$  0.23) &  -0.66 ($\pm$  \underline{4.22})   \\ 
		NIR ToF & &  \underline{-0.24} ($\pm$  \textbf{0.13}) &  +0.43 ($\pm$  0.11) &  -0.04 ($\pm$  0.10) &  \textbf{+0.00} ($\pm$  0.10) &  \underline{+0.64} ($\pm$  \underline{0.27}) &  \underline{+0.91} ($\pm$  0.12)   \\ 
		Active Stereo & &  +0.07 ($\pm$  0.21) &  \textbf{-0.16} ($\pm$  \textbf{0.09}) &  \underline{-0.06} ($\pm$  \textbf{0.08}) &  \underline{-0.09} ($\pm$  \textbf{0.09}) &  -0.20 ($\pm$  \textbf{0.10}) &  \textbf{-0.05} ($\pm$  \textbf{0.10})   \\ 
		Passive Stereo & &  \textbf{+0.03} ($\pm$  \textbf{0.13}) &  +0.35 ($\pm$  1.13) &  \textbf{+0.01} ($\pm$  \underline{0.15}) &  -0.07 ($\pm$  \underline{0.12}) &  +0.07 ($\pm$  0.13) &  +0.07 ($\pm$  0.20)   \\ 
		\bottomrule 
	\end{tabularx}  
	\newline 
	\newline 
	\begin{tabularx}{\textwidth}{@{}lcYYYYYY@{}} 
		\toprule 
		& \multirowcell{2}{Metric \\ Type} &  \multirowcell{2}{ S1 Hand Open }  & \multirowcell{2}{ S1 Hand Open \\ (Rev.) }  & \multirowcell{2}{ S2 Hand Open } & \multirowcell{2}{ S2 Hand Open \\ (Rev.) }  & \multirowcell{2}{ Hand Printed \\ Flat }  & \multirowcell{2}{ Corner \\ Reflector }  \\ 
		&&&&&&& \\ 
		\hline 
		\hline 
		RF ToF & \multirowcell{4}{\mone{}} 
		&   \underline{0.36} ($\pm$  \underline{0.38}) &  \underline{1.23} ($\pm$  \underline{1.42}) &  \underline{0.39} ($\pm$  \underline{0.37}) &  \underline{0.79} ($\pm$  \underline{0.79}) & \underline{0.71} ($\pm$  \underline{0.78}) &  \underline{3.48} ($\pm$  \underline{1.80})   \\ 
		NIR ToF & &  0.31 ($\pm$  0.14) &  0.27 ($\pm$  0.12) &  0.19 ($\pm$  \textbf{0.10}) &  0.21 ($\pm$  \textbf{0.11}) & 0.25 ($\pm$  0.12) &  1.81 ($\pm$  1.01)   \\ 
		Active Stereo & & \textbf{0.12} ($\pm$  \textbf{0.09}) &  0.16 ($\pm$  0.15) & \textbf{0.16} ($\pm$  0.13) &  0.21 ($\pm$  0.16) & \textbf{0.09} ($\pm$  \textbf{0.07}) &  0.48 ($\pm$  0.62)   \\ 
		Passive Stereo & & 0.20 ($\pm$  0.16) &  \textbf{0.12} ($\pm$  \textbf{0.09}) &  0.20 ($\pm$  0.16) &  \textbf{0.20} ($\pm$  0.15) & 0.21 ($\pm$  0.37) &  \textbf{0.30} ($\pm$  \textbf{0.29})   \\ 
		\hline 
		RF ToF & \multirowcell{4}{\mtwo{}} 
		&  0.22 ($\pm$  0.15) &  \textbf{0.17} ($\pm$  \textbf{0.11}) &  0.20 ($\pm$  \textbf{0.14}) &  \textbf{0.15} ($\pm$  \textbf{0.09}) &  0.17 ($\pm$  0.13) &  1.95 ($\pm$  \textbf{0.94})   \\ 
		NIR ToF & &  \underline{0.38} ($\pm$  \underline{0.26}) &  \underline{0.32} ($\pm$  0.24) &  \underline{0.25} ($\pm$  \underline{0.23}) & \underline{0.27} ($\pm$  \underline{0.24}) &  \underline{0.29} ($\pm$  0.20) &  \underline{1.97} ($\pm$  1.09)   \\ 
		Active Stereo & &  \textbf{0.13} ($\pm$  \textbf{0.10}) &  0.18 ($\pm$  \underline{0.73}) &  \textbf{0.17} ($\pm$  \textbf{0.14}) & 0.22 ($\pm$  0.15) & \textbf{0.09} ($\pm$  \textbf{0.06}) &  \textbf{0.55} ($\pm$  \underline{1.54})   \\ 
		Passive Stereo & &   0.26 ($\pm$  0.22) &  \textbf{0.17} ($\pm$  0.18) &  \underline{0.25} ($\pm$  0.20) &  0.25 ($\pm$  0.19) & 0.18 ($\pm$  \underline{0.34}) &  0.61 ($\pm$  1.22)   \\ 
		\hline 
		RF ToF & \multirowcell{4}{\mthree{}} 
		&  \textbf{0.22} ($\pm$  \textbf{0.25}) &  \textbf{0.16} ($\pm$  \textbf{0.21}) &  \textbf{0.20} ($\pm$  \textbf{0.24}) & \textbf{0.14} ($\pm$  \textbf{0.17}) &   \textbf{0.16} ($\pm$  \textbf{0.20}) &  \underline{3.31} ($\pm$  \textbf{1.53})   \\ 
		NIR ToF & & \underline{0.52} ($\pm$  0.43) &  \underline{0.47} ($\pm$  0.44) &  \underline{0.39} ($\pm$  0.57) &  \underline{0.35} ($\pm$  0.39) &  0.33 ($\pm$  0.29) &  2.87 ($\pm$  1.67)   \\ 
		Active Stereo & &  \textbf{0.22} ($\pm$  \underline{1.25}) &  0.30 ($\pm$  \underline{1.54}) &  0.29 ($\pm$  \underline{1.67}) & 0.33 ($\pm$  \underline{1.09}) & \textbf{0.16} ($\pm$  1.30) &  1.06 ($\pm$  \underline{3.92})   \\ 
		Passive Stereo & &  0.35 ($\pm$  0.41) &  0.22 ($\pm$  0.38) &  0.35 ($\pm$  0.85) & 0.32 ($\pm$  0.48) &   \underline{1.73} ($\pm$  \underline{8.95}) &  \textbf{0.99} ($\pm$  2.49)   \\ 
		\hline 
		RF ToF & \multirowcell{4}{\mfour{}} 
		&  0.22 ($\pm$  0.25) &  \textbf{0.16} ($\pm$  0.20) &  \textbf{0.20} ($\pm$  0.24) & \textbf{0.14} ($\pm$  \textbf{0.16}) & 0.16 ($\pm$  \underline{0.20}) &  \underline{3.14} ($\pm$  1.41)   \\ 
		NIR ToF & &  \underline{0.51} ($\pm$  0.27) &  \underline{0.38} ($\pm$  0.21) &  \underline{0.30} ($\pm$  \textbf{0.22}) & 0.25 ($\pm$  0.17) & \underline{0.30} ($\pm$  \textbf{0.09}) &  2.65 ($\pm$  1.29)   \\ 
		Active Stereo & &  \textbf{0.16} ($\pm$  \textbf{0.24}) &  0.20 ($\pm$  \underline{0.25}) &  \textbf{0.20} ($\pm$  0.23) & \underline{0.26} ($\pm$  \underline{0.30}) & \textbf{0.08} ($\pm$  0.10) &  \textbf{0.70} ($\pm$  \textbf{0.92})   \\ 
		Passive Stereo & &  0.25 ($\pm$  \underline{0.34}) &  \textbf{0.16} ($\pm$  \textbf{0.17}) &  0.21 ($\pm$  \underline{0.25}) & 0.25 ($\pm$  0.21) &  0.17 ($\pm$  0.15) &  1.26 ($\pm$  \underline{1.63})   \\
		\hline 
		RF ToF & \multirowcell{4}{\mthree{}*} 
		&  -0.15 ($\pm$  \textbf{0.25}) &  \textbf{-0.06} ($\pm$  \textbf{0.21}) &  -0.10 ($\pm$  \textbf{0.24}) & \textbf{-0.07} ($\pm$  \textbf{0.17}) &   -0.07 ($\pm$  \textbf{0.20}) &  \underline{+3.31} ($\pm$  \textbf{1.53})   \\ 
		NIR ToF & &  \underline{+0.49} ($\pm$  0.43) &  \underline{+0.42} ($\pm$  0.44) &  \underline{+0.32} ($\pm$  0.57) & \underline{+0.29} ($\pm$  0.39) & -0.30 ($\pm$  0.29) &  +2.86 ($\pm$  1.67)   \\ 
		Active Stereo & & \textbf{+0.06} ($\pm$  \underline{1.25}) &  -0.07 ($\pm$  \underline{1.54}) &  \textbf{-0.04} ($\pm$  \underline{1.67}) &  +0.12 ($\pm$  \underline{1.09}) & \textbf{+0.04} ($\pm$  1.30) &  \textbf{+0.82} ($\pm$  \underline{3.92})   \\ 
		Passive Stereo & &  +0.22 ($\pm$  0.41) &  +0.17 ($\pm$  0.38) &  \underline{+0.32} ($\pm$  0.85) & \underline{+0.29} ($\pm$  0.48) & \underline{+1.70} ($\pm$  \underline{8.95}) &  +0.88 ($\pm$  2.49)   \\ 
		\hline 
		RF ToF & \multirowcell{4}{\mfour{}*} 
		&  -0.14 ($\pm$  0.25) &  \textbf{-0.06} ($\pm$  0.20) &  \textbf{-0.10} ($\pm$  0.24) & \textbf{-0.07} ($\pm$  \textbf{0.16}) & -0.07 ($\pm$  \underline{0.20}) &  \underline{+3.14} ($\pm$  1.41)   \\ 
		NIR ToF & &  \underline{+0.51} ($\pm$  0.27) &  \underline{+0.38} ($\pm$  0.21) &  \underline{+0.28} ($\pm$  \textbf{0.22}) & \underline{+0.24} ($\pm$  0.17) &   \underline{-0.30} ($\pm$  \textbf{0.09}) &  +2.65 ($\pm$  1.29)   \\ 
		Active Stereo & & \textbf{+0.00} ($\pm$  \textbf{0.24}) &  -0.13 ($\pm$  \underline{0.25}) &  -0.14 ($\pm$  0.23) &  +0.11 ($\pm$  \underline{0.30}) & \textbf{+0.00} ($\pm$  0.10) &  \textbf{+0.45} ($\pm$  \textbf{0.92})   \\ 
		Passive Stereo & &  +0.05 ($\pm$  \underline{0.34}) &  +0.11 ($\pm$  \textbf{0.17}) &  +0.15 ($\pm$  \underline{0.25}) & +0.23 ($\pm$  0.21) & +0.16 ($\pm$  0.15) &  +1.23 ($\pm$  \underline{1.63})   \\ 
		\bottomrule 
	\end{tabularx} 	
\end{table*}
\begin{table*}[]
	\caption{\label{table:supp_metrics_3}%
		We measure the depth deviation with respect to \protect\mone{}\protect, \protect\mtwo{}\protect, \protect\mthree{}\protect, \protect\mfour{}\protect~and an additional signed version of \protect\mthree{}\protect,\protect\mfour{}\protect, which is denoted as \protect\mthree{}\protect*,\protect\mfour{}\protect*. All metrics are listed in the form $(\mu \pm \sigma)$, consisting of the mean $\mu$ and standard deviation $\sigma$ in centimeters, computed over the entire metric domain, respectively. The best results among all sensors of one metric type are highlighted in \bf{bold} and the worst results are \underline{underlined}.}
	\vspace*{-0.25 cm}
	\begin{tabularx}{\textwidth}{@{}lcYYYYYY@{}} 
		\toprule 
		& \multirowcell{2}{Metric \\ Type} & \multirowcell{2}{ Mirror }  & \multirowcell{2}{ Candle }  & \multirowcell{2}{ Flowerpot \\ (Transparent) }  & \multirowcell{2}{ V1 Metal Plate }  & \multirowcell{2}{ V2 Metal Plate }  & \multirowcell{2}{ Hand Printed F }  \\ 
		&&&&&&& \\ 
		\hline 
		\hline 
		RF ToF & \multirowcell{4}{\mone{}} 
		&  \textbf{0.87} ($\pm$  \textbf{0.26}) &  1.50 ($\pm$  \underline{1.12}) &  1.31 ($\pm$  \underline{1.21}) &  0.12 ($\pm$  \textbf{0.05}) &  \textbf{0.12} ($\pm$  \textbf{0.05}) &  \underline{0.69} ($\pm$  \underline{0.86})   \\ 
		NIR ToF & &  \underline{3.77} ($\pm$  \underline{1.97}) &  \underline{2.04} ($\pm$  0.40) &  \underline{2.73} ($\pm$  1.03) &  \underline{0.77} ($\pm$  \underline{0.42}) &  \underline{0.74} ($\pm$  \underline{0.45}) &  \textbf{0.12} ($\pm$  \textbf{0.09})   \\ 
		Active Stereo & &  2.13 ($\pm$  1.52) &  \textbf{0.26} ($\pm$  \textbf{0.29}) &  \textbf{0.74} ($\pm$  \textbf{0.53}) &  \textbf{0.08} ($\pm$  0.06) &  0.30 ($\pm$  0.29) &  0.17 ($\pm$  0.18)   \\ 
		Passive Stereo & &  2.31 ($\pm$  1.61) &  1.64 ($\pm$  0.78) &  2.01 ($\pm$  0.83) &  0.13 ($\pm$  0.07) &  0.15 ($\pm$  0.11) &  0.20 ($\pm$  0.14)   \\ 
		\hline 
		RF ToF & \multirowcell{4}{\mtwo{}} 
		&  \textbf{0.91} ($\pm$  \textbf{0.14}) &  \underline{5.57} ($\pm$  \underline{2.78}) &  1.86 ($\pm$  \underline{2.41}) &  0.13 ($\pm$  \textbf{0.06}) &  \textbf{0.13} ($\pm$  \textbf{0.07}) &  0.15 ($\pm$  \textbf{0.10})   \\ 
		NIR ToF & &  \underline{33.31} ($\pm$  9.07) &  1.71 ($\pm$  0.49) &  \underline{3.10} ($\pm$  1.22) &  \underline{0.81} ($\pm$  \underline{0.43}) &  \underline{15.66} ($\pm$  \underline{17.32}) &  \textbf{0.12} ($\pm$  \textbf{0.10})   \\ 
		Active Stereo & &  30.21 ($\pm$  \underline{14.59}) &  \textbf{0.25} ($\pm$  \textbf{0.26}) &  \textbf{1.27} ($\pm$  1.78) &  \textbf{0.09} ($\pm$  0.07) &  5.54 ($\pm$  12.95) &  0.21 ($\pm$  \underline{0.63})   \\ 
		Passive Stereo & &  27.02 ($\pm$  11.33) &  1.28 ($\pm$  0.65) &  1.86 ($\pm$  \textbf{0.93}) &  0.16 ($\pm$  0.11) &  0.20 ($\pm$  0.14) &  \underline{0.23} ($\pm$  0.16)   \\ 
		\hline 
		RF ToF & \multirowcell{4}{\mthree{}} 
		&  \textbf{0.93} ($\pm$  \textbf{0.12}) &  \underline{7.41} ($\pm$  \underline{3.79}) &  2.74 ($\pm$  \underline{3.66}) &  0.11 ($\pm$  \textbf{0.12}) &  \textbf{0.11} ($\pm$  \textbf{0.12}) &  \textbf{0.14} ($\pm$  \textbf{0.20})   \\ 
		NIR ToF & &  37.84 ($\pm$  14.84) &  2.78 ($\pm$  \textbf{0.35}) &  \underline{5.24} ($\pm$  2.04) &  \underline{0.95} ($\pm$  \underline{0.48}) &  \underline{16.23} ($\pm$  \underline{18.21}) &  0.17 ($\pm$  0.30)   \\ 
		Active Stereo & &  \underline{39.66} ($\pm$  \underline{24.75}) &  \textbf{0.42} ($\pm$  0.49) &  \textbf{2.08} ($\pm$  2.30) &  \textbf{0.10} ($\pm$  0.13) &  6.16 ($\pm$  13.78) &  \underline{0.56} ($\pm$  \underline{2.02})   \\ 
		Passive Stereo & &  30.82 ($\pm$  14.01) &  2.10 ($\pm$  0.98) &  3.50 ($\pm$  \textbf{1.37}) &  0.19 ($\pm$  0.15) &  0.22 ($\pm$  0.20) &  0.43 ($\pm$  0.73)   \\ 
		\hline 
		RF ToF & \multirowcell{4}{\mfour{}} 
		&  \textbf{0.93} ($\pm$  \textbf{0.12}) &  \underline{7.37} ($\pm$  \underline{3.85}) &  2.76 ($\pm$  \underline{3.66}) &  0.10 ($\pm$  0.11) &  \textbf{0.11} ($\pm$  \textbf{0.12}) &  \textbf{0.13} ($\pm$  0.20)   \\ 
		NIR ToF & &  39.68 ($\pm$  6.57) &  2.75 ($\pm$  \textbf{0.15}) &  \underline{6.18} ($\pm$  1.79) &  \underline{0.79} ($\pm$  \underline{0.39}) &  \underline{24.92} ($\pm$  \underline{17.59}) &  \textbf{0.13} ($\pm$  \textbf{0.16})   \\ 
		Active Stereo & &  \underline{43.84} ($\pm$  \underline{20.28}) &  \textbf{0.31} ($\pm$  0.44) &  \textbf{2.52} ($\pm$  1.77) &  \textbf{0.07} ($\pm$  \textbf{0.09}) &  7.46 ($\pm$  14.94) &  \underline{0.53} ($\pm$  \underline{1.07})   \\ 
		Passive Stereo & &  35.96 ($\pm$  7.83) &  2.15 ($\pm$  0.65) &  4.36 ($\pm$  \textbf{0.71}) &  0.15 ($\pm$  \textbf{0.09}) &  0.23 ($\pm$  0.19) &  0.34 ($\pm$  0.65)   \\ 
		\hline 
		RF ToF & \multirowcell{4}{\mthree{}*} 
		&  \textbf{+0.93} ($\pm$  \textbf{0.12}) &  \underline{+7.38} ($\pm$  \underline{3.79}) &  +2.68 ($\pm$  \underline{3.66}) &  -0.06 ($\pm$  \textbf{0.12}) &  \textbf{-0.06} ($\pm$  \textbf{0.12}) &  \textbf{-0.02} ($\pm$  \textbf{0.20})   \\ 
		NIR ToF & &  +37.84 ($\pm$  14.84) &  +2.78 ($\pm$  \textbf{0.35}) &  \underline{+5.24} ($\pm$  2.04) &  \underline{+0.95} ($\pm$  \underline{0.48}) &  \underline{+16.21} ($\pm$  \underline{18.21}) &  \underline{-0.11} ($\pm$  0.30)   \\ 
		Active Stereo & &  \underline{+39.58} ($\pm$  \underline{24.75}) &  \textbf{+0.37} ($\pm$  0.49) &  \textbf{+2.00} ($\pm$  2.30) &  \textbf{+0.00} ($\pm$  0.13) &  +5.77 ($\pm$  13.78) &  -0.03 ($\pm$  \underline{2.02})   \\ 
		Passive Stereo & &  +30.80 ($\pm$  14.01) &  +2.10 ($\pm$  0.98) &  +3.49 ($\pm$  \textbf{1.37}) &  +0.17 ($\pm$  0.15) &  +0.19 ($\pm$  0.20) &  \textbf{+0.02} ($\pm$  0.73)   \\ 
		\hline 
		RF ToF & \multirowcell{4}{\mfour{}*} 
		&  \textbf{+0.93} ($\pm$  \textbf{0.12}) &  \underline{+7.34} ($\pm$  \underline{3.85}) &  +2.70 ($\pm$  \underline{3.66}) &  -0.06 ($\pm$  0.11) &  \textbf{-0.06} ($\pm$  \textbf{0.12}) &  \textbf{-0.01} ($\pm$  0.20)   \\ 
		NIR ToF & &  +39.68 ($\pm$  6.57) &  +2.75 ($\pm$  \textbf{0.15}) &  \underline{+6.18} ($\pm$  1.79) &  \underline{+0.79} ($\pm$  \underline{0.39}) &  \underline{+24.90} ($\pm$  \underline{17.59}) &  -0.11 ($\pm$  \textbf{0.16})   \\ 
		Active Stereo & &  \underline{+43.84} ($\pm$  \underline{20.28}) &  \textbf{+0.27} ($\pm$  0.44) &  \textbf{+2.52} ($\pm$  1.77) &  \textbf{+0.01} ($\pm$  \textbf{0.09}) &  +7.08 ($\pm$  14.94) &  \underline{-0.36} ($\pm$  \underline{1.07})   \\ 
		Passive Stereo & &  +35.96 ($\pm$  7.83) &  +2.15 ($\pm$  0.65) &  +4.36 ($\pm$  \textbf{0.71}) &  +0.15 ($\pm$  \textbf{0.09}) &  +0.21 ($\pm$  0.19) &  -0.07 ($\pm$  0.65)   \\ 
		\bottomrule 
	\end{tabularx} 
	\newline 
	\newline 
	\begin{tabularx}{\textwidth}{@{}lcYYYYYY@{}} 
		\toprule 
		& \multirowcell{2}{Metric \\ Type} & \multirowcell{2}{ Hand Printed B }  & \multirowcell{2}{ Hand Printed U }  & \multirowcell{2}{ Metal Angle }  & \multirowcell{2}{ Plunger }  & \multirowcell{2}{ Silicone Cup }  & \multirowcell{2}{ V1 Christmas \\ Ball }  \\ 
		&&&&&&& \\ 
		\hline 
		\hline 
		RF ToF & \multirowcell{4}{\mone{}} 
		&  \underline{0.52} ($\pm$  \underline{0.70}) &  \underline{0.72} ($\pm$  \underline{0.80}) &  0.47 ($\pm$  0.28) &  0.62 ($\pm$  0.60) &  \underline{0.60} ($\pm$  \underline{0.47}) &  \underline{0.59} ($\pm$  \underline{0.40})   \\ 
		NIR ToF & &  \textbf{0.09} ($\pm$  \textbf{0.06}) &  \textbf{0.12} ($\pm$  \textbf{0.09}) &  \underline{0.90} ($\pm$  \underline{0.79}) &  \textbf{0.23} ($\pm$  \textbf{0.13}) &  0.16 ($\pm$  \textbf{0.10}) &  0.22 ($\pm$  0.14)   \\ 
		Active Stereo & &  0.12 ($\pm$  0.11) &  0.15 ($\pm$  0.12) &  \textbf{0.23} ($\pm$  0.18) &  0.24 ($\pm$  0.17) &  \textbf{0.13} ($\pm$  \textbf{0.10}) &  \textbf{0.13} ($\pm$  0.09)   \\ 
		Passive Stereo & &  0.19 ($\pm$  0.16) &  0.18 ($\pm$  0.14) &  0.24 ($\pm$  \textbf{0.17}) &  \underline{0.95} ($\pm$  \underline{0.64}) &  0.15 ($\pm$  \textbf{0.10}) &  \textbf{0.13} ($\pm$  \textbf{0.05})   \\ 
		\hline 
		RF ToF & \multirowcell{4}{\mtwo{}} 
		&  0.17 ($\pm$  \textbf{0.12}) &  0.18 ($\pm$  \textbf{0.13}) &  \underline{1.20} ($\pm$  \underline{0.87}) &  \underline{1.18} ($\pm$  \underline{2.39}) &  \underline{1.43} ($\pm$  \underline{1.49}) &  \underline{0.75} ($\pm$  \underline{0.95})   \\ 
		NIR ToF & &  \textbf{0.10} ($\pm$  \textbf{0.12}) &  \textbf{0.14} ($\pm$  0.15) &  1.02 ($\pm$  0.54) &  \textbf{0.28} ($\pm$  \textbf{0.19}) &  0.21 ($\pm$  0.22) &  0.41 ($\pm$  0.34)   \\ 
		Active Stereo & &  0.13 ($\pm$  \underline{0.25}) &  0.17 ($\pm$  \underline{0.51}) &  \textbf{0.29} ($\pm$  \textbf{0.32}) &  0.29 ($\pm$  0.32) &  \textbf{0.13} ($\pm$  \textbf{0.10}) &  \textbf{0.14} ($\pm$  \textbf{0.09})   \\ 
		Passive Stereo & &  \underline{0.20} ($\pm$  0.21) &  \underline{0.22} ($\pm$  0.22) &  0.49 ($\pm$  0.57) &  0.87 ($\pm$  0.63) &  0.17 ($\pm$  0.14) &  0.24 ($\pm$  0.15)   \\ 
		\hline 
		RF ToF & \multirowcell{4}{\mthree{}} 
		&  \textbf{0.17} ($\pm$  \textbf{0.23}) &  \textbf{0.19} ($\pm$  \textbf{0.28}) &  \underline{1.71} ($\pm$  \underline{1.47}) &  \underline{1.47} ($\pm$  \underline{2.94}) &  \underline{2.33} ($\pm$  \underline{2.79}) &  \underline{1.05} ($\pm$  \underline{1.41})   \\ 
		NIR ToF & &  \textbf{0.17} ($\pm$  0.33) &  0.23 ($\pm$  0.45) &  1.68 ($\pm$  1.29) &  \textbf{0.42} ($\pm$  \textbf{0.54}) &  0.30 ($\pm$  0.38) &  0.62 ($\pm$  0.64)   \\ 
		Active Stereo & &  0.26 ($\pm$  \underline{1.15}) &  \underline{0.46} ($\pm$  \underline{2.16}) &  \textbf{0.49} ($\pm$  \textbf{0.69}) &  0.55 ($\pm$  0.96) &  \textbf{0.20} ($\pm$  \textbf{0.28}) &  \textbf{0.22} ($\pm$  0.25)   \\ 
		Passive Stereo & &  \underline{0.33} ($\pm$  0.54) &  0.37 ($\pm$  0.62) &  0.76 ($\pm$  0.88) &  1.33 ($\pm$  1.35) &  0.27 ($\pm$  0.37) &  0.35 ($\pm$  \textbf{0.22})   \\ 
		\hline 
		RF ToF & \multirowcell{4}{\mfour{}} 
		&  0.17 ($\pm$  0.23) &  0.19 ($\pm$  0.28) &  \underline{1.79} ($\pm$  \underline{1.47}) &  1.62 ($\pm$  \underline{3.10}) &  \underline{2.55} ($\pm$  \underline{2.95}) &  \underline{1.05} ($\pm$  \underline{1.41})   \\ 
		NIR ToF & &  \textbf{0.08} ($\pm$  \textbf{0.11}) &  \textbf{0.16} ($\pm$  \textbf{0.22}) &  1.40 ($\pm$  0.76) &  \textbf{0.27} ($\pm$  \textbf{0.25}) &  0.23 ($\pm$  0.27) &  0.36 ($\pm$  0.37)   \\ 
		Active Stereo & &  0.18 ($\pm$  0.25) &  0.28 ($\pm$  0.45) &  \textbf{0.52} ($\pm$  0.74) &  0.58 ($\pm$  0.78) &  \textbf{0.16} ($\pm$  \textbf{0.16}) &  \textbf{0.15} ($\pm$  \textbf{0.19})   \\ 
		Passive Stereo & &  \underline{0.28} ($\pm$  \underline{0.40}) &  \underline{0.40} ($\pm$  \underline{0.58}) &  0.67 ($\pm$  \textbf{0.50}) &  \underline{1.71} ($\pm$  0.66) &  0.25 ($\pm$  0.21) &  0.44 ($\pm$  0.23)   \\ 
		\hline 
		RF ToF & \multirowcell{4}{\mthree{}*} 
		&  \textbf{-0.02} ($\pm$  \textbf{0.23}) &  \textbf{+0.00} ($\pm$  \textbf{0.28}) &  +1.56 ($\pm$  \underline{1.47}) &  \underline{+0.89} ($\pm$  \underline{2.94}) &  \underline{+2.08} ($\pm$  \underline{2.79}) &  \underline{+0.96} ($\pm$  \underline{1.41})   \\ 
		NIR ToF & &  \textbf{+0.02} ($\pm$  0.33) &  -0.01 ($\pm$  0.45) &  \underline{+1.66} ($\pm$  1.29) &  \textbf{+0.31} ($\pm$  \textbf{0.54}) &  +0.22 ($\pm$  0.38) &  +0.50 ($\pm$  0.64)   \\ 
		Active Stereo & &  -0.03 ($\pm$  \underline{1.15}) &  \underline{+0.15} ($\pm$  \underline{2.16}) &  \textbf{+0.40} ($\pm$  \textbf{0.69}) &  -0.32 ($\pm$  0.96) &  \textbf{+0.05} ($\pm$  \textbf{0.28}) &  \textbf{+0.18} ($\pm$  0.25)   \\ 
		Passive Stereo & &  \underline{+0.08} ($\pm$  0.54) &  \underline{+0.15} ($\pm$  0.62) &  +0.72 ($\pm$  0.88) &  \underline{-0.89} ($\pm$  1.35) &  +0.17 ($\pm$  0.37) &  +0.34 ($\pm$  \textbf{0.22})   \\ 
		\hline 
		RF ToF & \multirowcell{4}{\mfour{}*} 
		&  -0.01 ($\pm$  0.23) &  \textbf{+0.00} ($\pm$  0.28) &  \underline{+1.63} ($\pm$  \underline{1.47}) &  +0.96 ($\pm$  \underline{3.10}) &  \underline{+2.27} ($\pm$  \underline{2.95}) &  \underline{+0.96} ($\pm$  \underline{1.41})   \\ 
		NIR ToF & &  -0.02 ($\pm$  \textbf{0.11}) &  -0.04 ($\pm$  \textbf{0.22}) &  +1.37 ($\pm$  0.76) &  \textbf{+0.26} ($\pm$  \textbf{0.25}) &  +0.19 ($\pm$  0.27) &  +0.30 ($\pm$  0.37)   \\ 
		Active Stereo & &  \underline{-0.08} ($\pm$  0.25) &  -0.02 ($\pm$  0.45) &  \textbf{+0.39} ($\pm$  0.74) &  -0.55 ($\pm$  0.78) &  \textbf{+0.12} ($\pm$  \textbf{0.16}) &  \textbf{+0.14} ($\pm$  \textbf{0.19})   \\ 
		Passive Stereo & &  \textbf{-0.00} ($\pm$  \underline{0.40}) &  \underline{+0.05} ($\pm$  \underline{0.58}) &  +0.65 ($\pm$  \textbf{0.50}) &  \underline{-1.71} ($\pm$  0.66) &  +0.22 ($\pm$  0.21) &  +0.44 ($\pm$  0.23)   \\ 
		\bottomrule 
	\end{tabularx} 
\end{table*}
\begin{table*}[]
	\caption{\label{table:supp_metrics_4}%
		We measure the depth deviation with respect to \protect\mone{}\protect, \protect\mtwo{}\protect, \protect\mthree{}\protect, \protect\mfour{}\protect~and an additional signed version of \protect\mthree{}\protect,\protect\mfour{}\protect, which is denoted as \protect\mthree{}\protect*,\protect\mfour{}\protect*. All metrics are listed in the form $(\mu \pm \sigma)$, consisting of the mean $\mu$ and standard deviation $\sigma$ in centimeters, computed over the entire metric domain, respectively. The best results among all sensors of one metric type are highlighted in \bf{bold} and the worst results are \underline{underlined}.}
	\vspace*{-0.25 cm}
	\begin{tabularx}{\textwidth}{@{}lcYYYYYY@{}} 
		\toprule 
		& \multirowcell{2}{Metric \\ Type} & \multirowcell{2}{ V2 Christmas \\ Ball }  & \multirowcell{2}{ V3 Christmas \\ Ball }  & \multirowcell{2}{ Water Cube }  & \multirowcell{2}{ Flowerpot \\ (Brown) }  & \multirowcell{2}{ Brazen Rosette }  & \multirowcell{2}{ Pool Ball }  \\ 
		&&&&&&& \\ 
		\hline 
		\hline 
		RF ToF & \multirowcell{4}{\mone{}} 
		&  \underline{0.59} ($\pm$  \underline{0.40}) &  \underline{0.60} ($\pm$  \underline{0.41}) &  \textbf{0.16} ($\pm$  \textbf{0.11}) &  \underline{1.00} ($\pm$  \underline{0.90}) &  \textbf{0.11} ($\pm$  \textbf{0.08}) &  \underline{1.28} ($\pm$  \underline{0.78})   \\ 
		NIR ToF & &  \textbf{0.28} ($\pm$  0.18) &  0.47 ($\pm$  0.24) &  \underline{3.00} ($\pm$  \underline{0.44}) &  0.15 ($\pm$  \textbf{0.08}) &  \underline{0.94} ($\pm$  \underline{0.35}) &  0.64 ($\pm$  \textbf{0.17})   \\ 
		Active Stereo & &  0.51 ($\pm$  0.22) &  0.50 ($\pm$  0.20) &  0.56 ($\pm$  0.17) &  \textbf{0.12} ($\pm$  0.21) &  0.36 ($\pm$  0.24) &  \textbf{0.31} ($\pm$  0.25)   \\ 
		Passive Stereo & &  0.46 ($\pm$  \textbf{0.17}) &  \textbf{0.30} ($\pm$  \textbf{0.13}) &  0.46 ($\pm$  0.24) &  0.45 ($\pm$  0.25) &  0.18 ($\pm$  0.13) &  0.87 ($\pm$  0.30)   \\ 
		\hline 
		RF ToF & \multirowcell{4}{\mtwo{}} 
		&  \textbf{0.10} ($\pm$  \textbf{0.03}) &  \textbf{0.10} ($\pm$  \textbf{0.03}) &  \textbf{0.12} ($\pm$  \textbf{0.06}) &  \underline{0.53} ($\pm$  \underline{1.23}) &  \textbf{0.11} ($\pm$  \textbf{0.05}) &  \textbf{0.09} ($\pm$  \textbf{0.03})   \\ 
		NIR ToF & &  \underline{0.80} ($\pm$  \underline{0.80}) &  \underline{1.66} ($\pm$  \underline{1.65}) &  \underline{2.88} ($\pm$  \underline{0.54}) &  0.22 ($\pm$  0.27) &  \underline{8.11} ($\pm$  \underline{11.87}) &  0.69 ($\pm$  \underline{0.35})   \\ 
		Active Stereo & &  0.43 ($\pm$  0.23) &  0.39 ($\pm$  0.22) &  0.54 ($\pm$  0.27) &  \textbf{0.10} ($\pm$  \textbf{0.09}) &  0.46 ($\pm$  0.45) &  0.33 ($\pm$  0.27)   \\ 
		Passive Stereo & &  0.46 ($\pm$  0.21) &  0.39 ($\pm$  0.21) &  0.52 ($\pm$  0.28) &  0.49 ($\pm$  0.27) &  0.29 ($\pm$  0.29) &  \underline{0.71} ($\pm$  \underline{0.35})   \\ 
		\hline 
		RF ToF & \multirowcell{4}{\mthree{}} 
		&  \textbf{0.07} ($\pm$  \textbf{0.06}) &  \textbf{0.08} ($\pm$  \textbf{0.08}) &  \textbf{0.10} ($\pm$  \textbf{0.10}) &  \underline{0.77} ($\pm$  \underline{1.96}) &  \textbf{0.09} ($\pm$  \textbf{0.13}) &  \textbf{0.05} ($\pm$  \textbf{0.07})   \\ 
		NIR ToF & &  \underline{1.05} ($\pm$  \underline{1.29}) &  \underline{3.19} ($\pm$  \underline{6.90}) &  \underline{3.99} ($\pm$  \underline{0.84}) &  0.37 ($\pm$  0.46) &  \underline{9.03} ($\pm$  \underline{12.38}) &  1.06 ($\pm$  \underline{0.48})   \\ 
		Active Stereo & &  0.66 ($\pm$  0.24) &  0.64 ($\pm$  0.25) &  0.89 ($\pm$  0.35) &  \textbf{0.16} ($\pm$  \textbf{0.26}) &  0.72 ($\pm$  0.82) &  0.50 ($\pm$  0.38)   \\ 
		Passive Stereo & &  0.72 ($\pm$  0.22) &  0.59 ($\pm$  0.23) &  0.84 ($\pm$  0.35) &  0.67 ($\pm$  0.37) &  0.39 ($\pm$  0.48) &  \underline{1.15} ($\pm$  0.30)   \\ 
		\hline 
		RF ToF & \multirowcell{4}{\mfour{}} 
		&  \textbf{0.07} ($\pm$  \textbf{0.06}) &  \textbf{0.08} ($\pm$  \textbf{0.08}) &  \textbf{0.10} ($\pm$  0.10) &  0.78 ($\pm$  \underline{1.97}) &  \textbf{0.09} ($\pm$  \textbf{0.13}) &  \textbf{0.05} ($\pm$  \textbf{0.07})   \\ 
		NIR ToF & &  0.36 ($\pm$  \underline{0.57}) &  \underline{1.77} ($\pm$  \underline{2.11}) &  \underline{4.97} ($\pm$  \underline{0.27}) &  0.15 ($\pm$  \textbf{0.11}) &  \underline{15.00} ($\pm$  \underline{13.89}) &  0.96 ($\pm$  0.18)   \\ 
		Active Stereo & &  0.75 ($\pm$  0.21) &  0.73 ($\pm$  0.15) &  0.86 ($\pm$  0.19) &  \textbf{0.10} ($\pm$  0.12) &  0.76 ($\pm$  0.78) &  0.52 ($\pm$  \underline{0.37})   \\ 
		Passive Stereo & &  \underline{0.85} ($\pm$  0.16) &  0.70 ($\pm$  0.18) &  1.17 ($\pm$  \textbf{0.05}) &  \underline{0.81} ($\pm$  0.25) &  0.40 ($\pm$  0.44) &  \underline{1.29} ($\pm$  0.18)   \\ 
		\hline 
		RF ToF & \multirowcell{4}{\mthree{}*} 
		&  \textbf{-0.05} ($\pm$  \textbf{0.06}) &  \textbf{-0.05} ($\pm$  \textbf{0.08}) &  \textbf{+0.08} ($\pm$  \textbf{0.10}) &  \underline{+0.67} ($\pm$  \underline{1.96}) &  \textbf{-0.02} ($\pm$  \textbf{0.13}) &  \textbf{+0.01} ($\pm$  \textbf{0.07})   \\ 
		NIR ToF & &  +0.68 ($\pm$  \underline{1.29}) &  \underline{+3.08} ($\pm$  \underline{6.90}) &  \underline{+3.99} ($\pm$  \underline{0.84}) &  +0.36 ($\pm$  0.46) &  \underline{+9.01} ($\pm$  \underline{12.38}) &  +1.05 ($\pm$  \underline{0.48})   \\ 
		Active Stereo & &  +0.66 ($\pm$  0.24) &  +0.63 ($\pm$  0.25) &  +0.88 ($\pm$  0.35) &  \textbf{-0.07} ($\pm$  \textbf{0.26}) &  -0.62 ($\pm$  0.82) &  +0.48 ($\pm$  0.38)   \\ 
		Passive Stereo & &  \underline{+0.72} ($\pm$  0.22) &  +0.59 ($\pm$  0.23) &  +0.83 ($\pm$  0.35) &  +0.66 ($\pm$  0.37) &  -0.23 ($\pm$  0.48) &  \underline{+1.15} ($\pm$  0.30)   \\ 
		\hline 
		RF ToF & \multirowcell{4}{\mfour{}*} 
		&  \textbf{-0.05} ($\pm$  \textbf{0.06}) &  \textbf{-0.05} ($\pm$  \textbf{0.08}) &  \textbf{+0.08} ($\pm$  0.10) &  +0.68 ($\pm$  \underline{1.97}) &  \textbf{-0.02} ($\pm$  \textbf{0.13}) &  \textbf{+0.01} ($\pm$  \textbf{0.07})   \\ 
		NIR ToF & &  +0.17 ($\pm$  \underline{0.57}) &  \underline{+1.72} ($\pm$  \underline{2.11}) &  \underline{+4.97} ($\pm$  \underline{0.27}) &  +0.15 ($\pm$  \textbf{0.11}) &  \underline{+14.97} ($\pm$  \underline{13.89}) &  +0.96 ($\pm$  0.18)   \\ 
		Active Stereo & &  +0.75 ($\pm$  0.21) &  +0.73 ($\pm$  0.15) &  +0.86 ($\pm$  0.19) &  \textbf{-0.06} ($\pm$  0.12) &  -0.70 ($\pm$  0.78) &  +0.52 ($\pm$  \underline{0.37})   \\ 
		Passive Stereo & &  \underline{+0.85} ($\pm$  0.16) &  +0.70 ($\pm$  0.18) &  +1.17 ($\pm$  \textbf{0.05}) &  \underline{+0.81} ($\pm$  0.25) &  -0.30 ($\pm$  0.44) &  \underline{+1.29} ($\pm$  0.18)   \\ 
		\bottomrule 
	\end{tabularx} 
	\newline 
	\newline 
	\begin{tabularx}{\textwidth}{@{}lcYYY@{}} 
		\toprule 
		& \multirowcell{2}{Metric \\ Type} & \multirowcell{2}{ Polystyrene Plate }  & \multirowcell{2}{ Bunny Box }  & \multirowcell{2}{ Bunny }  \\ 
		&&&& \\ 
		\hline 
		\hline 
		RF ToF & \multirowcell{4}{\mone{}} 
		&  \textbf{2.10} ($\pm$  2.38) &  \textbf{0.32} ($\pm$  \textbf{0.28}) &  \underline{0.28} ($\pm$  \underline{0.22})   \\ 
		NIR ToF & &  \underline{3.28} ($\pm$  \textbf{2.15}) &  \underline{0.50} ($\pm$  0.29) &  0.26 ($\pm$  0.14)   \\ 
		Active Stereo & &  3.01 ($\pm$  \underline{3.05}) &  0.39 ($\pm$  0.40) &  0.12 ($\pm$  0.10)   \\ 
		Passive Stereo & &  2.99 ($\pm$  2.76) &  0.39 ($\pm$  \underline{0.62}) &  \textbf{0.08} ($\pm$  \textbf{0.05})   \\ 
		\hline 
		RF ToF & \multirowcell{4}{\mtwo{}} 
		&  \textbf{0.14} ($\pm$  \textbf{0.07}) &  \textbf{0.30} ($\pm$  \textbf{0.26}) &  \underline{0.50} ($\pm$  \underline{0.41})   \\ 
		NIR ToF & &  2.50 ($\pm$  0.81) &  0.48 ($\pm$  0.37) &  0.31 ($\pm$  0.20)   \\ 
		Active Stereo & &  1.74 ($\pm$  1.97) &  0.37 ($\pm$  0.35) &  \textbf{0.12} ($\pm$  \textbf{0.09})   \\ 
		Passive Stereo & &  \underline{2.92} ($\pm$  \underline{4.87}) &  \underline{0.61} ($\pm$  \underline{1.08}) &  0.13 ($\pm$  0.13)   \\ 
		\hline 
		RF ToF & \multirowcell{4}{\mthree{}} 
		&  \textbf{0.12} ($\pm$  \textbf{0.10}) &  1.32 ($\pm$  2.45) &  \underline{0.74} ($\pm$  0.69)   \\ 
		NIR ToF & &  \underline{17.95} ($\pm$  \underline{24.15}) &  \underline{1.91} ($\pm$  \underline{2.83}) &  0.45 ($\pm$  \underline{1.77})   \\ 
		Active Stereo & &  10.63 ($\pm$  15.21) &  \textbf{0.99} ($\pm$  \textbf{1.32}) &  \textbf{0.16} ($\pm$  \textbf{0.20})   \\ 
		Passive Stereo & &  10.08 ($\pm$  11.66) &  1.05 ($\pm$  1.88) &  0.17 ($\pm$  0.27)   \\ 
		\hline 
		RF ToF & \multirowcell{4}{\mfour{}} 
		&  \textbf{0.12} ($\pm$  \textbf{0.10}) &  1.12 ($\pm$  2.15) &  \underline{0.74} ($\pm$  \underline{0.69})   \\ 
		NIR ToF & &  \underline{17.86} ($\pm$  \underline{24.12}) &  \textbf{0.93} ($\pm$  \textbf{0.87}) &  0.38 ($\pm$  \textbf{0.12})   \\ 
		Active Stereo & &  10.48 ($\pm$  15.02) &  1.46 ($\pm$  1.54) &  0.11 ($\pm$  \textbf{0.12})   \\ 
		Passive Stereo & &  9.81 ($\pm$  11.41) &  \underline{2.68} ($\pm$  \underline{2.54}) &  \textbf{0.10} ($\pm$  \textbf{0.12})   \\ 
		\hline 
		RF ToF & \multirowcell{4}{\mthree{}*} 
		&  \textbf{+0.11} ($\pm$  \textbf{0.10}) &  +1.15 ($\pm$  2.45) &  \underline{+0.69} ($\pm$  0.69)   \\ 
		NIR ToF & &  \underline{+17.95} ($\pm$  \underline{24.15}) &  \underline{+1.73} ($\pm$  \underline{2.83}) &  -0.24 ($\pm$  \underline{1.77})   \\ 
		Active Stereo & &  +10.63 ($\pm$  15.21) &  -0.83 ($\pm$  \textbf{1.32}) &  -0.10 ($\pm$  \textbf{0.20})   \\ 
		Passive Stereo & &  +10.08 ($\pm$  11.66) &  \textbf{-0.78} ($\pm$  1.88) &  \textbf{+0.08} ($\pm$  0.27)   \\ 
		\hline 
		RF ToF & \multirowcell{4}{\mfour{}*} 
		&  \textbf{+0.11} ($\pm$  \textbf{0.10}) &  +1.03 ($\pm$  2.15) &  \underline{+0.69} ($\pm$  \underline{0.69})   \\ 
		NIR ToF & &  \underline{+17.86} ($\pm$  \underline{24.12}) &  \textbf{+0.92} ($\pm$  \textbf{0.87}) &  -0.38 ($\pm$  \textbf{0.12})   \\ 
		Active Stereo & &  +10.48 ($\pm$  15.02) &  -1.46 ($\pm$  1.54) &  -0.08 ($\pm$  \textbf{0.12})   \\ 
		Passive Stereo & &  +9.81 ($\pm$  11.41) &  \underline{-2.48} ($\pm$  \underline{2.54}) &  \textbf{+0.07} ($\pm$  \textbf{0.12})   \\ 
		\bottomrule 
	\end{tabularx}  
\end{table*}

	\fi

\end{document}